\documentclass[manuscript,screen]{acmart}

\AtBeginDocument{%
  \providecommand\BibTeX{{%
    \normalfont B\kern-0.5em{\scshape i\kern-0.25em b}\kern-0.8em\TeX}}}

\usepackage{makecell}
\usepackage{adjustbox}
\newcommand{\tabitem}{~~\llap{\textbullet}~~}
\usepackage{tabularx}
\usepackage{multirow}
\usepackage{multicol}

\usepackage{pifont}

\usepackage{color, xcolor}
\usepackage{enumitem}
\newcommand{\cmark}{\ding{51}}

\usepackage{graphicx}
\usepackage{subfigure}

\begin{document}

\title{Interest Flooding Attacks in Named Data Networking: Survey of Existing Solutions, Open Issues, Requirements and Future Directions (Extended version)}

\author{Ahmed Benmoussa}
\email{ah.benmoussa@lagh-univ.dz}
\author{Chaker Abdelaziz Kerrache}
\email{ch.kerrache@lagh-univ.dz}
\author{Nasreddine Lagraa}
\email{n.lagraa@lagh-univ.dz}
\affiliation{
  \institution{Laboratoire d'Informatique et de Math\'ematiques, Universit\'e Amar Telidji de Laghouat}
  %\institution{University of Laghouat}
  \city{Laghouat}
  \country{Algeria}}

\author{Spyridon Mastorakis}
\email{smastorakis@unomaha.edu}
\affiliation{
 \institution{Department of Computer Science, College of Information Science \& Technology, University of Nebraska}
 %\institution{University of Nebraska at Omaha}
 \city{Omaha}
 \state{Nebraska}
 \country{USA}}

\author{Abderrahmane Lakas}
\email{alakas@uaeu.ac.ae}
\affiliation{
  \institution{College of Information Technology, United Arab Emirates University}
  \state{Al Ain}
  \country{UAE}}

\author{Abdou el Karim Tahari}
\email{k.tahari@lagh-univ.dz}
\affiliation{
  %\institution{Laboratoire d'Informatique et de Math\'ematiques, Department of Computer Science, University of Laghouat}
  \institution{Universit\'e Amar Telidji de Laghouat}
  \city{Laghouat}
  \country{Algeria}}

\renewcommand{\shortauthors}{Benmoussa et al.}

\begin{abstract}
  Named Data Networking (NDN) is a prominent realization of the vision of Information-Centric Networking. The NDN architecture adopts name-based routing and location-independent data retrieval. Among other important features, NDN integrates security mechanisms and focuses on protecting the content rather than the communications channels. Along with a new architecture come new threats and NDN is no exception. NDN is a potential target for new network attacks such as Interest Flooding Attacks (IFAs). Attackers take advantage of IFA to launch (D)DoS attacks in NDN. Many IFA detection and mitigation solutions have been proposed in the literature. However, there is no comprehensive review study of these solutions that has been proposed so far. Therefore, in this paper, we propose a survey of the various IFAs with a detailed comparative study of all the relevant proposed solutions as counter-measures against IFAs. We also review the requirements for a complete and efficient IFA solution and pinpoint the various issues encountered by IFA detection and mitigation mechanisms through a series of attack scenarios. Finally, in this survey, we offer an analysis of the open issues and future research directions regarding IFAs. This manuscript consists of an extended version of the paper published in \textit{ACM Computing Surveys}: \url{https://dl.acm.org/doi/10.1145/3539730}.
\end{abstract}



\maketitle

\section{Introduction}
\label{sec:intro}
A long time has passed since the creation of the Internet. Back then, the goal was to interconnect pairs of hosts. With the constant growth of the connected devices that will reach almost 30 billion in 2023~\cite{ciscorep}, which represents roughly three times the global population, the actual Internet architecture was not designed for such massive numbers of hosts. In addition, Internet usage itself has changed. Users are interested in the content to retrieve rather than its source.
In addition, security was an afterthought when the Internet was created, which opened the door to security and privacy issues. 
Taking all these challenges in mind, the need for a new, suitable, and secure Internet architecture is essential. The research community started the discussion on a new architecture about three decades ago~\cite{rfc1287}. Since then, many Future Internet Architectures (FIAs) were proposed like Xtensible Internet Architecture (XIA)~\cite{anand2011xia}, Nebula~\cite{anderson2013nebula}, and others. However, the most promising FIA architecture candidate is Information Centric Networking (ICN)~\cite{ahlgren2012survey,xylomenos2013survey}. Several architectures were proposed under the umbrella of ICN, like Data-Oriented Network Architecture (DONA)~\cite{koponen2007data}, Publish-Subscribe Internet Technology (PURSUIT)~\cite{fotiou2010developing}, Scalable \& Adaptive Internet soLutions (SAIL), COntent Mediator architecture for content-aware nETworks (COMET)~\cite{garcia2011comet}, and MobilityFirst~\cite{seskar2011mobilityfirst}. But the most promising information-centric architecture is Named Data Networking (NDN). NDN is a project funded by the US Future Internet Architecture in 2010 and maintained at UCLA~\cite{zhang2010named}. NDN takes its roots from the Content-Centric Networking (CCN)~\cite{jacobson2007content}.  

NDN adopts a content-driven communication approach where packet forwarding is based on data names rather than IP addresses. NDN also provides features such as in-network caching, built-in multicast, mobility support, and native security mechanisms. 
NDN focuses on securing the content rather than the communication channels. NDN mandates the use of data signatures, which permits users to retrieve any available piece of content no matter where it comes from, as long as the signature can be verified. Although NDN integrates security mechanisms, it is still not immune to certain new security and privacy issues~\cite{tourani2017security,khelifi2018security}. One of these network threats is related to Interest Flooding Attacks (IFAs). IFA is a new type of attack that adversaries use to launch (D)DoS attacks in NDN. 

This survey provides an in-depth study of IFA and gives a broad analysis of IFA solutions that have been proposed so far before pointing out the open issues and providing the research directions that need to be considered in the future. Before we dive deep into the subject, we first review existing surveys and discuss the importance and novelty of our survey compared to the literature.

\subsection{Related Surveys}
Several surveys about NDN and ICN security issues have been authored. Some of them briefly discussed IFA, while others detailed IFA. In the following subsection, we summarize and explain the differences between each authored survey. 
The authors of \cite{abdallah2015survey} surveyed several ICN attacks and grouped them into four different categories: naming, routing, caching, and miscellaneous. The survey briefly discussed IFA and the number of cited solutions is very limited.
Similarly, the authors of \cite{tourani2017security} classified ICN threats into three categories: security, privacy, and access control. The authors defined IFA as a DoS attack and reviewed some countermeasures. However, the authors did not compare in detail the reviewed works.
The survey in \cite{mannes2019naming} discussed security threats and vulnerabilities in ICN. It classified the attacks into three categories: security in content, routers, and caches. It discussed IFA and presented several IFA solutions. Despite lacking many relevant works, this survey did not detail the cited solution, and the provided review does not mention their drawbacks.
The authors of \cite{aamir2015denial} presented a survey about DoS attacks in CCN and classified them into three categories: flooding, forced computation, and cache/content manipulation. The authors discussed IFA and gave a detailed review of some countermeasure solutions. However, this survey did not talk about their drawbacks. Also, this survey does not compare the cited works. 
The authors in \cite{lutz2016security} stated the benefits of the CCN paradigm on some security and privacy threats. Then, they outlined the actual challenges that need to be corrected. They talked about IFA and mentioned few countermeasure solutions without giving any comparison.
The survey in \cite{ambrosin2018security} compared four FIA architectures: Nebula, NDN, MobilityFirst, and XIA in terms of integrity, confidentiality, availability, authentication, trust, access control, and anonymity. This paper shortly discussed IFA and the number of cited works is very limited. Similarly, the survey \cite{chen2015survey} provides a short review about some NDN security issues before briefly discussing IFA without mentioning any solution. 
The authors of~\cite{buragohaindemystifying} classified NDN attacks according to the targeted layer. They classified IFA as an application layer attack and presented a few countermeasure solutions. However, the list of covered solutions is poor and misses recent and relevant works.
In another context, the authors of \cite{khelifi2018security} presented different security and privacy issues in Vehicular NDN (VNDN). This paper shortly discussed IFA and mentioned only two countermeasure solutions.   
The work presented in \cite{gasti2018content} focuses on IFA and privacy attacks in NDN/CCN. It classified solutions into simple techniques, Anomaly/Attack detection, and PIT-less routing. The provided analysis and the list of IFA work are limited. It lacks recent and relevant works.
The authors of \cite{rai2018survey} and \cite{rai2019survey} focused on IFA and gave a short comparison between some existing solutions. However, the authors did not review in detail cited works. Also, these two surveys miss many recent and relevant works.
The survey in \cite{kumar2019security} focuses on IFA and cache/content attacks. It described IFA and its variants before reviewing and classifying several solutions. Although it provides a long list of IFA solutions, this survey does not cover the full spectrum of IFA works and misses many recent and relevant works. Also, the authors did not review in detail cited solutions. We summarized the above surveys in Table~\ref{tab:surveys}.

\color{black}
\begin{table}
\centering
\scriptsize
\caption{Related Surveys}
\label{tab:surveys}
\begin{adjustbox}{width=\textwidth}
\begin{tabular}{lcllc}
\hline
\textbf{Ref} & \textbf{Year} & \textbf{Main focus} & \textbf{Limitations} & \textbf{\makecell{Covered IFA\\solutions}} \\ \hline

\cite{aamir2015denial} & 2015 & DoS attacks in CCN & \makecell[l]{
    \tabitem Does not compare cited IFA solutions.\\
    \tabitem Does not talk about the drawbacks of cited solutions.} & 09 \\ \hline
    
\cite{abdallah2015survey} & 2015 & ICN attacks & \makecell[l]{
    \tabitem Briefly talks about IFA.\\
    \tabitem The number of IFA solution is limited.} & 04 \\ \hline
    
\cite{chen2015survey} & 2015 & \makecell[l]{Security issues\\in NDN} & \makecell[l]{ \tabitem Do not mention IFA countermeasure solutions.} & 00 \\ \hline

\cite{lutz2016security} & 2016 & \makecell[l]{Security and privacy\\challenges in CCN} & \makecell[l]{
    \tabitem Does not talk about the drawbacks of cited solutions.\\
    \tabitem Did not provide a comparison between cited IFA solutions.} & 08 \\ \hline
    
\cite{tourani2017security} & 2017 & ICN attacks & \makecell[l]{\tabitem Does not compare in detail cited IFA solutions} & 08 \\ \hline

\cite{ambrosin2018security} & 2018 & \makecell[l]{Security and privacy\\in FIA architectures} & \makecell[l]{
    \tabitem IFA was very shortly discussed.\\
    \tabitem Very limited IFA solutions.} & 03 \\ \hline

\cite{gasti2018content} & 2018 & \makecell[l]{IFA and privacy\\in NDN/CCN} & \makecell[l]{
    \tabitem Does not compare cited solutions.\\
    \tabitem The paper misses relevant works.} & 06 \\ \hline

\cite{khelifi2018security} & 2018 & \makecell[l]{Security and privacy\\issues in VNDN} & \makecell[l]{
    \tabitem Briefly discuss IFA.\\
    \tabitem The number of cited solutions is very limited.} & 02 \\ \hline

\cite{rai2018survey} & 2018 & IFA in NDN & \makecell[l]{
    \tabitem The survey misses a lot of relevant works.\\
    \tabitem The comparison given is very limited.\\
    \tabitem Does not mention the drawbacks of solutions.\\} & 10 \\ \hline

\cite{kumar2019security} & 2019 & \makecell[l]{Security threats\\in NDN} & \makecell[l]{
    \tabitem The survey misses some recent relevant works.\\
    \tabitem Does not compare in details the cited solutions.\\
    \tabitem Does not talk about the drawbacks of every solutions.\\} & 21 \\ \hline

\cite{mannes2019naming} & 2019 & \makecell[l]{Security threats\\in ICN} & \makecell[l]{
    \tabitem Does not detail cited solution.\\
    \tabitem Does not mention solutions' drawbacks.\\
    \tabitem Misses many recent relevant research.} & 10 \\ \hline
    
\cite{rai2019survey} & 2019 & IFA in NDN & \makecell[l]{
    \tabitem Does not compare between the cited solutions.\\
    %\tabitem The number of covered solutions is low.\\
    %\tabitem Very limited for an IFA specific survey.\\
    \tabitem Misses many recent relevant research.} & 06 \\ \hline
        
\cite{buragohaindemystifying} & 2020 & \makecell[l]{Security threats\\in NDN} & \makecell[l]{ 
    \tabitem Briefly discuss IFA.\\
    \tabitem The number of cited solutions is very limited.} & 05 \\ \hline
    
Our & 2021 & IFA in NDN & & 43 \\ \hline
\end{tabular}
\end{adjustbox}
\end{table}

\subsection{Motivation and Goal of the paper}
Although the previously authored surveys talk about IFA and some countermeasure solutions, they only scratch the surface of the subject. These surveys fail to cover all the aspects of the attack. Therefore, the authors only focused on presenting the solutions without deep analysis and comparison. In addition, the mentioned surveys do not study the full spectrum of IFA variants and scenarios. They generally describe the conventional version of the attack. Moreover, the present surveys do not cover all the works present in the literature. For instance, the majority of recent (and relevant) works were not considered by these surveys.  
Considering all these limitations, there is a need for a survey that provides an in-depth study of IFA to bridge this gap. There is a need for a paper that provides a basic understanding of the attack, its variants, its characteristics, and the different techniques used to conduct such attacks. Furthermore, there is a need to provide a systematic and detailed review of all the countermeasure works present in the literature. In addition, a comparative study needs to be conducted to unravel the strength and weaknesses of each solution. Finally, there is a need to pinpoint the actual open issues and the lessons learned for future research directions. 

\subsection{Our Contributions}
Existing surveys in the literature are not IFA specific or do not discuss and compare in detail related IFA solutions. To the best of our knowledge, our work is the first attempt that comprehensively and systematically review all the aspects of the Interest Flooding Attack. The highlights of the various aspects covered in this work are summarized below:
\begin{itemize}
    \item Our survey provides a comprehensive discussion on state-of-the-art IFA research and an exhaustive research taxonomy of IFA by considering its working principles and approaches.

    \item Our survey reviews all the relevant IFA countermeasure solutions. It discusses the drawbacks of each one and presents a workflow that every IFA countermeasure solution follows.
    
    \item Our survey provides a comprehensive comparison of existing solutions. This comparison takes into consideration different aspects: the type of collected information, the calculated metrics, the detection parameters, and the defensive actions. 
    
    \item Our survey provides an extensive list of non-conventional attack scenarios that existing works cannot handle and existing literature has not considered. Il also presents directions for future research associated with designing efficient and robust IFA solutions. 
\end{itemize}

\subsection{Methodology of Survey on Interest Flooding Attack in Named Data Networking}
\begin{figure}
    \centering
    \begin{subfigure}[Distribution of cited works per year]{
    \includegraphics[width=.48\linewidth] {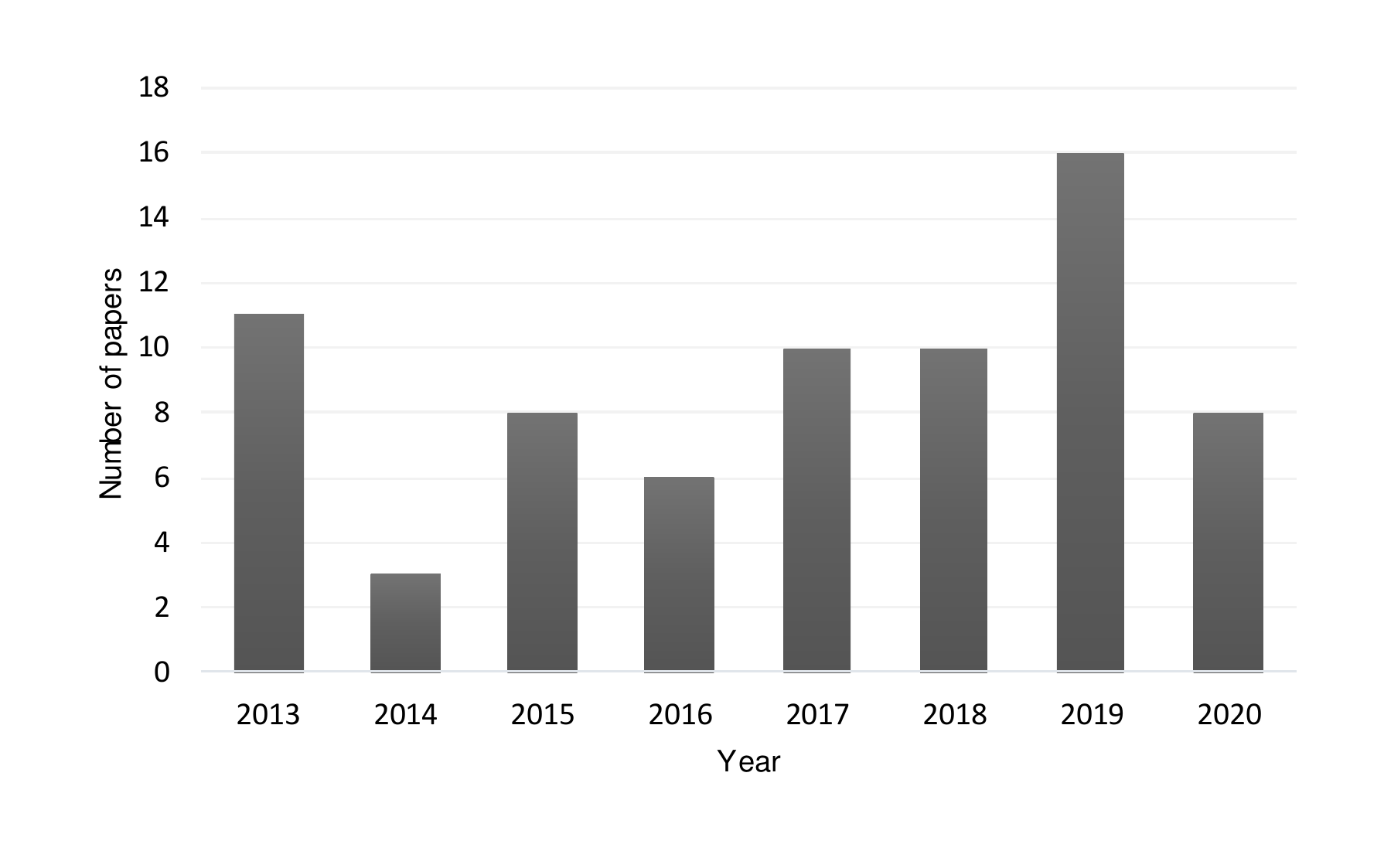}
    \label{fig:year_dist}}
    \end{subfigure}
    \begin{subfigure}[Distribution of cited works per publisher]{
    \includegraphics[width=.48\linewidth] {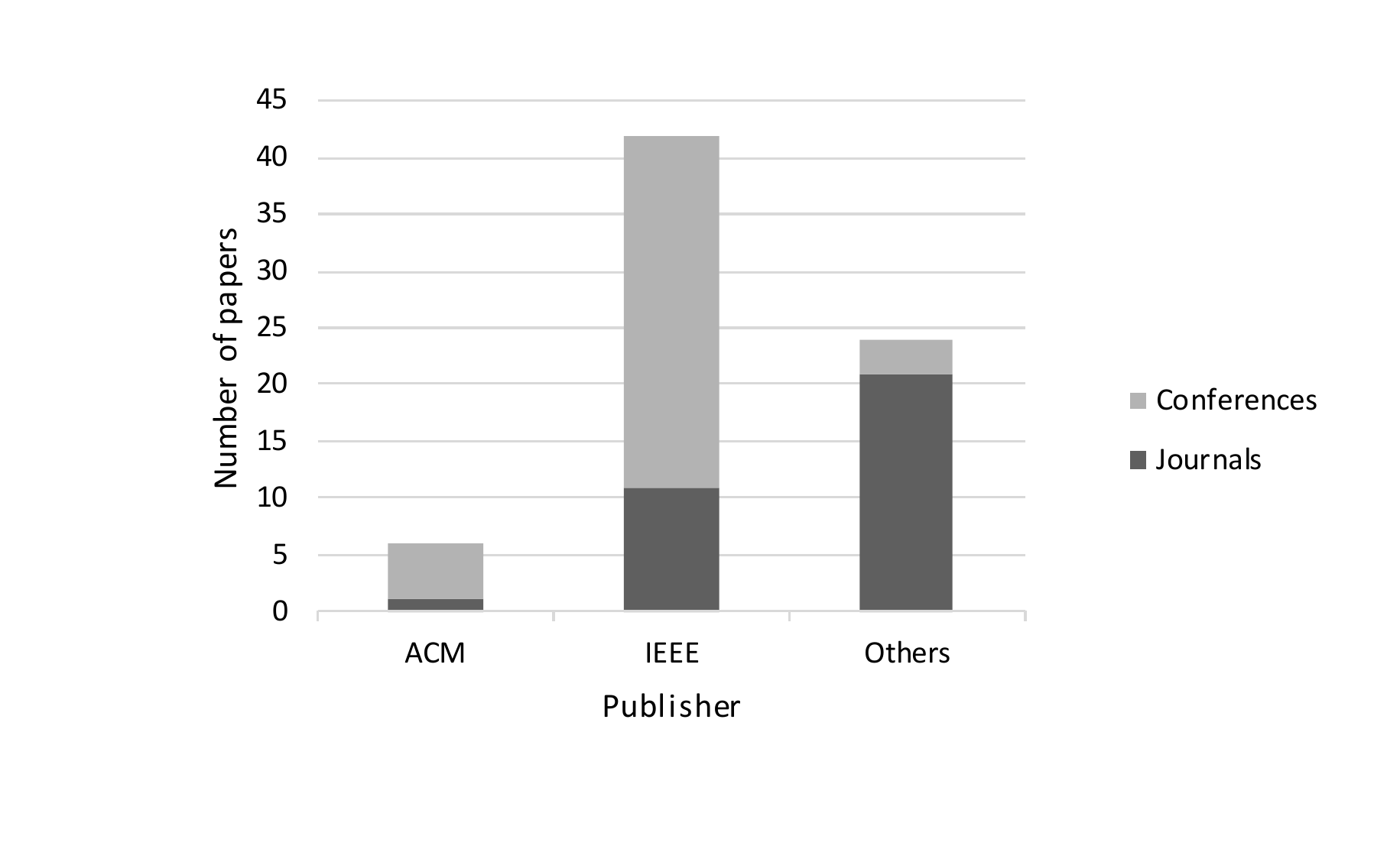}
    \label{fig:publisher_dist}}
    \end{subfigure}
    \caption{Statistics on cited IFA research papers}
\end{figure}

In this subsection, we present the methodology that we utilized during our review of the state-of-the-art Interest Flooding Attack. The process that we adopted is described below: 
\begin{itemize}
    \item \textbf{Surveyed Databases}
    To achieve our objective and cover all the relevant IFA-related research, we have collected papers on domain-relevant electronic databases, including \textit{ACM Digital Library}, \textit{IEEE Xplore}, \textit{Science Direct}, \textit{Springer Link}, \textit{Wiley Online Library}, and \textit{Hindawi}. Furthermore, we have collected articles related to this domain from \textit{arXiv}, \textit{Google Scholar}, and \textit{ResearchGate}. Figure~\ref{fig:publisher_dist} illustrates the distribution of collected IFA works depending on their publisher. 

    \item \textbf{Filtering and Paper Selection Process}
    The filtering process begins by categorizing each paper depending on its nature: a study paper, a survey paper, or a solution. The study papers category includes any IFA based study authored. It may include studies on the effects of IFA or a comparative study between some solutions, …etc. Regarding survey papers, we chose to cite any survey that mentions (briefly or extensively) IFA, whether it is NDN specific or an ICN survey. However, for the authored solutions, we chose a different approach. Before we cite a paper in this survey, we first read the proposed technique and then verify if we have already reviewed a similar solution. If a similar solution exists, we compare the authors of these two papers. If they are the same, we choose to cite only the most recent one (e.g., the authors presented it in a conference paper and then published the same solution in an article paper). However, if the authors are different, we choose to cite the oldest research paper. On the other hand, If a given paper presents an enhanced or a modified technique to another published solution, we cite them consecutively. 
    During our research, we collected 97 IFA based works. In the end, we chose 72 papers. Figure~\ref{fig:year_dist} illustrates the distribution of cited works since the research community started studying IFA. 
\end{itemize}

\subsection{Organization of the Survey}
We organize this survey in a top-down manner. Table \ref{tab:acronym} lists all the acronyms used in this paper. We begin with an overview of the NDN architecture in section \ref{sec:ndn}. We talk about the key features of this architecture before concluding this section with its security components. Following that, we introduce the NDN security requirements in section \ref{sec:requirements} before talking about the availability attacks that target the NDN architecture in section \ref{sec:available}. We discuss each attack and finish this section by briefly presenting the Interest Flooding Attack. In section \ref{sec:ifa}, we dive into the main subject of this survey and detail IFA. We show the characteristics of the attack and its variants before giving an overview of the IFA related studies present in the literature. Section \ref{sec:related_work} focuses on IFA solutions that were authored and gives a detailed discussion on each solution. Section \ref{sec:cda} introduces the Collect-Detect-Act workflow (CDA). This section presents an in-depth study of state-of-the-art solutions from different perspectives. Section \ref{sec:issues} discusses unconsidered attacking scenarios. This section proposes several attacking scenarios that were not considered by existing solutions. Section \ref{sec:future} highlights the lessons learned and future research directions that need to be considered when designing solutions. Section \ref{sec:conclusion} concludes this survey. This paper includes an appendix section in which we compare existing solutions in terms of simulation parameters and evaluation metrics the they adopted.  

\begin{table}
    \caption{Acronyms used in this paper}
    \scriptsize
    \label{tab:acronym}
    \centering
    \begin{adjustbox}{width=13cm}
    \begin{tabular}{llll}
    \hline
    \textbf{Acronym} & \textbf{Explanation} & \textbf{Acronym} & \textbf{Explanation} \\ \hline
    \hline
    NDN & Named Data Networking & SPOF & Single Point Of Failure \\ \hline
    CCN & Content-Centric Networking & AI & Artificial Intelligence \\ \hline
    ICN & Information-Centric Networking & NN & Neural Network \\ \hline
    FIA & Future Internet Architecture & RBF & Radial Basis Function \\ \hline
    PIT & Pending Interest Table & SVM & Support Vector Machine \\ \hline
    FIB & Forwarding Information Base & ndnSIM & NDN Simulator \\ \hline
    CS & Content Store & ipps & interest packet per second \\ \hline
    IFA & Interest Flooding Attack & FIFO & First In First Out \\ \hline
    DDoS & Distributed Denial of Service & LRU & Least Recently Used \\ \hline
    HMM & Hidden Markov Model & LFU & Least Frequently Used \\ \hline
    AS & Autonomous System & DC & Domain Controller \\ \hline
    Mon & Monitoring & CDA & Collect Detect Act \\ \hline
    LRT & Likelihood Radio Test & MLP & Multilayer Perceptron \\ \hline
    MANET & Mobile Ad-hoc Network & ARI & Autoregressive Integrated \\ \hline 
    NFD & NDN Forwarding Daemon & XIA & Xtensible Internet Architecture \\ \hline
    \end{tabular}
    \end{adjustbox}
\end{table}

\section{Named Data Networking: Architecture Design}
\label{sec:ndn}
NDN is a data-centric Internet architecture designed to replace the host-centric TCP/IP architecture \cite{zhang2014named, al2022promise, al2022reservoir}. NDN falls under the umbrella of Information Centric Networking (ICN) where the focus is on the data rather than its location. NDN defines two entities: \textit{Producer} and \textit{Consumer}. Producers generate and offer content for the consumers to request. To fetch for a content, consumers send the name of the desired data into an \textit{Interest} packet. NDN adopts a hierarchical naming scheme to identify contents~\cite{afanasyev2017ndns}. 

\subsection{NDN Protocol Stack}
The NDN protocol stack is composed of four different layers: application, network, link and physical layers \cite{zhang2018ndn}. The application layer supports the operation of NDN applications and it also embeds transport protocols as system libraries. The network layer’s role is to route \textit{Interest} and \textit{Data} packets. It uses the application layer's names to route the packets. The NDN link-layer supports a set of protocols like Ethernet and can use virtual links like IP and TCP overlays. Figure~\ref{fig:ndn_stack} shows the differences between the TCP/IP and NDN protocol stacks. 

\subsection{NDN Packet Types}
\begin{figure}
\centering
\begin{subfigure}[TCP/IP and NDN Protocol Stacks]{
    \includegraphics[width=.45\linewidth] {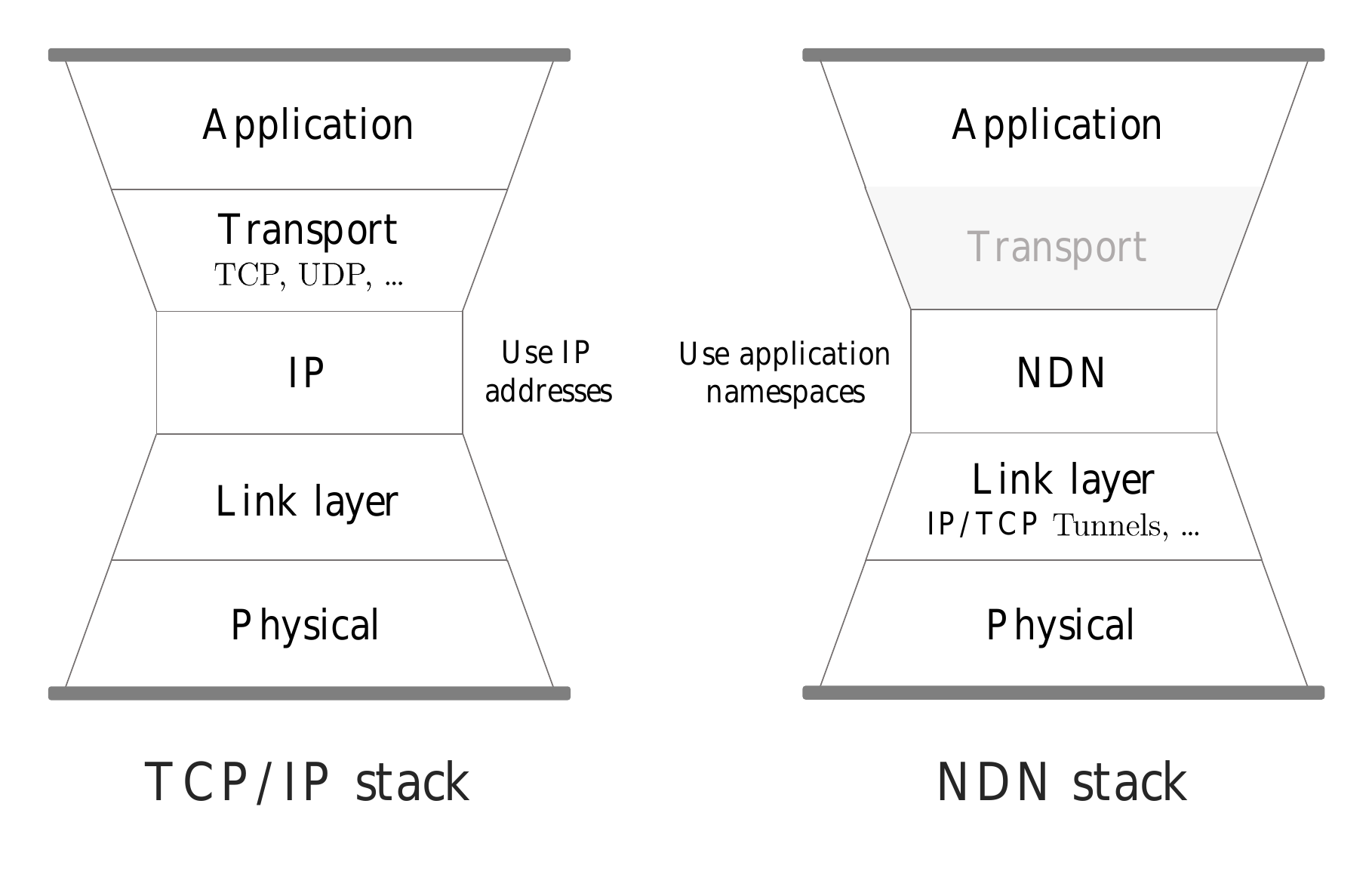}
    \label{fig:ndn_stack}}
    \end{subfigure}
    \hfill
\begin{subfigure}[NDN packets]{
    \includegraphics[width=.45\linewidth] {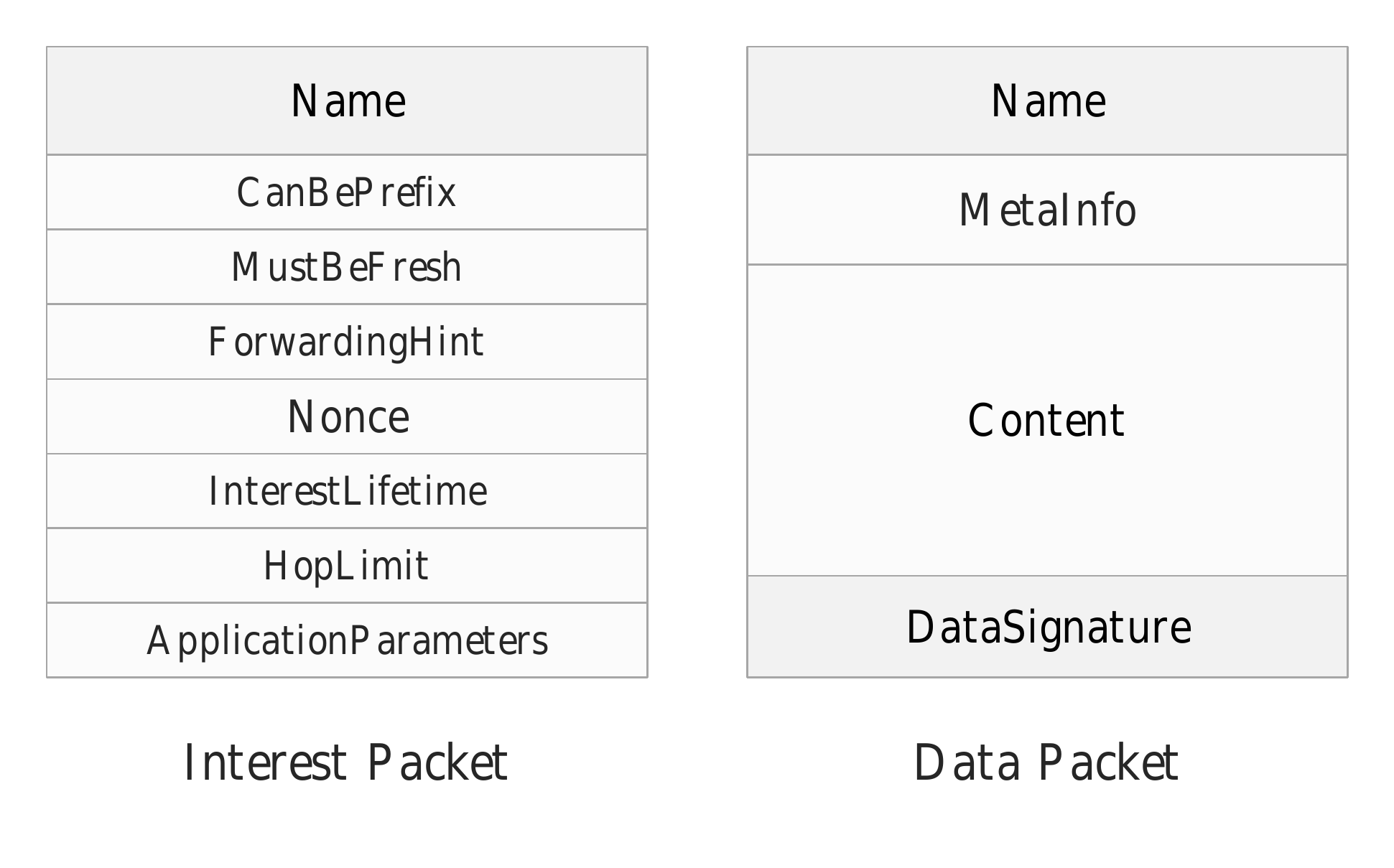}
    \label{fig:ndn_packets}}
    \end{subfigure}
    \hfil
\end{figure}

The NDN architecture uses two types of packets: \textit{Interest} packets and \textit{Data} packets. Figure \ref{fig:ndn_packets} illustrates the NDN packets.
\subsubsection{\textbf{The Interest Packet}}
According to the latest NDN packet specifications~\cite{ndnpktspec}, every \textit{Interest} packet is composed of a set of mandatory and optional parameters. Each \textit{Interest} packet must have a \textit{Name}. It represents the name of the content that the consumer is requesting. The \textit{Nonce} field is also mandatory when the \textit{Interest} packet is transmitted over the network. It consists of a randomly generated 4octets long string. The \textit{Nonce} is used to uniquely identify \textit{Interest} packets, hence preventing them from looping in the network. The optional parameters that \textit{Interest} packets may include are: \textit{CanBePrefix} for the \textit{Interest} packet's name, \textit{MustBeFresh} for the content of the requested data packet, \textit{InterestLifeTime} represents for how long an NDN router will maintain state for this \textit{Interest}, \textit{HopLimit} and \textit{ForwardingHint} are used in forwarding. The interest packet may also include application-specific parameters in \textit{ApplicationParameters}. Furthermore, interest packets can also be signed when needed~\cite{signinterest}.   

\subsubsection{\textbf{The Data Packet}}
\textit{Data} packets are the response that NDN consumers expect when they send \textit{Interest} packets. \textit{Data} packets carry the content requested by a consumer. Each \textit{Data} packet contains the following fields: the \textit{Name} of the data, the payload of the \textit{Data} packet held in the \textit{Content} field and the \textit{DataSignature}. \textit{Data} packets may also contain some optional information in the \textit{MetaInfo} fields: The \textit{ContentType}, the \textit{FreshnessPeriod}, which indicates how long the data can stay fresh, i.e., the producer did not produce newer data. The last parameter is the \textit{FinalBlockId} which identifies the last packet of a sequence of fragments~\cite{datapkt}. NDN \textit{Data} packets identify four principal content types. First is the \textit{BLOB} type, which represents the default content type. The second is the \textit{LINK} type, which is used to include a list of producers. This packet is used for forwarding purposes. The third is the \textit{KEY} type, used to represent a certificate. Finally, the \textit{NACK} type designates application-based \textit{NACK} packets.   

\subsection{NDN Node Model}
The NDN protocol stack is composed of four different layers: application, network, link and physical layers \cite{zhang2018ndn}. The application layer supports the operation of NDN applications and it also embeds transport protocols as system libraries. The network layer’s role is to route \textit{Interest} and \textit{Data} packets. It uses the application layer's names to route the packets. The NDN link-layer supports a set of protocols like Ethernet and can use virtual links like IP and TCP overlays. Each node with the NDN stack maintains three data structures: 

\subsubsection{Pending Interest Table (PIT)}
PIT is a data structure where are stored all the not yet satisfied \textit{Interest} packets.  Every pending \textit{Interest} packet is stored in PIT until a \textit{Data} packet returns or it times out. Each PIT entry contains the following fields: the name of requested \textit{Data} (\textit{Interest} packet’s name), the incoming interface(s), outgoing interface(s) and an expiry timer. 

\subsubsection{Content Store (CS)}
Each NDN node can store the passing \textit{Data} packets in a local cache~\cite{cao2016fetching}. This enables NDN to offer in-network caching. Requests can be satisfied by an intermediate cache without going down to the source of the \textit{Data} packet, which reduces time retrieval and saves links bandwidth.  Each CS need to implement a caching policy to maintain its size and keep the most relevant and popular data. Many caching policies can be used, including but not limited to, FIFO (First In First Out), LRU (Least Recently Used) and LFU (Least Frequently Used)~\cite{zhang2015survey, din2017caching}. 

\subsubsection{Forwarding Information Base (FIB)}
FIB is used to forward incoming \textit{Interest} packet to upstream nodes. Unlike IP networks, NDN's FIB entries are indexed with name prefixes instead of IP addresses~\cite{song2015scalable}. Every FIB entry is composed of a name prefix and a list of next hops. According to its forwarding strategy, routers can forward \textit{Interest} packets to one or multiple hopes, hence enabling multi-path forwarding~\cite{chan2017fuzzy, al2021cledge, mastorakis2020icedge, mtibaa2020ndntp}.

\subsection{Routing and Forwarding}
\begin{figure}
    \centering
    \includegraphics[scale=0.35]{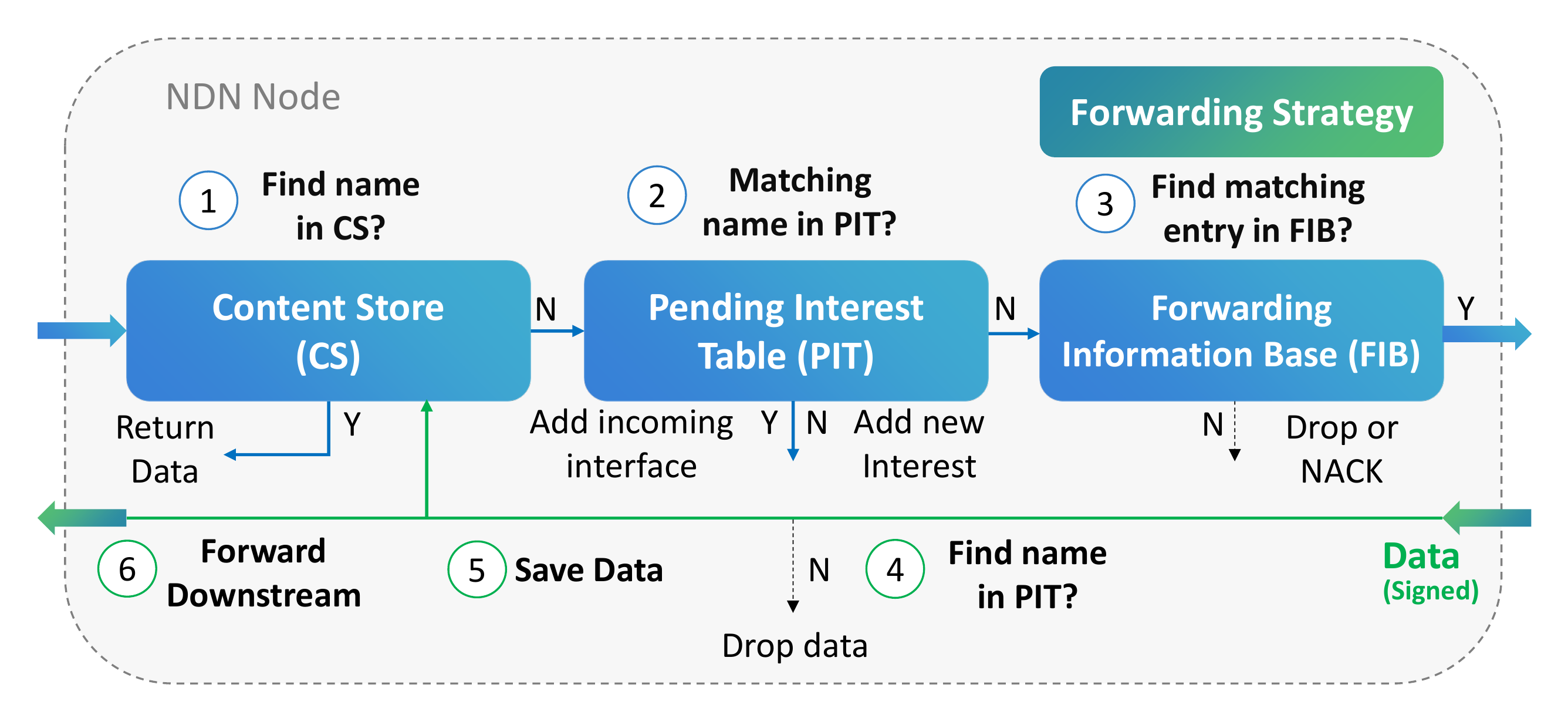}
    \caption{Interest and data packet forwarding process}    
    \label{fig:ndn_forwarding}
\end{figure}

NDN routers use application namespace instead of IP addresses to forward packets~\cite{li2018packet}. Routers update and announce their FIB entries using routing algorithms~\cite{lehman2016secure, voitalov2017geohyperbolic, ghasemi2018muca}, or a self-learning mechanism~\cite{shi2017broadcast}. Unlike IP routers, NDN routers use stateful forwarding, i.e., routers keep information about the received requests until they are satisfied or timed-out~\cite{yi2013case}. The forwarding strategy forwards \textit{Interest} packets according to the FIB entries, local measurements or other per-namespace forwarding policies~\cite{shi2017named}. The forwarding strategy is also responsible for choosing the destination interfaces. NDN routers could also use multi-path forwarding to ensure priority, load-balancing and avoid failed links. Every \textit{Interest} packet brings no more than one \textit{Data} packet, and each response takes the reverse path of its corresponding request. The NDN forwarding process is illustrated in Figure \ref{fig:ndn_forwarding}.

\subsubsection{Interest packet forwarding: Requesting data}
When a router receives an \textit{Interest} packet, first it checks if the requested data can be satisfied from the local CS (i.e., it matches the received data name with the existing content names in the CS). If the requested data already exists in the local CS, the router sends back the \textit{Data} packet to the source interface, i.e., from where the \textit{Interest} packet came from. 
If the CS does not hold the requested \textit{Data} packet, the router performs the following actions: first, it checks the PIT for any similar pending request, %i.e., the router checks if the name of the interest packet already exists in the PIT or not.
If an entry with a similar name already exists in the PIT, the router matches the source interface with all the interfaces associated with this pending entry. If the interface already exists, the incoming \textit{Interest} packet is considered as a duplicate packet and will be discarded. Otherwise, the router adds the source interface to the list of interfaces associated with the pending \textit{Interest} i.e., aggregates the incoming \textit{Interest} packets requesting the same data. On the other hand, when no pending \textit{Interest} with the same name exists in the PIT, the router creates a new entry for this \textit{Interest} packet. Once the PIT lookup process finished, the router sends the \textit{Interest} packet to its upstream neighbor(s) according to its forwarding strategy. 
NDN routers can also send a NACK packet to their downstream nodes when the \textit{Interest} packet cannot be satisfied, e.g., no matching entry in FIB, the upstream links are down~\cite{abraham2017controlling, vusirikala2016hop}.

\subsubsection{Data packet forwarding: Receiving data}
When a consumer’s request is satisfied from a data producer or in-network cache, a \textit{Data} packet is sent back. When a router receives a \textit{Data} packet, it checks if it was requested before by verifying the existence of a pending \textit{Interest} with the name of the received data. If no match exists in the PIT, the router drops the received \textit{Data}. Otherwise, and according to its caching strategy, the NDN router stores (or not) the received \textit{Data} before sending it downstream to all the interfaces associated with the pending \textit{Interest} in PIT. The aggregation of source interfaces gives NDN routers a built-in multicast mechanism. 

\subsection{NDN Security: Data-centric Security}
\begin{figure*}
    \centering
    \includegraphics[scale=0.45]{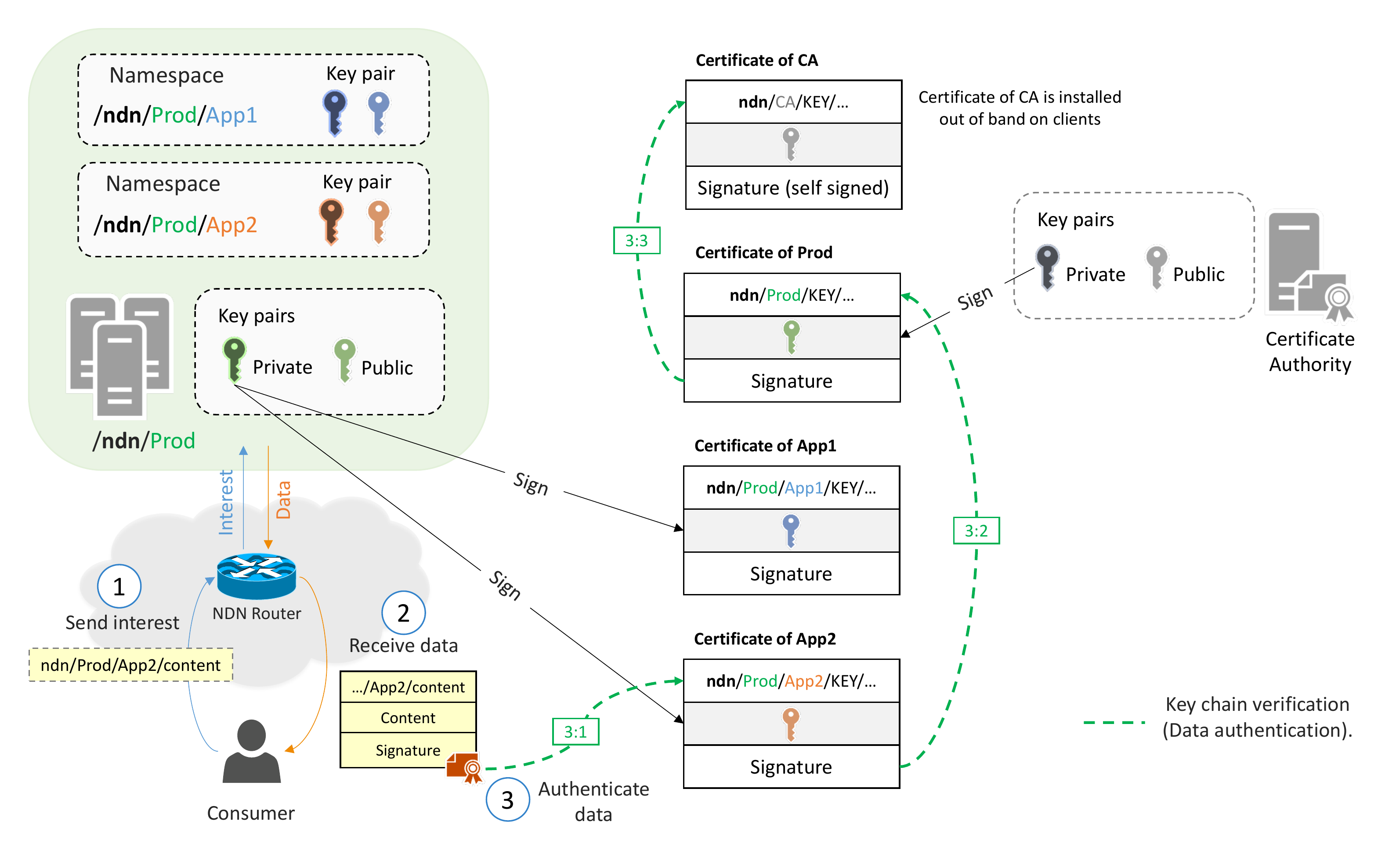}
    \caption{NDN security components}    
    \label{fig:trust_schema}
\end{figure*}

The NDN design natively embeds security mechanisms and mandates the use of signatures in \textit{Data} packets~\cite{ramani2019ndn}. Producers need to digitally sign every \textit{Data} produced. This will enable consumers to verify and validate the authenticity of the received \textit{Data}. Also, it permits to network nodes, i.e., routers and repositories, to store \textit{Data}~\cite{psaras2018mobile}. Furthermore, it allows consumers to retrieve and accept \textit{Data} packets regardless of their source~\cite{newberry2019power}. 
The NDN architecture integrates some security components to ensure data security~\cite{zhang2018overview}. Figure~\ref{fig:trust_schema} illustrates the below detailed security components.

\subsubsection{Packet signature}
Every \textit{Data} packet includes a signature field~\cite{zhang2016sharing}. The signature binds the content of the packet to its name~\cite{li2019secure}. The Signature field contains two components: \textit{SignatureInfo} and \textit{SignatureValue}. 
The \textit{SignatureInfo} component embeds the name of the producer's public key and the cryptographic algorithm used to sign the \textit{Data}. The \textit{SignatureValue} represents the bits of the generated signature. Althouth NDN does not mandate the use of signatures in \textit{Interest} packets, however, in some scenarios where the authenticity of \textit{Interest} packets is needed, signatures are required, e.g., a router sends a new route announcement, sending a command packet to an IoT device, etc. The \textit{Interest} packet’s signature field includes additional components compared to the \textit{Data} packet’s signature. It includes a \textit{SignatureNonce}, a \textit{SignatureTime} and/or a \textit{SignatureSeqNum}. These components are used to add uniqueness to the signature.

\subsubsection{Trust Schema and Access Control}
Although that \textit{Data} packet’s signature allows the consumers to validate the received \textit{Data} and proof its originator, it does not show whether the signer is authorized to produce the data or not~\cite{yu2015schematizing, nour2021access}. Consumers need a mechanism that allows them to check if a producer has the right to produce \textit{Data} under a given namespace, i.e. if a producer’s key has the right to sign a \textit{Data} packet under a given namespace~\cite{shang2017breaking}. Application-based trust policies are used to define which keys are authorized to sign which \textit{Data}. Besides, applications can also implement access control policies to protect the content of a \textit{Data} packet, through encryption, and permit only authorized nodes to decrypt it~\cite{zhang2018nac, lee2018supporting}. 

\subsubsection{Key pairs and Certificates}
Every data producer need at least one cryptographic key pair. Producers use private keys to sign \textit{Data} packets and consumers use public keys to verify them. Each public key is embedded in a digital certificate. The NDN certificate is a \textit{Data} packet signed by an issuer~\cite{afanasyev2016content}. It contains the producer’s public key encoded in X509 format~\cite{zhang2017ndncert}. The name of an NDN certificate follows a specific naming convention: \textit{/SubjectName/KEY/KeyId/IssuerId/Version}, where \textit{SubjectName} is the prefix to which the key is bound to, i.e., the namespace to which the key can operate under. Followed by the keyword \textit{KEY} is the \textit{KeyId}, which represents the value of the key. The value can be an 8-byte long random value, SHA-256 digest of the public key, a timestamp or a numerical identifier. After that comes the information about the issuer and the version of the certificate. Like any NDN \textit{Data} packet, certificates include also a signature field, i.e., signature of the issuer. 

\subsection{Named Data Networking vs TCP/IP}
The following subsection introduces a comparison between NDN and TCP/IP in terms of routing and forwarding, multicast, mobility, and security. Table \ref{tab:ndn_vs_ip} summarizes the differences between these two architectures. 
\subsubsection{\textbf{Routing and forwarding}}
\begin{figure*}
    \centering
    \includegraphics[width=\textwidth]{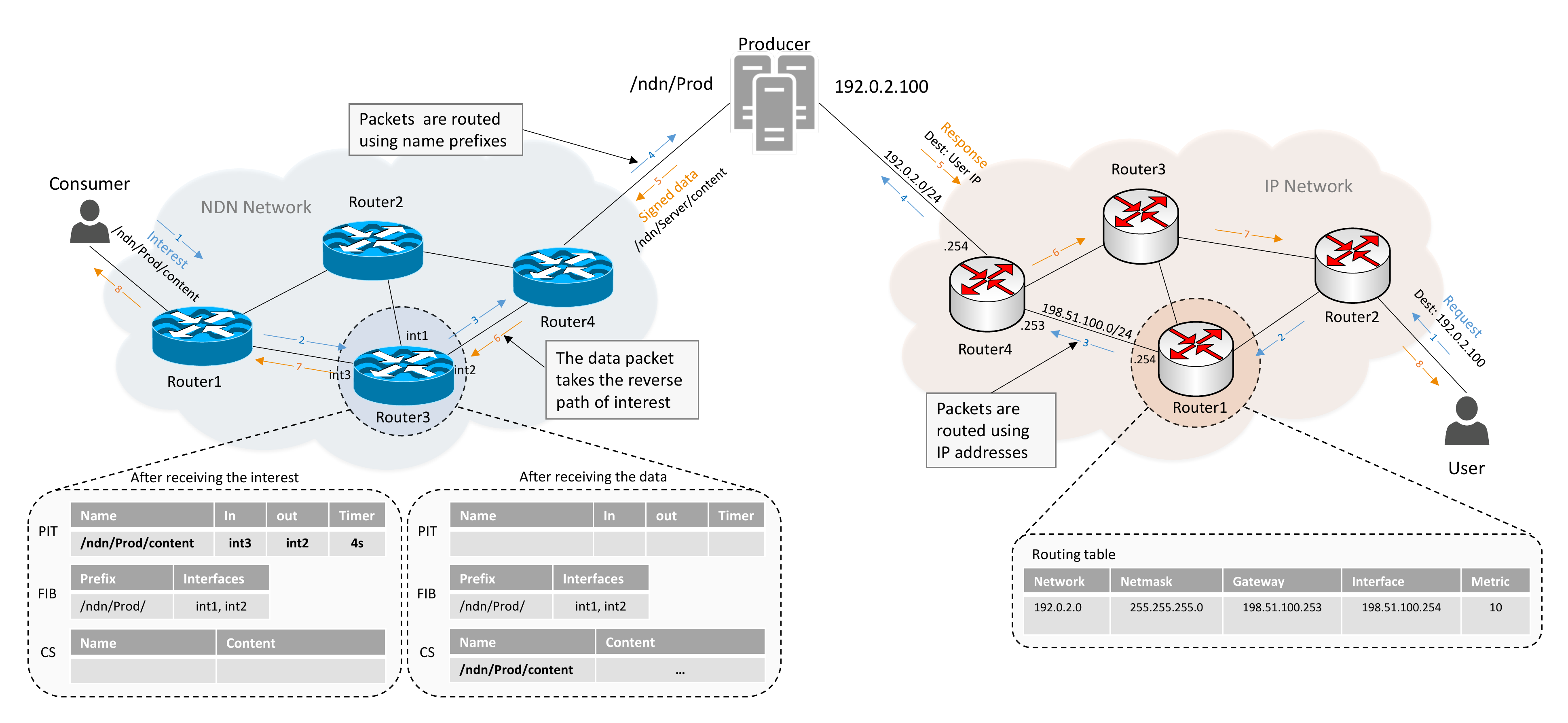}
    \caption{NDN vs TCP/IP packet forwarding}    
    \label{fig:ndn_vs_ip_fwr}
\end{figure*}

Using names instead of IP addresses comes with some advantages: first, compared to IP addresses, names are unbound and have no limit. Second, address translation NAT is no longer needed. Third, there is no need to assign and manage IP addresses in local networks. Moreover, NDN networks are immune against packet looping so nodes can take full advantage of multipath forwarding. Figure \ref{fig:ndn_vs_ip_fwr} depicts the differences of packet forwarding between NDN and TCP/IP.

\subsubsection{\textbf{Multicast}}
IP networks support multicast, but setting up multicast groups in IP networks is hard. Members need to know the group address to join it. Also, the number of nodes in a multicast group is limited. Moreover, multicast groups are difficult to deploy in large networks like the Internet. NDN comes with a built-in multicast forwarding mechanism with no pre-configuration required. NDN routers aggregate source interfaces requesting the same content in PIT. This allows routers to send the data packet to all requesting interfaces at ones, hence performing multicast forwarding~\cite{mastorakis2017ntorrent}.  

\subsubsection{\textbf{Mobility}}
Mobility is another application field that NDN handles better than IP networks~\cite{zhang2016survey,zhang2018kite}. The constantly changing network topology and paths make packet routing challenging in mobile networks. The data-centric approach that NDN adopts is beneficial to mobile networks like MANETs~\cite{mastorakis2020dapes, li2019distributed}. There is no need to maintain and manage nodes’ IP addresses. Besides, the cache-store that network nodes offer reduces the delivery time and gives alternative paths to requested data even if the source is not reachable (i.e., the data producer).

\subsubsection{\textbf{Security}}
IP networks rely on communication protocols like TLS~\cite{dierks2008rfc} and IPSec~\cite{krishnan2011ip} to securely send data packets. NDN networks, on the other hand, focalize on securing the network's most valuable asset, the data. Unlike IP networks, NDN secures the data instead of securing its path. By securing directly the data, NDN networks relieve the nodes from maintaining and monitoring secure channels and allow data producers to only focus on securing the data that they produce~\cite{zhang2018overview}.

\begin{table*}
\centering
\caption{Comparison between NDN and TCP/IP}
\label{tab:ndn_vs_ip}
\begin{tabularx}{\linewidth}{llXX}
\hline
\multicolumn{2}{l}{\textbf{Propriety}} & \textbf{NDN} & \textbf{TCP/IP} \\ \hline
\multicolumn{2}{l}{Architecture} & Data-centric & Host-centric \\ \hline
\multicolumn{2}{l}{Identification} & Names & IP addresses \\ \hline
\multicolumn{2}{l}{Forwarding} & Stateful forwarding & Stateless forwarding \\ \hline
\multirow{2}{*}{Forwarding paths} & Request & Multipath forwarding. The request packet could be forwarding through multiple faces & Single path forwarding \\ \cline{2-4} 
                                  & Response & The response packet takes the reverse path of the request & The response packet does not necessarily take the request’s path \\ \hline 
\multirow{2}{*}{Caching} & Intermediate nodes & All NDN routers offer in-network caching & Routers do not offer data caches\\ \cline{2-4}
                                & End nodes  & Every NDN node can offer a data cache  & Only specific nodes like file servers and P2P nodes        \\ \hline
\multicolumn{2}{l}{Multicast}                         & Built-in multicast mechanism & Uses multicast address range \\ \hline
\multicolumn{2}{l}{Mobility} & The NDN nodes are independent of their location & The nodes IP addresses are location dependent.  \\ \hline
\multicolumn{2}{l}{Security} & Secures data packets & Secures communication channels using protocols \\ \hline
\end{tabularx}
\end{table*}

\section{NDN Security Requirements}
\label{sec:requirements}
\subsection{Confidentiality}
Information confidentiality is essential to any security system. Network entities should restrict any unauthorized actors from accessing any confidential resources. NDN opens new challenges for data confidentiality. Data packets are more susceptible to breaches because of in-network caching. To provide data confidentiality, NDN nodes need to use encryption and implement strict access control schemes to prevent data leakage~\cite{shang2015ndn}. 

\subsection{Privacy}
\begin{figure}
        \centering
        \includegraphics[scale=0.4]{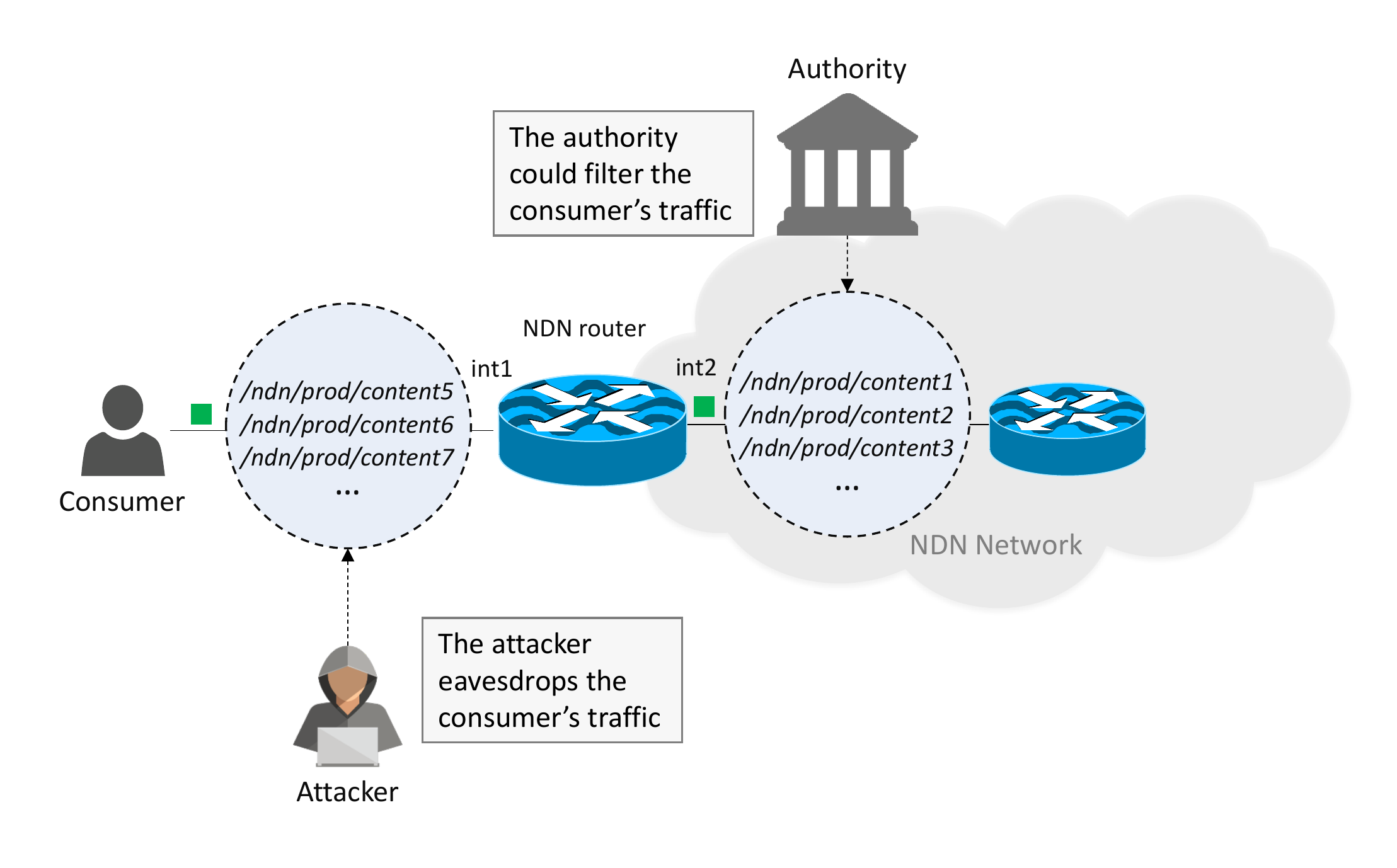}
        \caption{Example of a privacy attack}
        \label{fig:privacy}
\end{figure}

Privacy aims to prevent unauthorized actors from accessing private, personal, or confidential information which may be used to identify and track nodes or individuals. NDN uses application namespaces instead of IP address to route packets in the network layer. This can be challenging for privacy and anonymity. Users’ traffic could easily be monitored and filtered as showed in Fig. \ref{fig:privacy}. 

\subsection{Authentication}
The process of verifying the identity of a node is called authentication. In scenarios where nodes’ identity needs verification, authentication is necessary. It represents any additional information that can identify the authorized nodes like passwords and certificates. NDN mandates data packets signatures, which implies the authentication of every data packet before consumption.

\subsection{Integrity}
Data integrity prevents authorized and unauthorized nodes from making illegitimate modifications. The NDN architecture mandates the use of cryptographical signatures in every data packet produced. This enables a native data integrity mechanism in NDN. When the content of a data packet changes, consumers or network routers can easily detect it. However, this mechanism cannot provide immunity from threats. The data producer can suffer a key breach. The data content could also be altered before being signed by the producer.

\subsection{Non-repudiation}
Non-repudiation ensures that an entity cannot deny the production of a packet or its content. The data packet’s signature guarantees a non-repudiation mechanism in NDN. A data producer cannot deny the ownership of a data packet because it contains its signature. However, the private key of a producer could be stolen by a hacker or accidentally leaked because of human error. The certification authority could also be compromised. furthermore, interest packets are not signed by default. All these challenges are to consider when implementing a non-repudiation mechanism.

\subsection{Availability}
System availability consists of ensuring functional and uninterrupted access to authorized nodes. Multi-path forwarding and in-network caching are two strong assets that NDN offers to nodes to guarantee a high data availability level~\cite{yeh2014vip}. However, NDN Adversary nodes can launch (D)DoS attacks to affect and disturb systems availability. In NDN networks, routers and producers resources are potentials targets to (D)DoS attacks~\cite{gasti2013and}.

\section{Availability Attacks in NDN}
\label{sec:available}
Availability in NDN is susceptible to various types of attacks. In this section, we detail the availability attacks that target NDN routers and producers. 

\subsection{Availability Attacks Against Routers}
Every NDN router component is a potential target for attackers. This subsection explains the availability attacks that routers can deal with. 

\subsubsection{Availability attacks against the Content Store}

\begin{enumerate}[label=(\alph*)]
    \item \textit{Cache pollution attack}:
    To reduce time retrieval and network traffic, NDN nodes use the CS to cache the frequently demanded data.
    The cache pollution attack consists of altering the popularity of stored content. The attacker tries to evict popular content from caches by continuously requesting non-popular content. Figure \ref{fig:cache_pollution_attack} shows a scenario of cache pollution attack.
    
    \item \textit{Cache poisoning attack}: 
     Another CS-based attack is the cache poisoning attack. The goal of the attacker is to fill routers’ caches with invalid \textit{Data}. The attacker tries to inject \textit{Data} packets with valid names but altered or malicious content. Unlike cache pollution, cache-poisoning attacks are slightly hard to perform as it implies the modification of the content. The packet’s signature binds the name to the content, so any change on the content results in an invalid \textit{Data} packet. However, the attacker can still spread malicious \textit{Data} packets in the network, as routers tend to not verify \textit{Data} authenticity at wire rate~\cite{gasti2013and}.
    Attackers can perform cache poisoning attacks in two ways: (1) with malicious producers that send poisoned \textit{Data}. In this scenario, the malicious producers could also cooperate with malicious consumers to flood the network with invalid \textit{Data}~\cite{nguyen2017content}. (2) With malicious consumers. In this attacking scenario, the malicious nodes either perform a man-in-the-middle attack or control compromised routers (e.g., edge routers) as shown in figure.\ref{fig:cache_poisoning_attack}. 
\end{enumerate}

\subsubsection{Availability attacks against the FIB}
\begin{figure}
\centering
\begin{subfigure}[Example of a cache pollution attack]{
    \includegraphics[width=.48\linewidth] {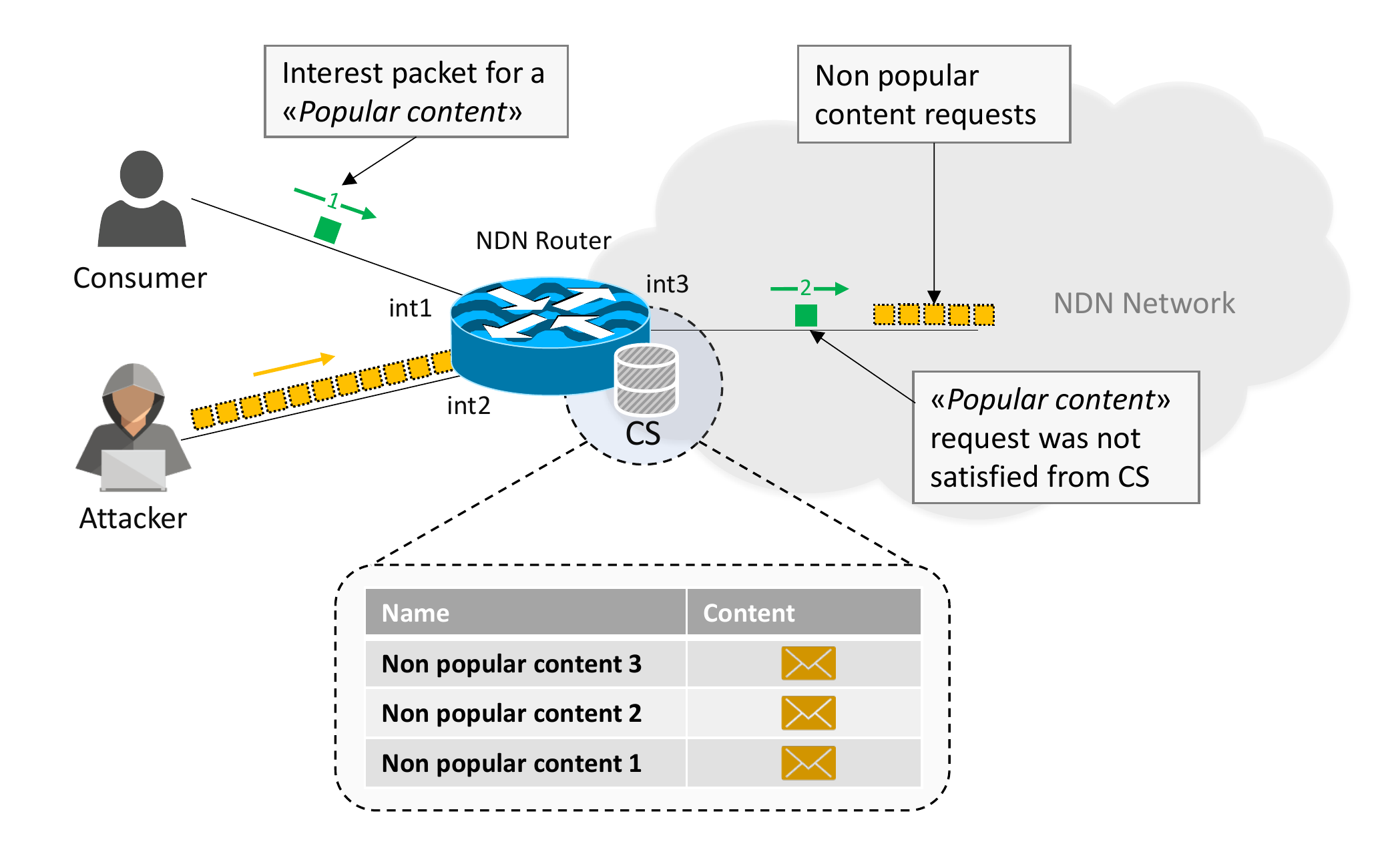}
    \label{fig:cache_pollution_attack}}
    \end{subfigure}
    \hfill
\begin{subfigure}[Example of a cache poisoning attack]{
    \includegraphics[width=.48\linewidth] {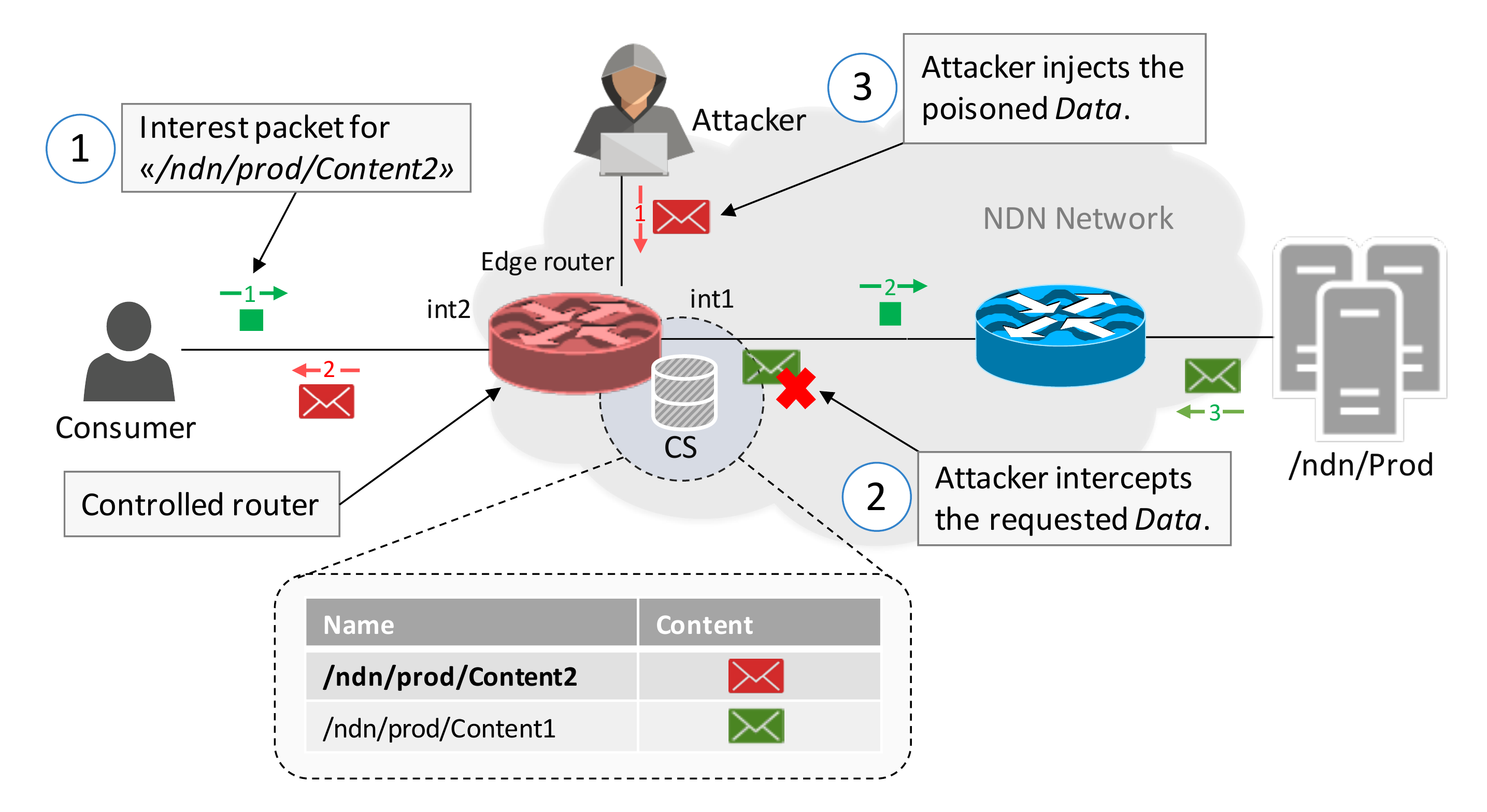}
    \label{fig:cache_poisoning_attack}}
    \end{subfigure}
    \hfill
\begin{subfigure}[Example of a false route announcement attack]{
    \includegraphics[width=.48\linewidth] {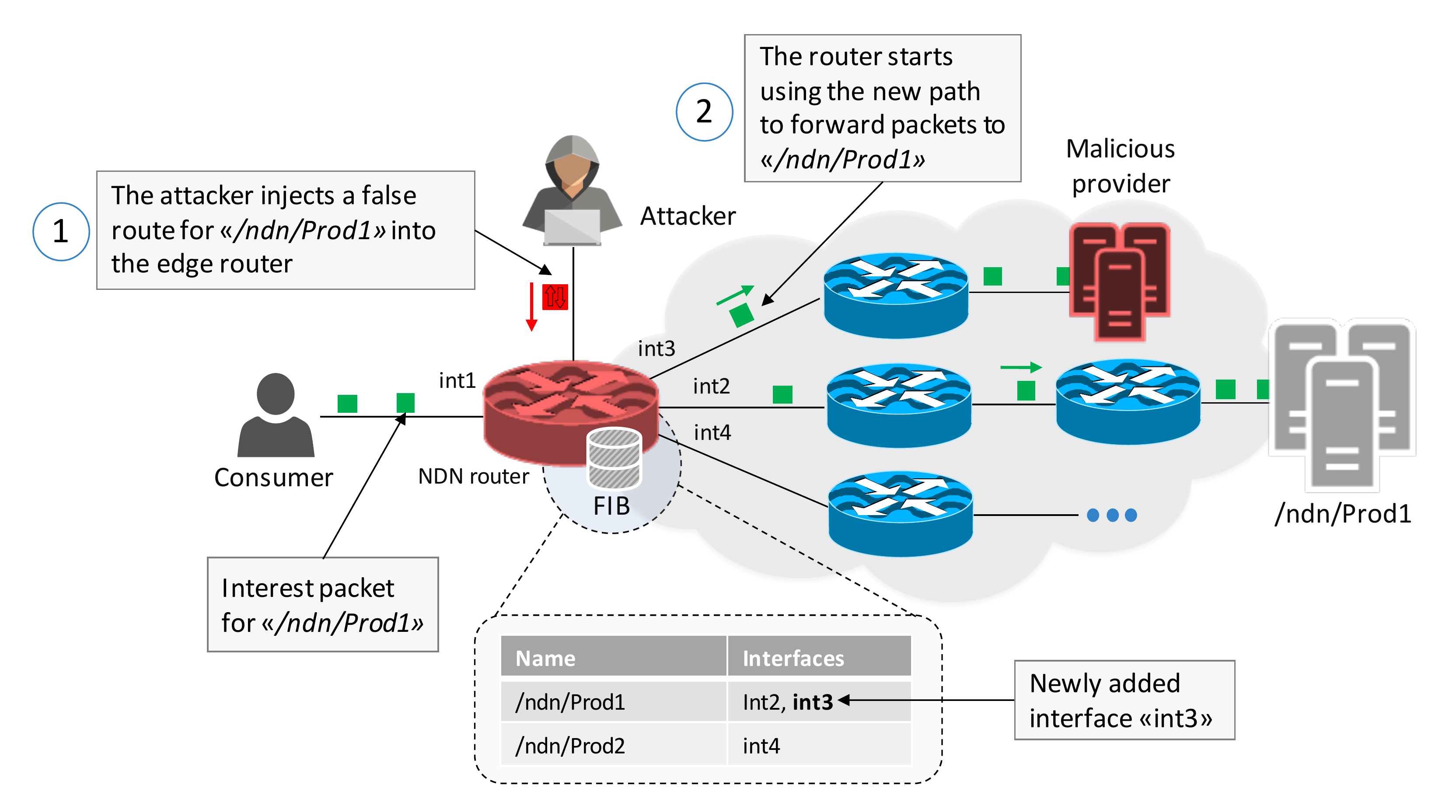}
    \label{fig:false_route}}
    \end{subfigure}
\begin{subfigure}[Example of an Interest Flooding Attack]{
    \includegraphics[width=.48\linewidth] {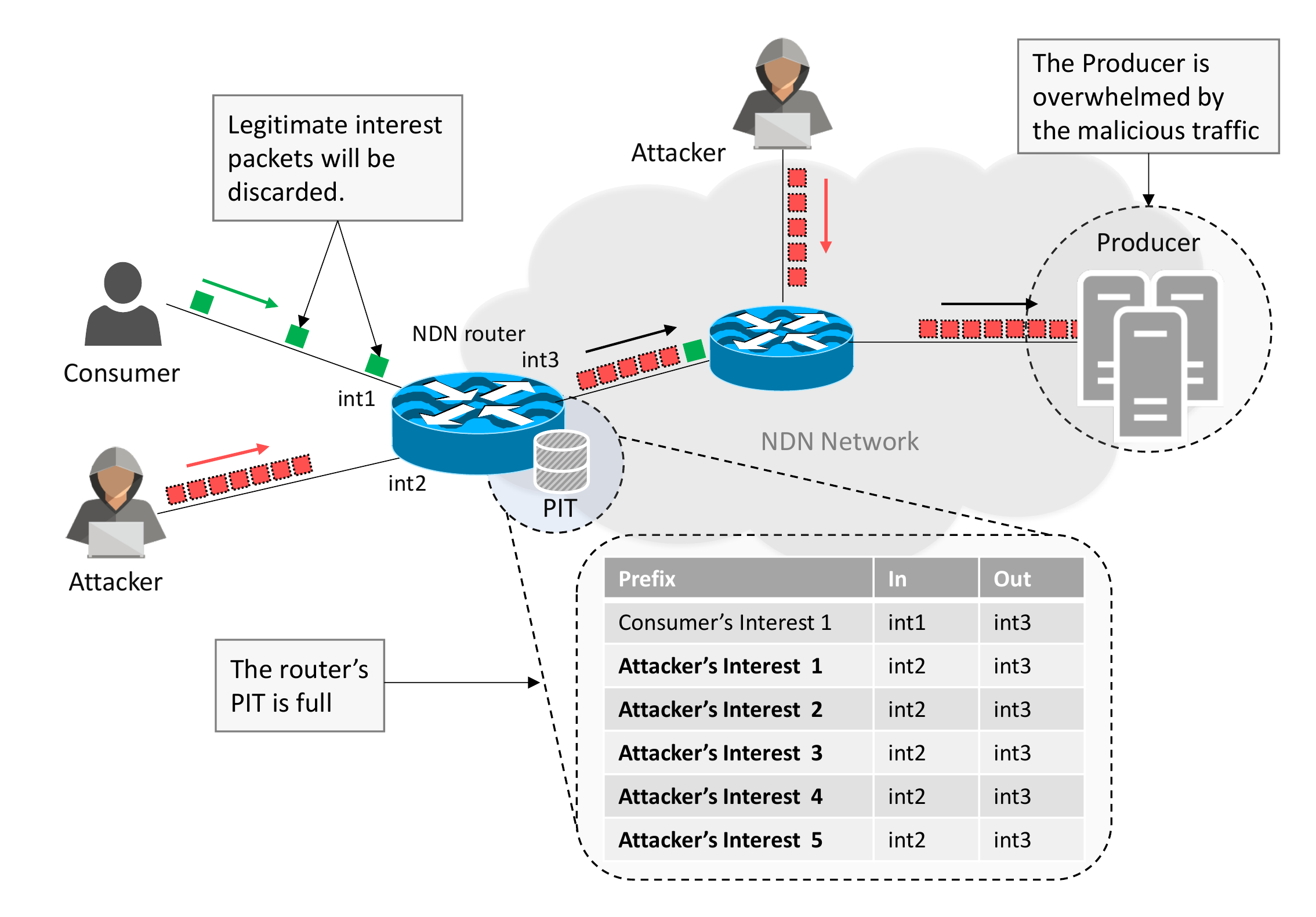}
    \label{fig:ifa}}
    \end{subfigure}

\caption{Availability Attacks in NDN}
\end{figure}

The availability attack that targets router's FIB is the false route announcements. The purpose of this attack is to inject false routes into the network. 
The attacker needs to take control of the router to perform such an attack. The adversary, which controls an edge router, injects false paths to redirect legitimate requests to a malicious provider, as shown in Figure.\ref{fig:false_route}. In a different scenario, adversarial consumers can also send false routes in a mobile network, like Vanet, to perform other attacks like black holing or packet interception~\cite{khelifi2018security}.

\subsubsection{Availability attacks against the PIT}
The main attack that targets PIT availability is the Interest Flooding Attack (IFA). An attacker target PIT’s availability by sending to it a large number of \textit{Interest} packets so it cannot accept legitimate requests. We discuss IFA in detail in Sec.\ref{sec:ifa}.

\subsection{Availability attacks against Producers}
Producer-based availability attacks are mainly associated with IFA. Attackers could target one or multiple producers by sending a big amount of \textit{Interest} packets towards them. Malicious nodes can lunch attacks against producers using all types of \textit{Interest} packets mentioned in \ref{ifa_tax}. The damage that IFA can inflect on producers depends on the nature of \textit{Interest} packets used during the attack. Depending on the severity of IFA, producers can suffer resources damages due memory and processing overhead.

\section{Interest Flooding Attack}
\label{sec:ifa}
The Interest Flooding Attack consists of flooding the network with a massive number of \textit{Interest} packets to drown network routers and/or data producers, as shown in Figure \ref{fig:ifa}. The main goal for attacking routers is to fill their PIT with unnecessary \textit{Interest} packets so there will be no remaining entries for incoming legitimate \textit{Interest} packets, which results in a denial of service. Additionally, data producers could also suffer a DoS from IFA. This includes memory exhaustion, service overload or any other hardware related overhead. 
attacks could be local, launched by one or a small group of nodes, or distributed, initiated by a large group of nodes often controlled by one master node.

\subsection{Local and Distributed Interest Flooding Attack}
IFA is considered local when one or a small group of locally connected attackers perpetuates them. Another form of IFA is distributed. A distributed IFA occurs when the victim suffers from massive traffic originating from a large number of distributed nodes. These groups of nodes, called botnets, are usually constituted of compromised systems managed by an attacker through one or multiple botnet controllers. Distributed IFA are difficult to stop and present a big threat as it floods the target with a large number of requests, resulting in a devastating attack. Botnet networks could implement hundreds of thousands to millions of nodes.  

\subsection{Taxonomy of IFA Requests}
\label{ifa_tax}
Attackers could perform IFA using two types of \textit{Interest} packets. The first type of \textit{Interest} packets carry valid data requests, and the second type carries invalid data requests.  

\subsubsection{Valid data requests}
The first type of requests that attackers could use to attack a target with IFA is valid data requests. This type of data requests is further divided into two classes: requesting static data or dynamic data. 

\begin{enumerate}
    \item \textit{Requesting static data}:
    The first type of valid requests that attackers could use to perform IFA on a target is using \textit{Interest} packets with valid names. Like any other valid \textit{Interest} packet, these requests are satisfiable by data producers or network caches. The attack consists of sending an enormous number of \textit{Interest} packets to the network to choke network routers and data producers. Static data are generally stored in intermediate caches or repositories, which reduces the traffic heading to producers, i.e., the request will be satisfied before reaching its producer. However, even that static data are likely to be found in network caches, the attack can still inflict serious harm to data producers especially in its debut, i.e., when data is not stored in caches. Besides, the malicious traffic induced by the attack can cause serious problems to the network especially in case of a large-scale attack. 
 
    \item \textit{Requesting dynamic data}:
    Dynamic data represent any content that producers generate on-demand. Unlike static content, dynamic data changes over time and needs to be produced when a request arrives. The optional \textit{FreshnessPeriod} field in \textit{Data} packets indicates for how long the content could be considered as fresh. Dynamic \textit{Data} can range from the current record of a sensor in an engine to the actual stock price values. As their content are time-dependent, dynamic \textit{Data} are usually not stored in caches. Attackers can take advantage of this aspect to lunch IFA against producers. In fact, dynamic data requests are normally not satisfied by network caches and are likely to be routed to their producers. This may cause high damage to producers as they use hardware resources to process and sign the content before sending it back.
\end{enumerate}

\subsubsection{Invalid data requests}
An invalid data request corresponds to any \textit{Interest} packet with an invalid content name, i.e., forged \textit{Interest} packet. Attackers can lunch IFA using fake \textit{Interest} packets. There are two different ways of forging an \textit{Interest} packet’s name: the first method is to choose a completely random name for the \textit{Interest} packet. As their names are random, the \textit{Interest} packets will not reach any producer and will affect only network routers. Nevertheless, the first type of forged \textit{Interest} packet could heavily affect the network, especially the routers near the source of the attack. The second method of forging \textit{Interest} names is to append a random series of characters to a real producer’s prefix. This ensures that the forged \textit{Interest} packet will reach the targeted producer. For example, if an adversary wants to target a producer \textit{"/Prod"}, the \textit{Interest} packets’ names that the attacker may use are similar to \textit{"/Prod/nonce"} where nonce is a random value. This type of forged \textit{Interest} packet will affect producers and all the network routers traversed by the forged \textit{Interest} packet. IFA with invalid data requests is more harmful to the network compared to an IFA with valid data requests. This is because routers will not aggregate the \textit{Interest} packets and will stay in PIT until they time-out. NACK packets help reduce the number of pending invalid PIT entries. However, the attack can still do harm to the network. Additionally, attackers can use this mechanism to flood the network with NACK packets.

\subsection{IFA Studies}
IFA was first mentioned in \cite{jacobson2009networking}. The authors of this paper talked about how consumers could overwhelm the network by generating a huge number of \textit{Interest} packets. Following that, several papers studied the IFA phenomenon and its impacts on the network. The authors of \cite{gasti2013and} discussed IFA and cache poisoning attacks before giving some tentative countermeasures. The authors of \cite{choi2013threat} conducted a study on the impact of IFA, through a series of simulation tests, on the throughput and the delay. The authors in \cite{virgilio2013pit} evaluated three PIT architectures against IFA according to their memory consumption: the default PIT, a hashed names-based PIT, and a bloom filter-based PIT. Similarly, the authors of \cite{so2013named} presented a hash-based design to reduce memory occupancy. On the other hand, authors in \cite{pang2017research} conducted a study on the collusive IFA, where attacker cooperate with a malicious provider to overwhelm the network.  In \cite{compagno2015nack}, the authors discussed the benefits and issues of using network-based and application-based NACKs to help mitigate IFA.  Similarly, the authors of \cite{wang2017urgency} argued that NACK packets help routers from keeping pending \textit{Interests} in PIT until they time out. Another similar study was presented in \cite{wang2019analyzing}. The authors also recommend the usage of NACKs to remove pending requests and hence mitigate attacks.  
In \cite{signorello2017advanced}, the authors evaluated three collaborative solutions against two variants of IFA: The first uses different name prefixes. The second uses a mix of valid and invalid names. Similarly, in \cite{al2015revisiting}, the authors compared five techniques according to their satisfaction ratio.    
In another context, The authors of \cite{kumar2019feature} conducted a comparison study between several machine learning algorithms to detect IFA.

\section{Related Works on Interest Flooding Attacks }
\label{sec:related_work}
\begin{figure*}
    \centering
    \includegraphics[scale=0.65]{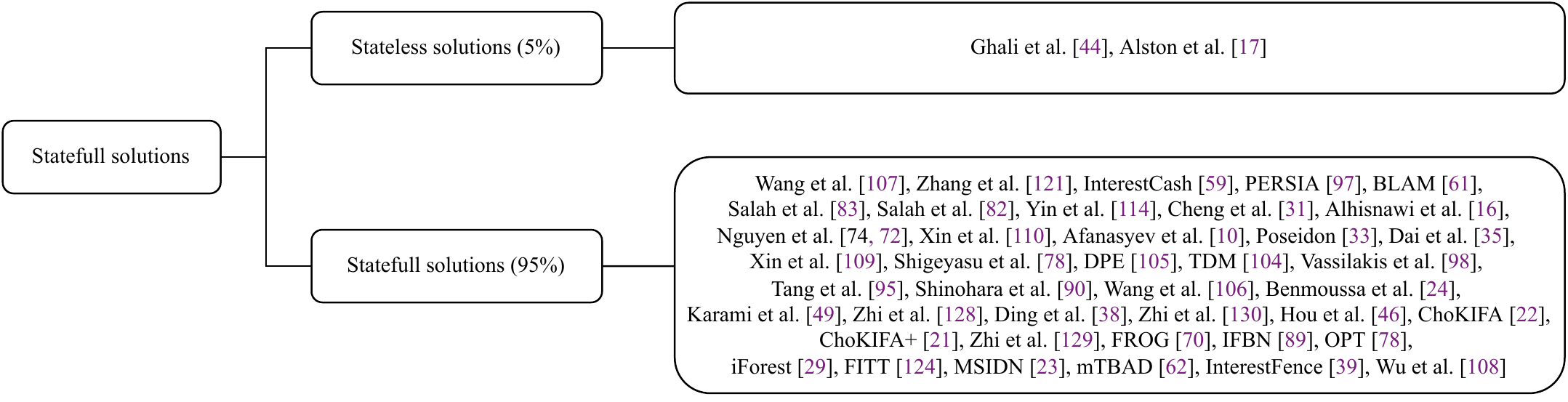}
    \caption{Taxonomy of IFA solutions}
    \label{fig:ifa_solutions_classification}
\end{figure*}

Many IFA solutions are present in the literature. In this section, we categorize the IFA related works into two main categories, as illustrated in Figure.\ref{fig:ifa_solutions_classification}. The first category, named  the stateless category mentions the solutions that took a stateless architecture approach to deal with IFA. The second is the stateful solutions category. This category is further classified into proactive and reactive solutions subcategories.

%Stateless solutions
%*******************
\subsection{Stateless solutions}
The solutions in this category chose a stateless approach to deal with IFA (i.e., do not store traffic-related information in PIT). The authors of \cite{ghali2017closing} proposed a new architecture named \textit{Stateless CCN}. The solution modifies the \textit{Interest} and \textit{Data} packets to incorporate a new component, called \textit{Supporting Name}, which includes the consumer's prefix. Routers use the consumer’s name to forward back the content packet. In this architecture, every consumer needs to be identified, which raises prefix announcements and management problems. Besides, this architecture introduces some TCP/IP based threats, like reflective and privacy attacks.

In a different approach, the solution proposed in \cite{alston2016neutralizing} uses cryptographic tokens to route packets. Each receiving router updates these token, called route tokens, with additional information to forward the \textit{Data} backward. Symmetric keys are used to ensure the integrity and confidentiality of route tokens. Although this solution prevents memory consumption due to PIT overhead, the proposed technique may consume a lot of processing resources, especially with large traffic. Attackers can use it as a tool to inflict high damage to routers, and the damage gets even higher as it gets closer to the core network. 

\subsection{Stateful solutions}
\begin{figure*}
    \centering
    \includegraphics[scale=0.65]{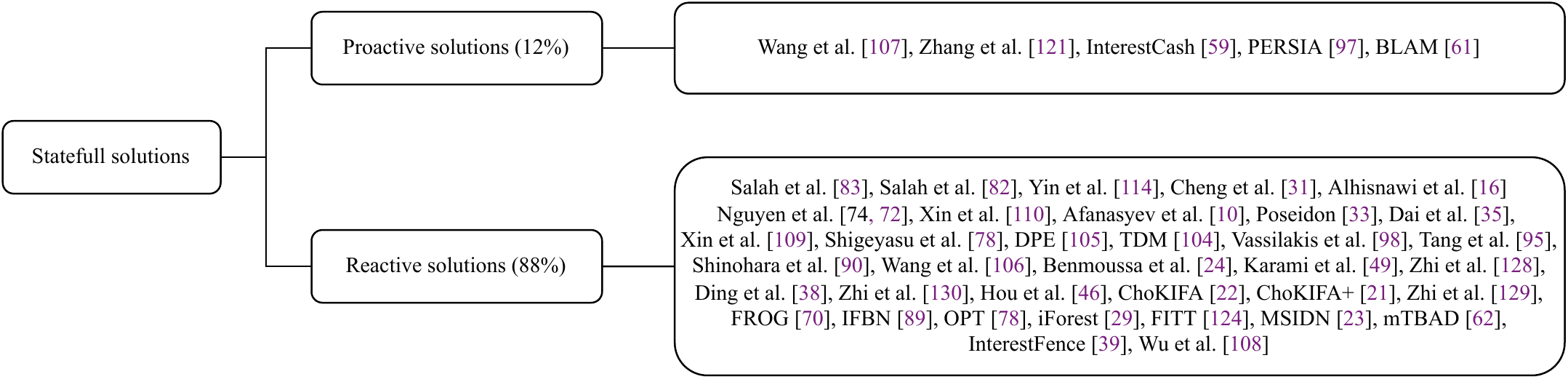}
    \caption{Taxonomy of stateful solutions}
    \label{fig:stateful_taxonomy}
\end{figure*}

The last main category of our classification groups the stateful solutions authored in the literature. The stateful solutions category is divided into: proactive and reactive solutions. Figure \ref{fig:stateful_taxonomy} illustrate the different stateful IFA solutions. 

%Proactive solutions
%*******************
\subsubsection{Proactive solutions}
A small group of proactive countermeasure solutions are present in the literature. The authors of \cite{wang2017economic} introduced micro-payments into the network to regulate IFA. Authority nodes in the network act as banks to conduct any virtual money-related actions. To request a \textit{Data} packet, consumers need to add a prepayment to the \textit{Interest} packet, which includes a PIT delay and content delivery fees. When the request reaches a destination, the node checks its content delivery fee, deducts the necessary amount, and sends the content downstream. Every traversed router takes an amount from the content delivery fee. The consumer receives the requested \textit{Data} packet only if the fees are sufficient to cross all the nodes. The micro-payment system limits legitimate traffic. Also, legitimate consumers can be penalized in scenarios like unsatisfied requests or unavailable producers. Additionally, deploying and maintaining such a system is extremely hard in large networks, given that the system relies on central nodes to act as banks. To overcome this problem, the authors in \cite{zhang2019ari} proposed a payment solution that relies on auto-regressive integrated (ARI) and the hidden Markov models (HMM). When a router receives an \textit{Interest} packet, the price that the router charges to forward the incoming \textit{Interest} depends on the number of pending requests associated with this interface. The router uses the satisfaction ratio as a parameter for the Hidden Markov Model to predict the consumer's state, legal or bad. Besides the fact that legitimate traffic is penalized by the charge/reward mechanism, relying only on the satisfaction ratio to predict the consumer’s state may lead to false detection.

Another proactive solution was presented in \cite{li2014interest}. It employs a proof-of-work mechanism to countermeasure the signing overhead resulting from IFA requesting dynamic content. When a consumer expresses the need for specific data, he needs to answer a puzzle sent by the producer and re-sends the \textit{Interest} packet with the computed value (result). If it is correct, the producer sends back the \textit{Data}. To reduce the communication delay resulting from retrieving/answering puzzles, the solution in \cite{tourani2020persia} relies on tokens that consumers get when they solve computational puzzles. When a consumer sends an \textit{Interest} packet, the edge router verifies if the token exists in its table. If it does, it forwards the \textit{Interest} and updates the count number of this token. Furthermore, core routers monitor the loss rate of each interface. When it reaches a threshold, the router suspects malicious traffic and starts using a bloom filter, instead of PIT, to store incoming \textit{Interest} from this interface.
Similarly, the authors of \cite{liu2018blam} proposed BLAM, a lightweight bloom filter-based mitigation technique for IoT devices. In this solution, each IoT node uses a bloom filter array during the forwarding process. When an \textit{Interest} arrives, the node matches its name with the bloom filter array. If an entry exists, the node continues the forwarding process. Otherwise, the node considers the received \textit{Interest} as malicious and discards it. The proposed solution cannot be deployed on large networks as it requires knowing all the \textit{Data} packets that producers offer. Similar to the last three techniques, this solution does not prevent IFA. Attackers can still flood the network with \textit{Interest} packets.

%Reactive solutions
%******************
\subsubsection{Reactive solutions}
\begin{figure*}
    \centering
    \includegraphics[scale=0.65]{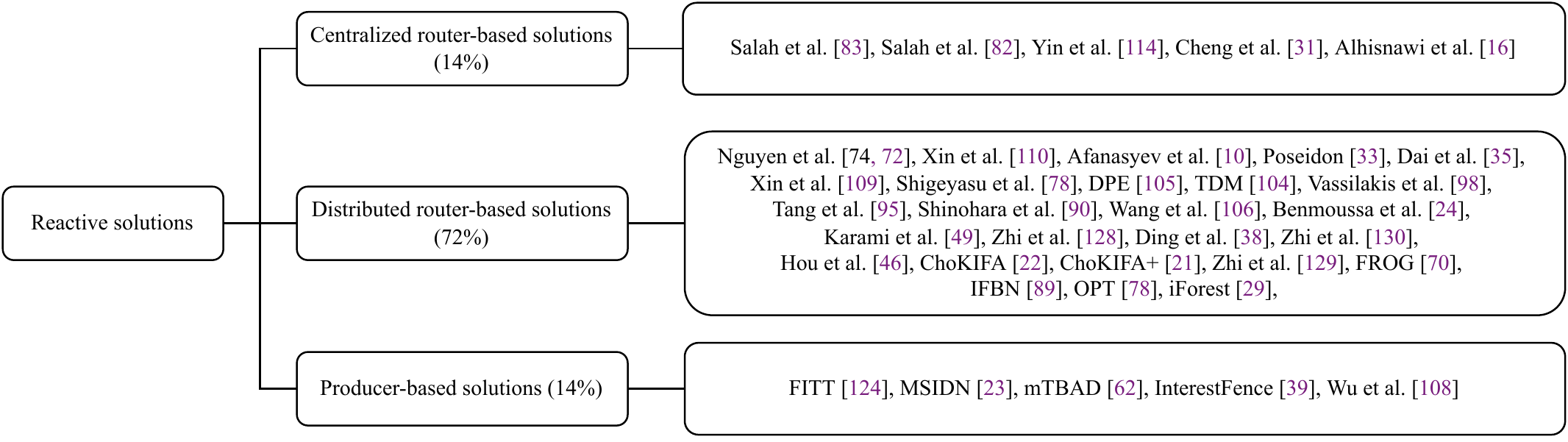}
    \caption{Taxonomy of reactive solutions}
    \label{fig:reactive_classification}
\end{figure*}

Several IFA reactive solutions have been proposed. Reactive solutions are classified into router-based and producer-based solutions. Router-based solutions are further grouped into centralized and distributed solutions as showed in Fig~\ref{fig:reactive_classification}. \newline

\textbf{Centralized router-based solutions}
Several centralized router-based IFA mitigation solutions exist in the literature. 
The solution presented in \cite{salah2015coordination} relies on a Domain Controller (DC) and a group of selected routers named Monitoring Routers (MR) to detect IFA. The MRs calculate the PIT usage and the expired \textit{Interests} associated with each interface. When these metrics exceed a certain threshold, the MRs send the name prefixes and their respective expiration rates to the DC. The DC will then decide on the infected name prefixes and inform the MRs. However, the metrics used can lead to false detection in the case of high legitimate traffic or unavailable producers. In addition, the legitimate requests going to infected prefixes can be affected. The authors proposed an enhanced version in \cite{salah2016evaluating} capable of detecting collusive IFA. Compared to the previous solution, this mechanism relies only on PIT usage, which leads to false-positive situations affecting legitimate requests.

Unlike the previous two, the solution in \cite{yin2019controller} relies on every router to collects and sends to the domain controller the number of \textit{Interest}, \textit{Data}, and NACK packets. The controller calculates the satisfaction ratio associated with each router. If the metrics exceed their respective thresholds, the controller considers that this router is under attack and sends back the malicious prefixes. However this solution relies on the controller to calculate the metrics, which makes it a SPOF. Also, the traffic that routers send may overwhelm the network. 
To overcome this problem, the solution proposed in \cite{cheng2019detecting} relies only on edge routers. when the rate and the number of timed-out \textit{Interests} of a router's interface exceed their respective thresholds, it informs the controller. The latter inspects the received traffic information and checks whether the links will reach their capacity. If one or more links are likely to reach their capacity, the controller determines that an IFA is happening and informs the routers on the interfaces to block. The proposed solution may consider high traffic rates as malicious.

The solution presented in \cite{alhisnawidetecting} has the particularity to rely not only on a domain controller but also a content provider. When the content provider router (CPR) receives an \textit{Interest}, it verifies the name of the \textit{Interest} against the FIB and the Quotient-based Cuckoo filter (QCF), which represents an updated list of fake \textit{Interest} names. If the name exists in the QCF and no entry is found in FIB, it considers this \textit{Interest} as malicious and sends its name within a warning message to the controller. The controller then updates the QCF before informing the edge routers. When an edge router receives the warning message, it restricts the originating interface. However, the QCF table could be used by attackers. Also, the content provider router needs to process all \textit{Interests} of the AS, which is resource-consuming.
\newline

\textbf{Distributed router-based solutions}
The majority of IFA solutions that exists in the literature are distributed router-based solutions. In \cite{nguyen2015detection}, the authors presented a solution that uses hypothesis testing theory to detect IFA. Routers calculate a packet-loss rate associated with each interface to model the legitimate traffic. Then this model is used to detect IFA. Regardless of being a detection-only mechanism, the proposed solution relies only on the number of \textit{Interest} and \textit{Data} packets statistics to identify IFA, which may give false-positive results. Routers can alleviate this problem if they cooperate to detect IFA. The authors also employed this technique to detect collusive IFA in~\cite{nguyen2019reliable}. Similarly, the proposed solution in \cite{xin2017detection} analyses the traffic using a discrete wavelet transform in an uncooperative way to detect collusive IFA. The detection process can be resource-consuming in some scenarios. These two solution do not act to mitigate attacks.

The authors of \cite{afanasyev2013interest} proposed a solution based on the token bucket. It consists of distributing the tokens according to interfaces’ satisfaction ratio. When a router calculates the limit values of an interface, it announces this limit to the neighboring routers connected to it. The receiving routers will then adapt their sending rates according to the received value.  
In addition to the satisfaction ratio, the solutions proposed in \cite{compagno2013poseidon}, use the PIT usage of each interface. The router suspects malicious traffic when these two metrics exceed their thresholds. As a mitigation action, the router limits the incoming traffic and penalizes the interface with reduced thresholds. Then, it sends an alert message to the router connected to this interface to inform it about the reduced rate. Similar to the previous solution, the rate limit employed can impact legitimate consumers.
Similarly, routers in \cite{dai2013mitigate} monitor the size of their PIT. When the size exceeds a certain threshold, the router generates a spoofed  \textit{Data} packet and sends it back to the originator of the spoofed \textit{Interest}. Then, the edge router limits the rate of the source interface.
Another pushback-based mechanism was proposed in \cite{xin2016novel}. In this solution, routers monitor the name of incoming \textit{Interests} and compute the interface's cumulative entropy. When This latter exceeds an upper bound, the router confirms that an attack is undergoing. Afterward, the router sends a spoofed \textit{Data} to the originators of the malicious prefix. When an edge router receives the spoofed \textit{Data}, it applies rate limiting on source interfaces. Besides the resource exhaustion of the detection process, legitimate interfaces can be penalized by rate-limiting in some scenarios. Another drawback is that attackers can use the spoofed \textit{Data} as a tool to drown the network. 

The authors of \cite{wang2013decoupling} proposed a solution that decouples the malicious traffic from legitimate traffic to prevent PIT exhaustion. Routers monitor the number of timed-out \textit{Interest} packets under each prefix. When the router considers the prefix under attack, it stores this prefix in a list called "m-list". The router forwards the \textit{Interest} packet without storing it in PIT if the prefix exists in the m-list. This solution limits the effects of an IFA but does not stop it. Also, it affects legitimate requests. Furthermore, attackers can target the router’s resources by forging names to fill the m-list, which makes the solution inefficient.
Similarly, the mechanism presented in \cite{wang2014detecting}  monitors the timed-out \textit{Interest} packet to detect IFA. It expands the FIB to include four additional fields. When an \textit{Interest} times out before a \textit{Data} returns, the router changes the mode value of this prefix to malicious and increments a related counter Ci in FIB. If the Ci value goes above a predefined threshold, the router applies rate limiting on this prefix. The rate-limiting used in this solution is associated with FIB entries, which affects all the traffic of this entry. 
The solution in \cite{vassilakis2015mitigating} also relies on the number of expired PIT entries. Edge routers classify each interface, according this metric, into three categories: normal, suspicious, or malicious. The edge router can reduce or block the traffic depending on the interface's behavior. It also notifies other routers if it blocks a consumer. However, identifying NDN consumers is not an easy task, which makes this solution hard to apply.

The authors in \cite{tang2013identifying} presented a two-phase detection mechanism. A router calculates the satisfaction ratio of each interface. If it exceeds its threshold, the router counts the number of expired \textit{Interests} under each prefix. If the number exceeds its threshold, the router considers the prefix as malicious and starts blocking \textit{Interests} going to this prefix. Similarly, in \cite{shinohara2016cache}, routers calculate a reputation value for each interface depending on its satisfaction ratio. If an interface has a low reputation value, the router checks if the number of PIT entries is high. In this case, it discards incoming \textit{Interests}. These two solutions do not prevent attackers. Additionally, they may lead to blocking legitimate \textit{Interest} as they rely only on the satisfaction ratio. 

The solution presented in \cite{wang2014cooperative} relies on fuzzy logic to detect IFA. Every router on the network monitors the PIT occupancy and PIT expiration rates. The router takes these values as entries for the detection algorithm. If an IFA is detected, an alert message containing the malicious prefix is sent to downstream routers. When an edge router receives the pushback message, it limits the traffic going to the malicious prefix. The detection mechanism is resource-consuming. Besides, namespace-based rate-limiting can also affect legitimate requests and may lead to false-positive detection in some scenarios.
To avoid false detection, the work presented in \cite{benmoussa2019novel} takes network congestion as a parameter when detecting IFA. Edge routers classify a consumer as suspicious when its incoming rate and satisfaction ratio are above and below their respective thresholds. After that, the router checks the network congestion to avoid false detection. It compares the number of timed-out \textit{Interests} with NACK packets. If the network is not congested, the router classifies the consumer as malicious and blocks it. However, the congestion detection process employed is not cooperative, which may give false results in some scenarios.

Unlike the previous solutions, the authors in \cite{karami2015hybrid} proposed an AI-based mechanism to mitigate IFA and cache poisoning using Radial Basis Function (RBF) neural networks. It uses a set of statistics as inputs, such as the number of satisfied and timed-out \textit{Interests}. When the detector module signals malicious traffic, the router sends an alert message to source interfaces. The alert message contains the new reduced rate, the generation timestamp, and the reduction period. However, the rate-limiting can affect legitimate consumers. Another machine learning-based detection mechanism was proposed in \cite{zhi2019resist}. In this solution, a router constantly collects the entropy of \textit{Interest} names, the satisfaction ratio, and the PIT usage of interfaces before using them as entries for the support vector machine (SVM) classifier. If the detector classifies it as an anomaly, the router declares that an IFA is happening. Following that, the router extracts the malicious prefixes using the Jensen-Shannon divergence. Then, it informs downstream routers about the malicious prefixes. This solution affects also the legitimate \textit{Interest} packets going to blocked prefixes. Similar to the previous solution, the detection process may consume a lot of resources.

The authors of \cite{ding2016cooperative} presented a solution based on the vector space model and Markov chains. When the PIT size reaches an alarming level, the router determines a state for each received \textit{Interest} (normal, unknown, and risk). Upon receiving an \textit{Interest}, a router first checks its state. If the previous (i.e., received from the upstream router) and the actual (i.e., calculated by the router) are both equal to risk, the router discards the \textit{Interest}. To prevent legitimate requests from being discarded, the proposed solution stores a copy of unsatisfied \textit{Interests} in a specific cache, so if an \textit{Interest} with the same name arrives, the router considers it as legitimate. However, attackers can use this propriety to target routers. Additionally, the process of state checking consumes router’s resources.

A statistical solution, that uses Gini impurity, was proposed in \cite{zhi2018gini}. A router monitors the name of received \textit{Interests} and calculates their probability. Then, it uses Gini impurity to measure the disparity of the set and compares it with a threshold to determine if an IFA is happening. After that, the router uses Gini impurity to compare the actual set with a previously recorded set to detect the malicious prefix. The proposed solution may lead to false-positive detection. Similarly, An entropy-based solution was proposed in \cite{hou2019theil}. Based on the fact that when IFA happens, the occurrence of forged names increases, and the Thiel entropy decreases. Every router records the names of incoming \textit{Interest} packets and calculates their entropy to decide whether IFA is happening. The proposed solution can mistakenly consider legitimate traffic as malicious as it relies only on the statistical distribution of \textit{Interest} names. Furthermore, storing \textit{Interest} names may consume a lot of resources, especially in the case of a massive attack.  

The authors of \cite{benarfa2019chokifa} presented a solution that relies on Active Queue Management (AQM). For each received \textit{Interest}, the router randomly picks a pending \textit{Interest} and matches its name with the received \textit{Interest}. Then, it compares their source interfaces and checks the satisfaction ratio of the source interface. If it is under a threshold, the router drops both \textit{Interests}. Otherwise, it drops the received \textit{Interest} with a probability. However, attackers can timely orchestrate attacks to fill the target’s PIT so the router will start dropping incoming \textit{Interests}. The authors modified this mechanism to include blocking~\cite{benarfa2020chokifap}. When an edge router detect that the PIT occupancy and the satisfaction ratio of an interface are above thresholds, it blocks it. However, legitimate traffic is still penalized by the mechanism and it increases as it gets close to the core network. Similarly, the authors of \cite{zhi2020reputation} use a reputation-based early detection mechanism to counter collusive IFA. When an interface's PIT usage is between the minimum and the maximum thresholds, the router drops incoming \textit{Interests} with probability. The router defines the minimum and maximum thresholds according to the satisfaction ratio and the average RTT of an interface. The metrics used may lead to dropping a large portion of legitimate \textit{Interest}, especially in core routers. Similar to the previous two, this solution is a traffic control mechanism, which does not stop attackers. Another solution that aims to counter collusive IFA was proposed in \cite{nakatsuka2018frog}. It relies on the fact that malicious traffic is satisfied only by producers, which makes the mean hop count of each received \textit{Data} high and its variance low. Hence, edge routers store the hop count of each \textit{Data} and calculate the mean and variance of the recorded set. Following these values, the router affirms if a consumer is malicious or not. This solution may lead to mistakenly blocking legitimate users in scenarios where the legitimate requests need to be satisfied by the producer. Being a non-cooperative solution, the proposed solution is unable to detect attacks against routers. Another mechanism against collusive IFA was presented in \cite{shigeyasu2018distributed}. Each router monitors its PIT usage. If the rate reaches a threshold, the router examines the number of times the \textit{Data} was satisfied from the router’s cache. If it equals zero, the router recognizes an ongoing attack. However, attackers can avoid detection since this solution relies on the fact that attackers send \textit{Interests} with random names. Besides being a detection-only mechanism, it can also lead to false-positive detection in some scenarios (e.g., live streaming). 

In \cite{pu2019self}, the authors presented a mechanism that uses a new structure called Operation Trace Table. The router uses it to record the traffic statistics associated with each interface. When the unsatisfaction ratio of an interface is above the average satisfaction ratio, the router reduces the interface’s share and distributes it equally among all interfaces. If the router receives an \textit{Interest} from an interface that reached its allowed share, it forwards it to another outgoing interface to discover unknown paths. The proposed solution does not mitigate attackers. In addition, sending \textit{Interests} to other routes may overwhelm upstream links and routers.

Each router in \cite{chen2019isolation} randomly selects some prefixes from a recorded list containing calculated metrics for each prefix. Then, it constructs a group of search binary trees from the minimum and maximum values of the selected prefixes. After that, it calculates the average traverse path of chosen prefixes and uses it to calculate an abnormal score for each prefix. Finally, it notifies downstream routers with the malicious prefixes. The namespace-based rate-limiting penalizes legitimate \textit{Interest}. Besides, the detection process can consume a lot of resources, especially for core routers. Additionally, this technique can take time to detect attacks.\newline

\textbf{Producer-based solutions}
Some producer-based reactive solutions were proposed in the literature.
The solution proposed in \cite{zhang2019expect} relies on producers’ messages to initiate IFA mitigation. When a producer needs to release its resource, it sends a specific NACK packet to the gateway router. The generated NACK carries essentially the nature of the traffic, fake or valid, and all the fake names received under the affected prefix. When a router receives the NACK, it notifies neighboring routers of the fake names. If an edge router receives a NACK with traffic equal to fake, it blocks the source interfaces for some time. The fake names list used can burden routers in case of massive distributed attacks. Additionally, the proposed technique relies on producers’ feedback to initiate the mitigation process, which leaves routers vulnerable to attacks. To overcome this limitation, The authors of \cite{benmoussa2020msidn} proposed a solution that relies on both producers and routers to mitigate IFA. Besides detecting conventional distributed IFA, it also detects distributed IFA where attackers adopt regular sending rates. When a producer is overwhelmed with requests, it sends a signed \textit{Interest} asking routers to limit forwarding requests to the affected prefix. When received, the router stops forwarding \textit{Interests} before informing neighboring routers. An edge router classifies a consumer as harmful when it keeps sending requests to the infected prefix and has a low satisfaction ratio along with a high number of timed-out \textit{Interests}. Although the proposed mechanism can mitigate different types of distributed IFA without affecting legitimate traffic, it still may unintentionally block legitimate consumers.
 
A new version of \cite{wang2013decoupling} was proposed in \cite{liu2018accuracy}. In this solution, a new producer-based monitor mechanism was introduced to reduce the detection delay. When a producer receives a malicious \textit{Interest}, it responds with a NACK to inform downstream routers. When a router receives the NACK, it checks whether a similar entry exists or not in the m-list. Although it reduces detection delays, this solution still does not stop attackers. Additionally, producers may consider some legitimate requests malicious. Similarly, each router in \cite{dong2020interestfence} maintains an m-list where it stores the affected prefixes received from producers. When the router receives an \textit{Interest}, it checks its existence in the m-list. If it matches the information received from the producer, the router forwards the \textit{Interest}. Otherwise, it drops it. Routers need to process every received \textit{Interest}, which may consume resources due to hash verification. Also, the m-list can occupy a lot of memory in case of a massive attack with a large number of prefixes. Additionally, routers need to process application-based NACK packets.

Unlike the previous solutions, the mechanism presented in \cite{wu2020mitigation} mitigates collusive IFA. Routers monitor the interfaces'  throughput and PIT usage during fixed time windows to detect malicious traffic. When these metrics go above their thresholds, the router starts to delete the oldest PIT entries. However, this solution may delete legitimate requests.

\section{CDA Workflow: Collect-Detect-Act}
\label{sec:cda}
\begin{figure}
    \centering
    \includegraphics[scale=0.35]{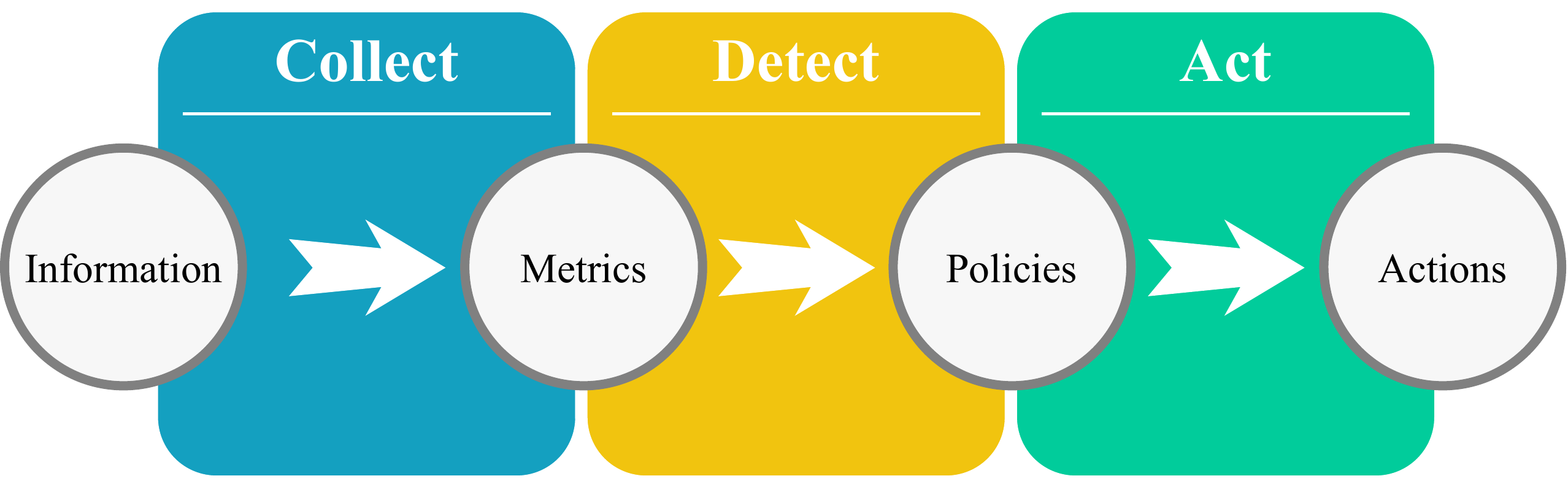}
    \caption{CDA workflow}   
    \label{fig:cda}
    \vspace{-0.6cm}
\end{figure}

In this section, we provide an in-depth analysis of each authored solution. Our comparison depends on the information solutions collect, the metrics and parameters they use during detection, and the actions they take to counter IFA. To this end, we present CDA, which is a workflow that every solution follows. The CDA workflow consists of three activities \textit{Collect}, \textit{Detect} and \textit{Act} as illustrated in Fig.~\ref{fig:cda}.

\subsection{Collect}
The first activity of the CDA workflow is \textit{Collect}. During this activity, solutions collect a set of traffic-related information to detect IFA. The nature of this information varies from one solution to another.  Some solutions rely only on router-based information \cite{afanasyev2013interest}, \cite{salah2015coordination}, \cite{benmoussa2019novel}, while others rely on producer-based information \cite{zhang2019expect},\cite{liu2018accuracy}. On the other hand, to detect IFA, some solutions rely on both router-based and producer-based information like \cite{benmoussa2020msidn}. 

\subsubsection{Collection parameters}
Besides the source of collected traffic, router-based or/and producer-based solutions use a variety of parameters during the \textit{collect} phase. We conducted a quantitative and qualitative comparison based on the parameters used by existing solutions to collect traffic information. The first column of Table~\ref{tab:collect} specifies proposed solutions. The next two columns indicate if a solution collects router or/and producer information during this phase. The fourth column shows whether a solution stores or not the collected traffic information. The fifth column provides the routers categories used by a solution. Routers are classified depending on their role in the system. The default value equals one. The following two columns specify whether a solution modifies the PIT and FIB tables. The "New structure" field indicates if a solution introduces a new data structure. The last column specifies the solutions that rely on a central node during the Collect phase.           
\begin{table*}
\centering
\scriptsize
\caption{Collection parameters used by existing solutions}
\label{tab:collect}
\begin{adjustbox}{width=\textwidth}
\begin{tabular}{cllllllllll}
\hline
\textbf{Ref} & \makecell{\textbf{Router}\\\textbf{statistics}} & \makecell{\textbf{Producer}\\\textbf{statistics}} & \makecell{\textbf{Statistics}\\\textbf{storage}} & \textbf{Router classes} & \textbf{Modified PIT} & \textbf{Modified FIB} & \textbf{New structure} & \makecell{\textbf{Central}\\\textbf{node}}\\ \hline

\cite{afanasyev2013interest} & Yes & No & No & 01 & \makecell[l]{Flags:\\Inqueue, Fwd} & No & No & No \\ \hline

\cite{compagno2013poseidon} & Yes & No & No & 01 & No & No & No & No \\ \hline

\cite{dai2013mitigate} & Yes & No & No & 01 & No & No & No & No \\ \hline

\cite{wang2013decoupling} & Yes & No & Yes & 01 & No & No & \makecell[l]{Affected\\prefixes list} & No \\ \hline

\cite{liu2018accuracy} & Yes & Yes & Yes & 01 & No & No & \makecell[l]{Affected\\prefixes list} & No \\ \hline
 
\cite{wang2014cooperative} & Yes & No & Yes & 01 & No & No & No & No \\ \hline

\cite{wang2014detecting} & Yes & No & Yes & 01 & No & Yes & No & No \\ \hline

\cite{karami2015hybrid} & Yes & No & No & 01 & No & No & No & No \\ \hline

\cite{zhi2019resist} & Yes & No & Yes & 01 & No & No & No & No \\ \hline

\cite{vassilakis2015mitigating} & Yes & No & No & 01 & No & No & No & No \\ \hline

\cite{salah2015coordination}, \cite{salah2016evaluating} & Yes & No & Yes & \makecell[l]{
    \tabitem Monitoring\\
    \tabitem Normal} & No & No & No & Yes \\ \hline

\cite{yin2019controller} & Yes & No & Yes & 01 & No & No & ADT and AIT & Yes \\ \hline

\cite{nguyen2015detection} & Yes & No & Yes & 01 & No & No & No & No \\ \hline

\cite{xin2017detection} & Yes & No & No & 01 & No & No & No & No \\ \hline

\cite{shigeyasu2018distributed} & Yes & No & Yes & 01 & No & No & No & No \\ \hline

\cite{ding2016cooperative} & Yes & No & Yes & 01 & No & No & \makecell[l]{Unsatisfied\\\textit{Interests} cache} & No \\ \hline

\cite{xin2016novel} & Yes & No & Yes & 01 & No & No & No & No \\ \hline

\cite{hou2019theil} & Yes & No & Yes & 01 & No & No & No & No \\ \hline

\cite{zhi2018gini} & Yes & No & Yes & 01 & No & No & No & No \\ \hline

\cite{nakatsuka2018frog} & Yes & No & Yes & 01 & No & No & No & No \\ \hline

\cite{benarfa2019chokifa}, \cite{benarfa2020chokifap} & Yes & No & No & 01 & No & No & No & No \\ \hline

\cite{zhi2020reputation} & Yes & No & No & 01 & No & No & No & No \\ \hline

\cite{zhang2019expect} & Yes & Yes & Yes & 01 & No & No & No & No \\ \hline

\cite{pu2019self} & Yes & No & Yes & 01 & No & No & \makecell[l]{Operation\\Trace Table} & No \\ \hline

\cite{cheng2019detecting} & Yes & No & Yes & \makecell[l]{
    \tabitem Monitoring\\
    \tabitem Normal} & No & No & No & Yes \\ \hline

\cite{chen2019isolation} & Yes & No & Yes & 01 & No & No & No & No \\ \hline

\cite{benmoussa2019novel} & Yes & No & No & 01 & No & No & No & No \\ \hline

\cite{dong2020interestfence} & No & Yes & Yes & 01 & No & No & \makecell[l]{Affected\\prefixes list} & No \\ \hline

\cite{tang2013identifying} & Yes & No & No & 01 & No & No & No & No \\ \hline

\cite{shinohara2016cache} & Yes & No & No & 01 & No & No & No & No \\ \hline

\cite{benmoussa2020msidn} & Yes & Yes & No & 01 & No & No & No & No \\ \hline

\cite{alhisnawidetecting} & Yes & No & Yes & 01 & No & No & \makecell[l]{Quotient based\\Cuckoo filter} & Yes \\ \hline 

\cite{wu2020mitigation} & Yes & No & No & 01 & No & No & No & No \\ \hline

\cite{tourani2020persia} & Yes & No & No & 01 & No & No & Bloom filter & No \\ \hline

%PROACTIVE SOLUTIONS
\cite{wang2017economic} & No & No & No & 01 & No & No & No & No \\ \hline

\cite{zhang2019ari} & Yes & No & No & 01 & No & No & No & No \\ \hline

%Alston et al.\cite{alston2016neutralizing} & No & No & No & 01 & Not used & No & No & No \\ \hline

\cite{liu2018blam} & No & No & No & 01 & No & No & \makecell[l]{Bloom filter\\array} & No \\ \hline

\textbf{Distribution} & 87\% & 10\% & 51\% & NA & 2\% & 2\% & 23\% & 10\% \\ \hline

\end{tabular}
\end{adjustbox}
\end{table*}

\subsubsection{Key observations of our comparison}
The fundamental observations of our comparative study on the information that solutions collect to detect attacks are as follows: The first thing that can be noted from Table~\ref{tab:collect} is that the majority of authored solutions rely on router information to detect IFA. Routers are the heart of a network. Each connection session passes through a router. That explains the reason of the deployment of the solutions on the routers. It permits gathering a maximum of traffic information, which helps detect ongoing attacks. Routers are favored over producers when deploying solutions, as shown in this study. Producers do not have a global view of a network. They are limited to the communication details that they receive. Solutions that rely only on producers’ information are not efficient to detect attacks against routers. That explains why only one solution chose to counter IFA using session information that producers collect.
As NDN uses application-layer names at the network level, routers and producers can cooperate to share information. Some solutions took the hybrid approach. These solutions rely on both routers and producers to gather information. It is the case of three authored solutions.%, which represent 7\% of all IFA solutions. 
Another key observation regards information storage. Our comparison shows that half of the solutions record connection statistics. Stored information help in detecting attacks with an additional precision level as it implicates the use of more data. It also permits detecting behavioral changes in some cases. However, this process implies using additional storage capacity. In addition, it slows routers at high network peaks, especially those close to the core network. That explains why researchers are divided on the use of information storage.
Our comparative study found that only two solutions adopted personalized PIT and FIB tables. Personalized tables occupy more space in memory as they store additional information, and the memory space they occupy grows considerably for core routers. Furthermore, connection processing times increase as routers perform more checks before forwarding packets, which leads to higher network delays. That explains why solutions do not use personalized tables. Nevertheless, solutions may employ a completely new structure, as shown in this comparative study. We notice that almost a quarter of solutions uses a new data structure. They are usually associated with PIT. Using a new data structure comes with a trade-off. It supplies additional information for routers but at the same time occupies more space and introduces processing and computational overhead.  
Finally, our study shows that some solutions rely on a central node to collect traffic information. They represent 10\% of authored solutions. Router classes are mainly associated with centralized solutions. They describe the role that routers occupy in a system. We will discuss centralized solutions further in the next subsection.

\subsubsection{Collection metrics}
Routers and producers collect a variety of traffic-related statistics. Table~\ref{tab:detect_metric} summarizes the metrics that existing solutions gather during the Collect phase. The first column of the table specifies proposals. The second column represents the rate of incoming \textit{Interest} packets of each interface. When the sending rate of an interface reaches a threshold, the router suspects malicious behavior. Solutions usually couple the traffic rate with other metrics. The third column corresponds to the ratio of received \textit{Data} packets to the total number of requested \textit{Interest} packets in a given period. Routers/producers calculate the satisfaction ratio of each interface separately. The next column denotes the occupancy ratio of each interface in PIT, i.e., the number of pending \textit{Interest} packets generated by an interface to the total number of PIT entries. The following four columns of the table show the nature of packets being counted during this phase. Solutions may collect the number of received \textit{Interest} and \textit{Data} packets, the number of timed-out pending \textit{Interest} packets, and the number of NACK packets. The ninth column shows whether a solution stores the name of received \textit{Interest} packets. The last column of the table shows whether the solution collects producer-based metrics, including but is not limited to processing overhead, memory consumption, and service overload.

%Collection metrics table
\begin{table*}
\setlength{\tabcolsep}{4pt}
\centering
\scriptsize
\caption{Collected metrics by existing solutions}
\label{tab:detect_metric}
\begin{adjustbox}{width=\textwidth}
\begin{tabular}{cccccccccc}
\hline
\multicolumn{9}{c}{\textbf{Router-based metrics}} & \makecell{\textbf{Producer-based}\\\textbf{metrics}} \\ \hline
\textbf{Ref} & \makecell{\textbf{Satisfaction}\\\textbf{ratio}} & \textbf{PIT usage} & \textbf{Traffic rate} & \makecell{\textbf{Timed-out}\\\textbf{\textit{Interests}}} & \makecell{\textbf{Number of}\\\textbf{\textit{Data} packets}} & \makecell{\textbf{Number of}\\\textbf{\textit{Interests}}} & \makecell{\textbf{Number of}\\\textbf{NACK}} & \makecell{\textbf{\textit{Interest}}\\\textbf{prefixes}} &  \\ \hline

%\multicolumn{10}{c}{\textbf{Reactive solutions}} \\ \hline
\cite{afanasyev2013interest} & \cmark &  &  &  &  &  &  &  & \\ \hline

\cite{compagno2013poseidon} &  & \cmark &  & \cmark &  &  &  &  & \\ \hline

\cite{dai2013mitigate} &  & \cmark &  &  &  &  &  &  & \\ \hline

\cite{wang2013decoupling} &  &  &  & \cmark &  &  &  &  & \\ \hline

\cite{liu2018accuracy} &  &  &  &  &  &  &  &  & \makecell{Malicious\\\textit{Interests}} \\ \hline 

\cite{wang2014cooperative} &  & \cmark &  & \cmark &  &  &  &  & \\ \hline

\cite{wang2014detecting} &  &  &  & \cmark &  &  &  &  & \\ \hline

\cite{karami2015hybrid} & \cmark &  & \cmark & \cmark & \cmark & \cmark &  &  & \\ \hline

\cite{zhi2019resist} & \cmark & \cmark &  &  &  &  &  & \cmark &  \\ \hline

\cite{vassilakis2015mitigating} &  &   &  & \cmark &  &  &  &  & \\ \hline

\cite{salah2015coordination} &  & \cmark &  & \cmark &  &  &  &  & \\ \hline

\cite{salah2016evaluating} &  & \cmark &  &  &  &  &  &  & \\ \hline

\cite{yin2019controller} & \cmark &  &  &  &  &  & \cmark &  &  \\ \hline

\cite{nguyen2015detection} &  &  &  &  & \cmark & \cmark &  &  & \\ \hline

\cite{xin2017detection} &  &  &  &  &  & \cmark &  &  & \\ \hline

\cite{shigeyasu2018distributed} &  & \cmark & \cmark &  &  &  &  &  &  \\ \hline

\cite{ding2016cooperative} &  & \cmark &  & \cmark &  & \cmark &  &  & \\ \hline

\cite{xin2016novel} &  &  &  &  &  &  &  & \cmark & \\ \hline

\cite{hou2019theil} &  &  &  &  &  &  &  & \cmark & \\ \hline

\cite{zhi2018gini} &  &  &  &  &  &  &  & \cmark &  \\ \hline

\cite{nakatsuka2018frog} &  &  &  &  &  &  &  &  &  \\ \hline

\cite{benarfa2019chokifa},\cite{benarfa2020chokifap} & \cmark & \cmark &  &  &  &  &  &  &  \\ \hline

\cite{zhi2020reputation} & \cmark &  &  &  &  &  &  &  &  \\ \hline

\cite{zhang2019expect} &  &  &  &  &  &  &  &  & \makecell{Resource\\exhaustion}\\ \hline

\cite{pu2019self} & \makecell{Unsatisfaction\\ratio} &  &  &  &  &  &  &  &  \\ \hline

\cite{cheng2019detecting} &  &  & \cmark & \cmark &  &  &  &  &  \\ \hline

\cite{chen2019isolation} &  & \cmark &  & \cmark & \cmark & \cmark &  &  &  \\ \hline

\cite{benmoussa2019novel} & \cmark &  & \cmark & \cmark &  &  & \cmark &  &  \\ \hline

\cite{dong2020interestfence}&  &  &  &  &  &  &  &  & \makecell{Number of\\forged \textit{Interests}} \\ \hline 

\cite{tang2013identifying} & \cmark &  &  & \cmark &  &  &  &  &  \\ \hline

\cite{shinohara2016cache} & \cmark & \cmark &  &  &  &  &  &  &  \\ \hline

\cite{benmoussa2020msidn} & \cmark &  &  & \cmark &  &  &  &  & \makecell{Resource\\exhaustion} \\ \hline

\cite{alhisnawidetecting} &  & \cmark &  & \cmark &  &  &  &  & \\ \hline  

\cite{wu2020mitigation} &  & \cmark & \cmark &  &  &  &  &  &  \\ \hline 

\cite{tourani2020persia} &  &  & \cmark &  &  &  &  &  & \\ \hline
%PROACTIVE SOLUTIONS
%\multicolumn{10}{c}{\textbf{Proactive solutions}} \\ \hline
%\cite{wang2017economic} &  &  &  &  &  &  &  &  &  \\ \hline

\cite{zhang2019ari} & \cmark &  &  &  &  &  &  &  &  \\ \hline

\cite{liu2018blam} &  &  &  &  &  &  &  &  &  \\ \hline

\textbf{Distribution} & 30\% & 33\% & 15\% & 35\% & 7\% & 12\% & 5\% & 10\% & NA\\ \hline
\end{tabular}
\end{adjustbox}
\end{table*}

\subsubsection{Key observations of our comparison}
The key observations of the performed comparative study on metrics used by solutions are as follows: First, we notice that the most used detection metric is the number of timed-out \textit{Interest} packets. It was used by 35\% of existing solutions. Many solutions consider timed-out pending PIT entries as an alarm for an ongoing attack. Attackers usually use non-valid requests to overwhelm networks, and malicious \textit{Interests} stay in PIT until they time out. That is why solutions usually consider the high number of timed-out \textit{Interests} as a potential attack. However, this metric may lead to false positives, e.g., non-available producers. Another observation from Table~\ref{tab:detect_metric} states that the second most frequently used metric is PIT occupancy. The goal of adversary nodes when they perform IFA is to fill the target’s PIT with invalid requests so it cannot accept legitimate \textit{Interests}. That justifies the use of PIT occupancy as a metric to detect attacks. However, this metric is not reliable in every scenario, as we will see in the next section.
Our comparative study found that the satisfaction ratio is the third most used metric with 30\% of solutions. Attackers usually perform IFA with invalid requests. That is why solutions consider low satisfaction-ratio levels as a sign of ongoing attacks. We also notice that solutions adopt either the satisfaction ratio or the numbers of timed-out \textit{Interest}, as they both rely on the fact that the adversary floods the network with invalid requests. However, coupling these two metrics can help reduce false-positive detections that result from the use of the number of timed-out \textit{Interest} packets. The next most used metric is the traffic rate. It was used by 15\% of authored solutions. Adversary nodes send malicious requests at a high pace to overwhelm a network. That is why IFA is usually associated with high traffic rates. We also found that solutions always associate the traffic rate with another metric.
Our comparison shows that some solutions count the number of received packets. This includes the number of \textit{Interest} packets, \textit{Data} packets, and NACK packets. For instance, we notice that the solutions, which use the number of \textit{Interest} packets, usually compare it with timed-out \textit{Interests}. We also remark that the solutions which collect the number of \textit{Data} packets always compare it with the number of \textit{Interests}. On the other hand, only two solutions keep statistics of NACK packets. One used it to detect network congestion and the other to exchange information. The last router-based metric of our comparative study regards \textit{Interest} prefixes. This metric was essentially used by statistical-based solutions. They use this metric to calculate an entropy to detect prefixes under attack.  
Finally, our last observation regards the number of metrics used by each solution. We found that 39\% of authored solutions rely only on one metric, and 34\% of solutions use two metrics to detect IFA. It is considered insufficient to detect IFA in different situations. We will discuss this point further in the following sections. 

\subsection{Detect}
The second activity of the CDA workflow is \textit{"Detect"}, in which solutions process the collected information to detect anomalies and ongoing attacks. There are two main IFA detection architecture designs, centralized, like \cite{salah2016evaluating} and \cite{cheng2019detecting}, and distributed, like \cite{compagno2013poseidon} and \cite{dai2013mitigate}. %Figure~\ref{fig:detect} shows the different detection approaches that IFA solutions may adopt. 
Before we detail the detection parameters that solutions may adopt during the \textit{Detect} phase, we first discuss detection architecture designs. There are two architecture designs: centralized and distributed.

%\begin{figure*}
%    \centering
%    \includegraphics[scale=0.53]{Figures/Detect.pdf}
%    \caption{Classification of detection methods}   
%    \label{fig:detect}
%\vspace{-0.5cm}
%\end{figure*}

\subsubsection{Centralized detection design}
\begin{figure}
\centering
\begin{subfigure}[Cluster-based centralized detection]{
    \includegraphics[width=.32\linewidth] {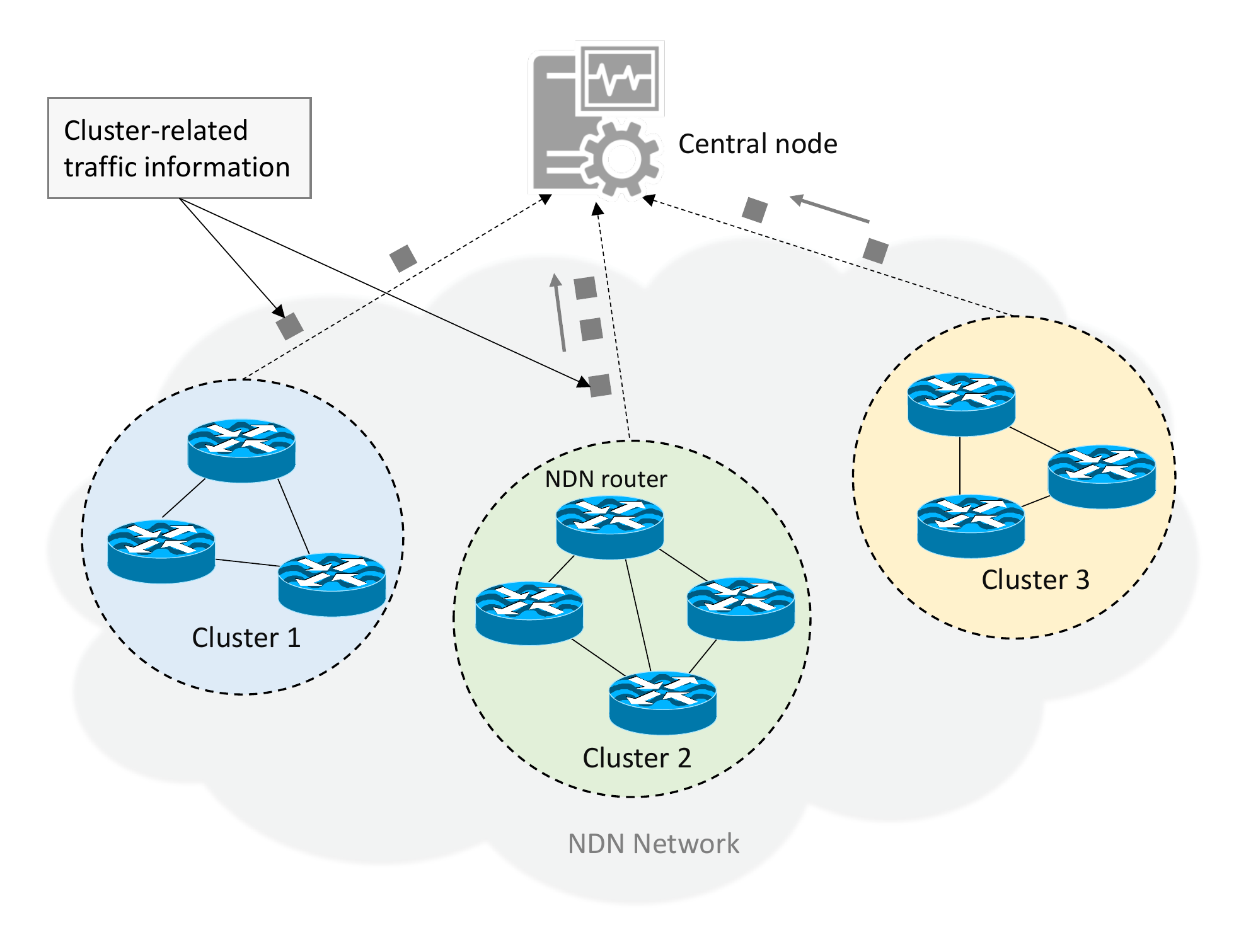}
    \label{fig:cluster_detect}}
    \end{subfigure}
    \hfill
\begin{subfigure}[Selective nodes centralized detection]{
    \includegraphics[width=.31\linewidth] {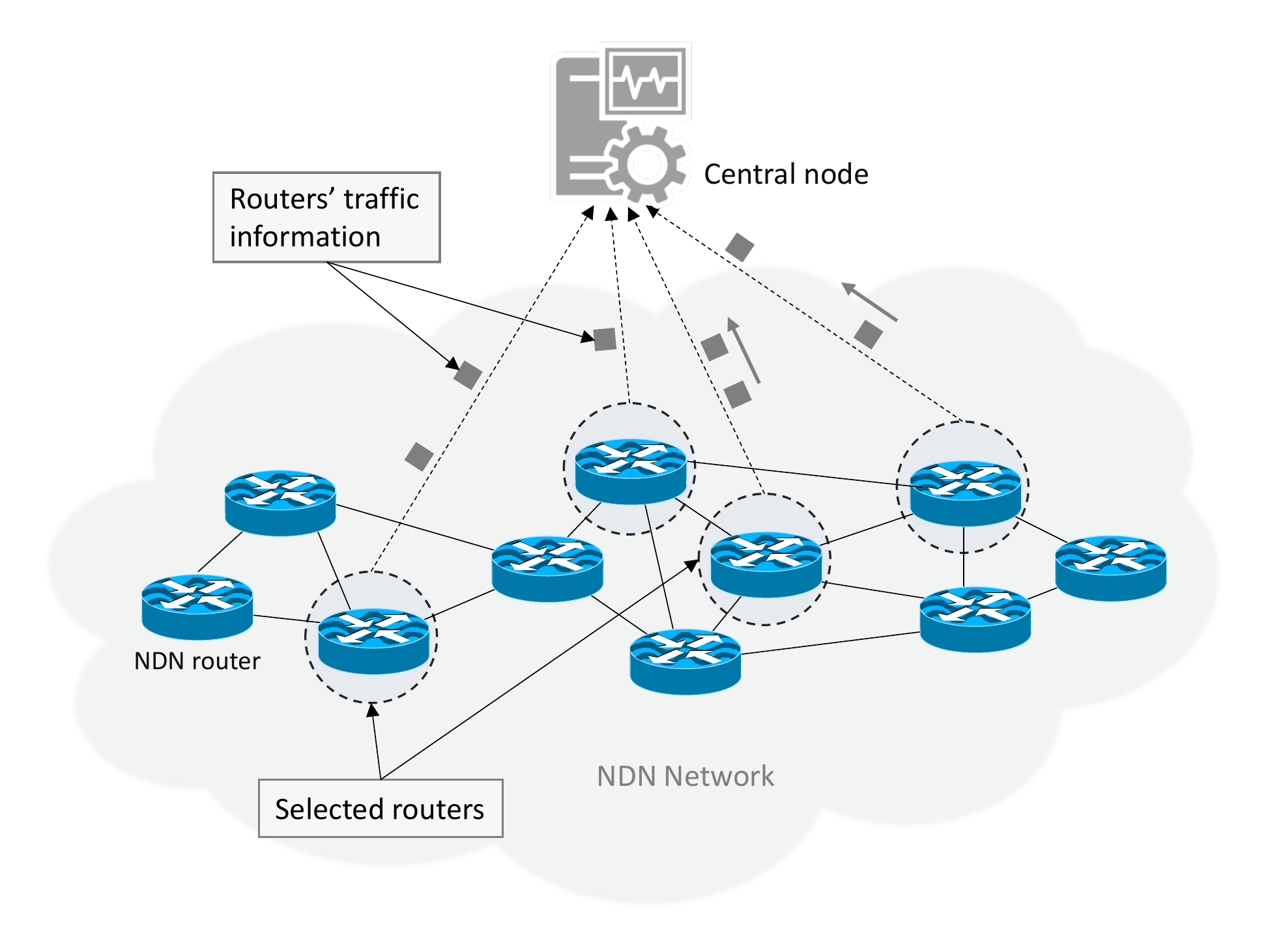}
    \label{fig:selective_detect}}
    \end{subfigure}
    \hfil
\begin{subfigure}[Global nodes centralized detection]{
    \includegraphics[width=.31\linewidth] {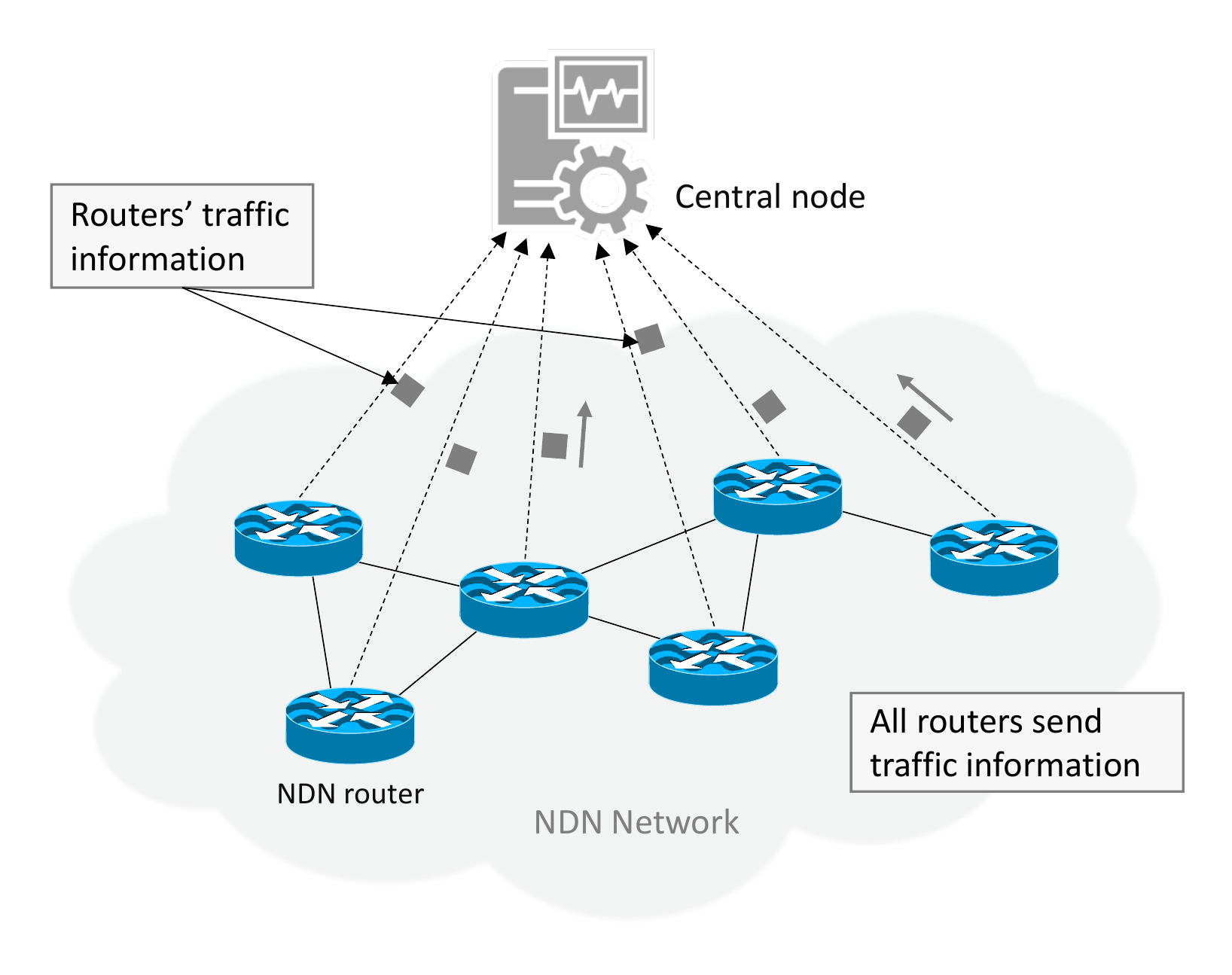}
    \label{fig:global_detect}}
    \end{subfigure}
\vspace{-0.2cm}
\caption{Distributed detection architectures}
\vspace{-0.5cm}
\end{figure}

The centralized detection architecture relies on a central node to detect an IFA. The central node periodically receives traffic-related information from routers to analyze and decide whether an IFA is happening or not. The centralized detection is further categorized into cluster-based detection, selective nodes detection and global detection.
\begin{enumerate}[label=(\alph*)]
    \item \textit{Cluster-based}: 
    The first centralized detection topology is cluster-based centralized detection. The central node groups the network routers into different clusters. The cluster will then collect its traffic information before sending it to the central node.  After that, the central node analyzes the received information to detect attacks. When an IFA is detected, it sends back to the clusters the actions to take in order to mitigate the attack. Figure~\ref{fig:cluster_detect} illustrates the cluster-based centralized detection. 
    None of the existing solution adopted a cluster-based detection approach. 

    \item \textit{Selective nodes}: 
    The second centralized-based detection topology is using selective nodes. The central node selects a group of routers from the network. The selection criteria are essential to ensure a global view of the network. That is why the central node needs to select the most relevant routers to cover the maximum of the network’s traffic. The selected routers are responsible for sending the information that the central node needs in order to evaluate whether the traffic is malicious or not. This topology is illustrated in Fig.~\ref{fig:selective_detect}. 
    After that, the central node sends back to the selected routers the proper actions to take. Solutions \cite{salah2015coordination},\cite{salah2016evaluating}, and \cite{cheng2019detecting} adopted the selective centralized detection mode.
    
    \item \textit{Global}:
    The third and last centralized-based detection topology is the global topology. Every router in the network is part of the system. That is, every network router collects and sends the information of the traffic passing through it to the central node, as showed in Fig.~\ref{fig:global_detect}. 
    Similar to other centralized detection topologies, the central node analyzes the information that it receives from the network routers before deciding the proper actions to take. The solution presented in \cite{yin2019controller} chose the global centralized approach as its detection mode.
\end{enumerate} 

\subsubsection{Distributed detection design}
\begin{enumerate}[label=(\alph*)]
    \item \textit{Autonomous detection}: 
    The first distributed detection approach is the autonomous detection. It means that network routers and data producers collect and detect IFA locally. In this approach, the network nodes do not share traffic or attack information with each other. The majority of existing works adopted the autonomous detection approach like \cite{benmoussa2019novel},\cite{benarfa2019chokifa}, and \cite{nakatsuka2018frog}.  
    
    \item \textit{Cooperative detection}:
    The second distributed detection approach that the network nodes can adopt is the cooperative detection. Network nodes share traffic-related information and take cooperative actions to mitigate IFA. The cooperative detection is further classified into partial cooperation and full cooperation. 
    \begin{enumerate}[label=(\arabic*)]
         \item \textit{Partial cooperation}: 
        In a partial cooperation design, the network nodes collaborate only on the mitigation action, like \cite{afanasyev2013interest}, \cite{zhi2019resist} and \cite{benmoussa2020msidn}. Each node analyzes its local traffic to detect an IFA and then takes the proper action to mitigate it. After that, the node informs its neighbors about the action and cooperates to block the attack and stop its growth.  
        \item \textit{Full cooperation}: 
        In a full cooperation distributed detection, the nodes fully cooperate to detect and mitigate IFA, i.e., the nodes share information and jointly decide on the actions to take. Solutions presented in \cite{liu2018accuracy} and  \cite{wang2017economic} adopted the full cooperation detection mode. 
    \end{enumerate}
\end{enumerate}

\subsubsection{Detection parameters}
Proposed solutions employ multiple parameters to detect attacks. Table~\ref{tab:detect} details the parameters used by each mechanism. The first column of the table specifies proposals. The second column defines the entities that are responsible for detecting attacks. The following two columns indicate the architecture approach that solutions use to detect IFA, centralized or distributed. Centralized detection may employ the entire routers or just a selection of them. In distributed detection, routers work solemnly or cooperatively to detect attacks. The fifth column shows whether a solution employs AI to detect attacks. The sixth column specifies if a solution considered network congestion when detecting IFA. The "Check interval" field represents the time that a solution waits for before it performs another detection check.  The eight parameters determines the nature of IFA that a solution detect. The following two fields indicate exchanged packets during the Detect phase. The first parameter specifies if a solution introduces a new packet. The second parameter shows whether a solution modifies a packet (essentially an \textit{Interest} packet). "The number of thresholds" field represents the set of thresholds that a solution uses to detect attacks. The final parameter shows the categories that a solution uses to classify consumers.

%Detect classification table
\begin{table*}
\centering
\caption{Detection parameters used by existing solutions}
\label{tab:detect}
\setlength{\tabcolsep}{4pt}
\begin{adjustbox}{width=\textwidth}
\begin{tabular}{cllllllllllll}
\hline
 \textbf{Ref} & \makecell{\textbf{Detection}\\\textbf{actors}} & \makecell{\textbf{Centralized}\\\textbf{detection}} & \makecell{\textbf{Distributed}\\\textbf{detection}} & \textbf{AI-based} & \makecell{\textbf{Congestion}\\\textbf{aware}} & \textbf{Check interval} & \textbf{IFA type} & \makecell{\textbf{Specific}\\\textbf{packets}} & \makecell{\textbf{Modified}\\\textbf{packets}} & \makecell{\textbf{Number of}\\\textbf{thresholds}} & \makecell{\textbf{Consumer}\\\textbf{classes}} \\
 \hline

%\multicolumn{10}{c}{\textbf{Reactive solutions}} \\ \hline
\cite{afanasyev2013interest} & All routers & No & Autonomous & No & No & $1sec$ & Non existent & No & No & 01 & - \\ \hline

\cite{compagno2013poseidon} & All routers & No & Autonomous & No & No & $60ms$ & Non existent & No & No & 02 & - \\ \hline

\cite{dai2013mitigate} & All routers & No & Autonomous & No & No & - & Non existent & No & No & 01 & - \\ \hline

\cite{wang2013decoupling} & All routers & No & Autonomous & No & No & - & Non existent & No & No & 01 & - \\ \hline

\cite{liu2018accuracy} & \makecell[l]{All routers\\Producers} & No & Cooperative & No & No & - & Non existent & No & No & 01 & - \\ \hline 

\cite{wang2014cooperative} & All routers & No & Cooperative & Fuzzy logic & No & real-time & - & No & No & 02 & \makecell[l]{Normal\\Malicious} \\ \hline

\cite{wang2014detecting} & All routers & No & Autonomous & No & No & - & Non existent & No & No & 02 & - \\ \hline

\cite{karami2015hybrid} & All routers & No & Autonomous & RBF NN & No & - & All types & No & No & 02 & - \\ \hline

\cite{zhi2019resist} & All routers & No & Autonomous & SVM & No & time period & Non existent & No & No & - & -\\ \hline

\cite{vassilakis2015mitigating} & Edge routers & No & Autonomous & No & No & - & Non existent & No & No & 02 & \makecell[l]{Legitimate\\Suspicious\\Malicious} \\ \hline

\cite{salah2015coordination} & \makecell[l]{Mon routers\\and DC} & \makecell[l]{Selective\\nodes} & Autonomous & No & No & Window q & Non existent & Yes & No & 02 & - \\ \hline

\cite{salah2016evaluating} & \makecell[l]{Mon routers\\and DC} & \makecell[l]{Selective\\nodes} & Autonomous & No & No & Window q & Collusive & Yes & No & 01 & - \\ \hline

\cite{yin2019controller} & DC & \makecell[l]{Global\\nodes} & No & No & No & - & Non existent & Yes & No & 02 & - \\ \hline

\cite{nguyen2015detection} & All routers & No & Autonomous & No & No & - & Non existent & No & No & 01 & - \\ \hline

\cite{xin2017detection} & Edge routers & No & Autonomous & No & No & - & Collusive & No & No & 01 & - \\ \hline

\cite{shigeyasu2018distributed} & All routers & No & Autonomous & No & No & - & Collusive & No & No & 02 & - \\ \hline

\cite{ding2016cooperative} & Edge routers & No & Autonomous & No & No & Windows L & Non existent & No & \makecell[l]{\textit{Interest} state\\ and router ID} & 01 & - \\ \hline

\cite{xin2016novel} & All routers & No & Autonomous & No & No & - & Non existent & No & No & 01 & - \\ \hline

\cite{hou2019theil} & All routers & No & Autonomous & No & No & - & Non existent & No & No & 01 & - \\ \hline

\cite{zhi2018gini} & All routers & No & Autonomous & No & No & Time $\Delta t$ & Non existent & No & No & 01 & - \\ \hline

\cite{nakatsuka2018frog} & Edge routers & No & Autonomous & No & No & Time window & Existent & No & No & 02 & - \\ \hline

\cite{benarfa2019chokifa}, \cite{benarfa2020chokifap} & All routers & No & Autonomous & No & No & - & Non existent & No & No & 03 & - \\ \hline

\cite{zhi2020reputation} & All routers & No & Autonomous & No & No & - & Collusive & No & No & 02 & - \\ \hline

\cite{zhang2019expect} & Producers & No & Autonomous & No & No & - & All types & No & No & \makecell[l]{Producer\\based} & \makecell[l]{Normal\\Suspicious} \\ \hline

\cite{pu2019self} & Edge routers & No & Autonomous & No & No & Window $\omega$ & Non existent & No & No & 01 & - \\ \hline
 
\cite{cheng2019detecting} & \makecell[l]{Mon routers\\DC} & \makecell[l]{Selective\\nodes} & No & No & No & - & Non existent & Yes & No & 03 & - \\ \hline

\cite{chen2019isolation} & All routers & No & Autonomous & No & No & time interval & Non existent & No & No & 02 & - \\ \hline

\cite{benmoussa2019novel} & Edge routers & No & Autonomous & No & Yes & - & Non existent & No & No & 04 & \makecell[l]{Legitimate\\Suspicious\\Malicious} \\ \hline

\cite{dong2020interestfence} & Producers & No & Autonomous & No & No & - & Non existent & No & No & \makecell[l]{Producer\\based} & - \\ \hline

\cite{tang2013identifying} & All routers & No & Autonomous & No & No & - & Non existent & No & No & 02 & - \\ \hline 

\cite{shinohara2016cache} & All routers & No & Autonomous & No & No & - & Non existent & No & No & 02 & - \\ \hline

\cite{benmoussa2020msidn} & \makecell[l]{All routers\\Producers} & No & Cooperative & No & No & - & All types & No & No & 03 & \makecell[l]{Normal\\Suspicious\\Harmful}\\ \hline

\cite{alhisnawidetecting} & All routers & No & Cooperative & No & No & - & Non-existent & No & No & 02 & - \\ \hline 

\cite{wu2020mitigation} & All routers & No & Autonomous & No & No & Time window & Collusive & No & No & 02 & - \\ \hline

\cite{tourani2020persia} & Core routers & No & Autonomous & No & No & - & Non existent & No & No & 01 & - \\ \hline

%PROACTIVE SOLUTIONS
\cite{wang2017economic} & All routers & No & Cooperative & No & No & - & All types & No & \makecell[l]{PIT delay\&\\\ content fees} & - & - \\ \hline

\cite{zhang2019ari} & Edge routers & No & Autonomous & No & No & - & Non existent & No & No & - & \makecell[l]{Legal\\Bad} \\ \hline

\cite{liu2018blam} & All nodes & No & Autonomous & No & No & - & Non existent & No & No & - & - \\ \hline

\textbf{Distribution} & NA & 10\% & NA & 7\% & 2\% & NA & NA & 10\% & 5\% & NA & NA \\ \hline
\end{tabular}
\end{adjustbox}
\end{table*}

\subsubsection{Key observations of our comparison}
The key observations of our comparative study on the parameters that solution use during the detection phase are as follows: First, we found that 89\% of solutions use routers as the detection actor. It includes both router-based and hybrid solutions. This number was predictable, as we previously noticed, in Table~\ref{tab:collect}, that the majority of solutions rely on routers' information. We also observed that all centralized solutions rely on the central node to decide whether IFA exists or not. 
Second, we notice that 94\% of authored solutions adopted the distributed detection design, and only 10\% chose the centralized architecture. Centralized solutions require additional deployment phases. It implies the exchange of control information with system nodes. Additionally, deployment complexity grows, as the network gets bigger. Furthermore, the traffic information that system nodes send to the central node may burden the network, especially in large deployments. That explains why a small portion of existing countermeasure solutions adopted the centralized architecture. Nevertheless, we found that two solutions chose a hybrid approach to detect IFA. Another key observation regards the use of AI to detect IFA. Although it can help increase detection precision and reduce false positives, only 10\% of authored solutions used an AI-based algorithm. AI-based solutions require additional computational and memory resources, which may penalize some nodes. That explains why researchers usually do not choose to rely on AI. 
Our comparative study found that most proposals are solutions against the third type of IFA. As discussed earlier, it represents the most harmful IFA type, as it implies the use of invalid requests. The first type of IFA, which uses valid requests, causes less harm compared to the third type. That explains why 80\% of solutions chose to counter IFA with non-existent content, and only one was authored for the first type. However, some solutions counter both types. It is the case of 10\% of solutions. These solutions usually rely on both router-based and producer-based information to detect attacks. On the other hand, solutions against collusive IFA represent only 13\%. It shows that the research community did not study in detail this IFA variant.
Our last observation from this comparative study points to control packets that solutions exchange during detection. Solutions adopted two approaches: the first one implies the creation of new specific packets. This approach was exclusively used by centralized solutions. The second approach consists of sending control information within exchanged \textit{Interest} packets. It modifies \textit{Interest} packets, usually their name, to include additional information for neighboring routers. Using control packets may introduce delays to the network, as routers need to process them. Solutions that rely on control packets during detection represents 15\% of the overall solutions.

\subsection{Act}
The last activity of the CDA workflow is "Act". After collecting the traffic related information, and detecting the existence on an IFA, solutions act by taking mitigation actions to stop the attack.
Table \ref{tab:act} summarizes the mitigation parameters used by existing solutions. Before we detail the mitigation actions that solutions use to stop attacks, we first discuss the mitigation parameters that solutions may employ during this phase.

\subsubsection{Mitigation parameters}
Table~\ref{tab:act} lists the parameters that solutions may use during the mitigation process. The first column specifies proposals. The second column corresponds to the way a solution work to mitigate attackers, autonomous or cooperative. System nodes may cooperate or work independently to stop IFA. The third column specifies the nodes that take the action against attacks. The following field represents the action that a solution adopts to counter IFA. Mitigation actions are detailed in the following subsection. The next two columns indicate whether a solution uses special packets during the mitigation process. The "Control information" field lists the information that a solution shares during the mitigation process. The last column specifies whether a solution uses signed \textit{Interest} packets or not.

\subsubsection{Router-based mitigation actions}
\begin{enumerate}[label=(\alph*)]
    \item \textit{Rate-limiting}:
    The first mitigation action that routers usually take after detecting an IFA is rate limiting. It consists of reducing the amount of traffic allowed from a given interface. When a router detects a malicious traffic, it first checks the source interface of this traffic. Following this, the router reduces the incoming traffic from this interface. The router may apply rate-limiting on the whole interface's traffic or just for a given namespace. The vast majority of existing IFA countermeasure solution act by limiting the rate of incoming traffic. 
    
    \item \textit{Block}: 
    The second router-based mitigation action consists of blocking the traffic of an interface. The network router, when necessary, can block an interface for a period. The blocking action is usually considered as a second-level reaction. When rate limiting does not suffice to throttle an attack, the router considers blocking the interface for a period. Interface blocking is particularly used by edge routers. The blocking periods may vary during time. Routers could increase the blocking period when the previous amount of time did not help to stop the attack. Blocking actions could be prefix-based (i.e.,  applied on a specific prefix) as used in \cite{tang2013identifying} and \cite{zhi2019resist}, or on interface-base (i.e., block all traffic of an interface) as used in \cite{benmoussa2019novel} and \cite{hou2019theil}.  
\end{enumerate}

\subsubsection{Producer-based mitigation actions}
The producer-based mitigation actions are essentially associated with blocking. Data producers, when needed, may block the incoming network traffic. Producer-based blocking action can be temporary or permanent.
\begin{enumerate}[label=(\alph*)]
    \item \textit{Temporary block}:
    Temporary blocking consists of blocking the network traffic for a limited period. To throttle attacks, data producers use temporary blocking against malicious traffic. The blocked traffic could be the whole traffic of an interface or just a portion of the traffic (e.g., the network traffic heading to a particular service) as used in \cite{benmoussa2020msidn}.
    
    \item \textit{Permanent block}:
    Data producers may also use permanent blocking against consumers. Producers can implement a security policy to permanently block consumers when needed (i.e., blacklisting consumers). Producers do not use permanent block against network interfaces.   
\end{enumerate}

%Act classification table
\begin{table*}
\centering
\footnotesize
\caption{Mitigation parameters used by existing solutions}
\label{tab:act}
\begin{adjustbox}{width=\textwidth}
\begin{tabular}{cllllllll}
\hline
\textbf{Ref} & \makecell{\textbf{Mitigation}\\\textbf{method}} & \makecell{\textbf{Mitigation}\\\textbf{actors}} & \makecell{\textbf{Mitigation}\\\textbf{actions}} & \makecell{\textbf{Specific}\\\textbf{packets}} & \textbf{Specific namespace} & \makecell{\textbf{Control}\\\textbf{information}} & \makecell{\textbf{Signed}\\\textbf{\textit{Interests}}} \\
 \hline
\cite{afanasyev2013interest} & Autonomous & All routers & Rate-limit & No & No & No & No \\ \hline
 
\cite{afanasyev2013interest} & Cooperative & All routers & Rate-limit & Yes & - & \makecell[l]{Rate limit\\announcements} & No\\ \hline

\cite{compagno2013poseidon} & Cooperative & All routers & Rate-limit & Yes & \texttt{/pushback/alerts/} & Reduced rate & No \\ \hline

\cite{dai2013mitigate} & Cooperative & All routers & Rate-limit & Yes & No & Spoofed \textit{Data} & No \\ \hline

\cite{wang2013decoupling} & Cooperative & All routers & \makecell[l]{Forwards \textit{Interest}\\without using PIT} & Yes & No & Interfaces list & No \\ \hline

\cite{liu2018accuracy} & Cooperative & \makecell[l]{All routers\\Producers} & \makecell[l]{Sends Nack} & Yes & No & \makecell[l]{Malicious\\\textit{Interest}} & No \\ \hline 

\cite{wang2014cooperative} & Cooperative & Edge routers & \makecell[l]{Rate-limit (prefix)} & Yes & \texttt{/ALERT/IFA/} & Malicious prefix & No \\ \hline

\cite{wang2014detecting} & Autonomous & All routers & \makecell[l]{Rate-limit (prefix)} & No & No & No & No \\ \hline

\cite{karami2015hybrid} & Cooperative & All routers & Rate-limit & Yes & \texttt{/pushbackmessage/alert/} & Reduced rate & No \\ \hline

\cite{zhi2019resist} & Cooperative & All routers & \makecell[l]{Block (prefix)} & Yes & Empty & \makecell[l]{Malicious\\prefixes} & No \\ \hline 

\cite{vassilakis2015mitigating} & Autonomous & Edge routers & \makecell[l]{Rate-limit\\Block} & Yes & - & Blocked user & No \\ \hline

\cite{salah2015coordination}, \cite{salah2016evaluating} & Cooperative & Mon routers & \makecell[l]{Rate-limit (prefix)} & Yes & - & Infected prefixes & No \\ \hline

\cite{yin2019controller} & Cooperative & Edge routers & Block & Yes & \texttt{/CTRL/} & {Malicious prefixes} & No \\ \hline

%\cite{nguyen2015detection} & - & - & - & - & - & - & - \\ \hline

%\cite{xin2017detection} & - & - & - & - & - & - & - \\ \hline

%\cite{shigeyasu2018distributed} & - & - & - & - & - & - & - \\ \hline 

\cite{ding2016cooperative} & Cooperative & All routers & \makecell[l]{Rate-limit (prefix)} & No & No & No & No \\ \hline

\cite{xin2016novel} & Cooperative & All routers & Rate-limit & Yes & No & Spoofed \textit{Data} & No \\ \hline

\cite{hou2019theil} & Cooperative & All routers & Block & Yes & No & Spoofed \textit{Data} & No \\ \hline

\cite{zhi2018gini} & Cooperative & All routers & \makecell[l]{Rate-limit (prefix)} & Yes & No & Malicious prefix & No \\ \hline

\cite{nakatsuka2018frog} & Autonomous & Edge routers & Block & No & No & No & No \\ \hline 

\cite{benarfa2019chokifa} & Autonomous & All routers & Rate-limit & No & No & No & No \\ \hline

\cite{zhi2020reputation} & Autonomous & All routers & Rate-limit & No & No & No & No \\ \hline

\cite{benarfa2020chokifap} & Autonomous & Edge routers & \makecell[l]{Rate-limit\\Block} & No & No & No & No \\ \hline

\cite{zhang2019expect} & Cooperative & \makecell[l]{All routers\\Producers} & \makecell[l]{Rate-limit\\Block} & Yes & No & \makecell[l]{RSN, PREF, C\\FakeList} & No \\ \hline

\cite{pu2019self} & Autonomous & Edge routers & Rate-limit & No & No & No & No \\ \hline

\cite{cheng2019detecting} & Cooperative & Edge routers & Block & Yes & \texttt{/ndn/ddos/flooding/} & Interfaces list & Yes \\ \hline

\cite{chen2019isolation} & Cooperative & All routers & \makecell[l]{Rate-limit (prefix)} & Yes & - & Malcious prefix & No \\ \hline

\cite{benmoussa2019novel} & Autonomous & Edge routers & Block & No & No & No & No \\ \hline

\cite{dong2020interestfence} & Cooperative & \makecell[l]{All routers\\Producers} & Rate-limit & Yes & No & Affected prefixes & No \\ \hline

\cite{tang2013identifying} & Autonomous & All routers & \makecell[l]{Block (prefix)} & No & No & No & No \\ \hline

\cite{shinohara2016cache} & Autonomous & All routers & Rate-limit & No & No & No & No \\ \hline

\cite{benmoussa2020msidn} & Cooperative & Edge routers & \makecell[l]{Rate-limit (prefix)\\and Block} & Yes & \makecell[l]{\texttt{/ndn/PCIP}\\\texttt{/ndn/RCIP}} & Affected prefix & Yes \\ \hline

\cite{alhisnawidetecting} & Cooperative & Edge routers & Rate-limit & Yes & No & No \\ \hline

\cite{wu2020mitigation} & Autonomous & All routers & Delete \textit{Interests} & No & No & No & No \\ \hline

\cite{tourani2020persia} & Autonomous & Core routers & Bloom filter & No & No & No & No \\ \hline

%PROACTIVE SOLUTIONS 
\cite{wang2017economic} & Cooperative & All routers & Rate-limit & Yes & No & Insufficient fees & No \\ \hline 

\cite{zhang2019ari} & Autonomous & Edge routers & Rate-limit & No & No & No & No \\ \hline

\cite{liu2018blam} & Autonomous & All nodes & Rate-limit & No & No & No & No \\ \hline
\end{tabular}
\end{adjustbox}
\end{table*}

\subsubsection{Key observations of our comparison}
The fundamental observations of our comparative study on mitigation parameters are as follows: We first note that the number of solutions that cooperate to mitigate attacks represents 54\%. The number may seem low as cooperation helps to better contain attacks, especially in the case of distributed IFA. However, cooperation implies a pre-configuration phase and the exchange of solution-based information. Scalability is also a challenge for cooperative solutions, especially in large deployments compared to the autonomous approach. We also notice from our comparison that autonomous mitigation is usually associated with edge routers, and solutions that employ core routers are always cooperative.
The next observation regards the actions that authored solutions chose to mitigate attacks. Our study found that 66\% of them apply rate-limiting. Countermeasure solutions chose this action for two reasons: (1) When applied on a core interface, it does not stop all the traffic. (2) Avoid penalizing legitimate consumers connected to edge routers after behavioral changes. (3) Avoid penalizing the traffic of a specific producer in case of prefix-based rate limiting. Routers usually employ rate limiting for a period. On the other hand, 29\% of proposals adopted blocking as a mitigation action. We first notice that interface-based blocking is always associated with edge routers, which is understandable. Core routers do not block the whole interface's traffic. However, all routers adopted prefix-based blocking.   
Our comparative study found that cooperative solutions employ control packets during mitigation. Solutions use them to exchange information. We found that the most shared information regards rate limiting. Routers, especially core routers, inform neighboring nodes after the application of traffic limitation. Some solutions also send the applied rate. The second most shared information within cooperative solutions is the affected/malicious prefix. It is associated with prefix-based actions. Routers use this information to ask other routers to limit/stop forwarding packets to that prefix. 
Our last observation regards the nature of packets used by cooperative solutions to share information. System nodes usually use \textit{Interest} packets to exchange information between them. However, we discovered that only 14\% of countermeasure solutions used signed \textit{Interest} packets to share information, which is very low. Using non-signed \textit{Interest} packets constitutes a threat to the reliability of a solution, as we will see in the following section.

\section{Unconsidered IFA Scenarios}
\label{sec:issues}
All existing IFA solutions may lack the detection of attacks in some particular scenarios, which attackers can take advantage of to flood the network and/or penalize legitimate consumers. In this section, we present and explain several unconsidered IFA scenarios. 

\subsection{Attacking scenarios against non cooperative solutions}
This subsection groups several attacking scenarios that target non cooperative IFA solutions.   

\subsubsection{Targeting neighboring consumers in  non-cooperative solutions}
Targeting neighboring consumers in non-cooperative solutions. The attacking scenario illustrated in Fig.~\ref{fig:non-coop-scenario1} shows how an attacker could take advantage of a non-cooperative solution to affect legitimate consumers. In this scenario, Router \textit{R1} takes a defensive action against its interfaces \textit{int1} and \textit{int2} because they reached their respective thresholds due to the malicious traffic. As a result, Router \textit{R1} will penalize legitimate consumers connected to Router \textit{R2} and those behind the switch.

\subsubsection{Targeting distant consumers in  non-cooperative solutions}
\begin{figure}
\centering
\begin{subfigure}[Attacking scenario against neighboring consumers in non-cooperative solutions]{
    \includegraphics[width=.48\linewidth] {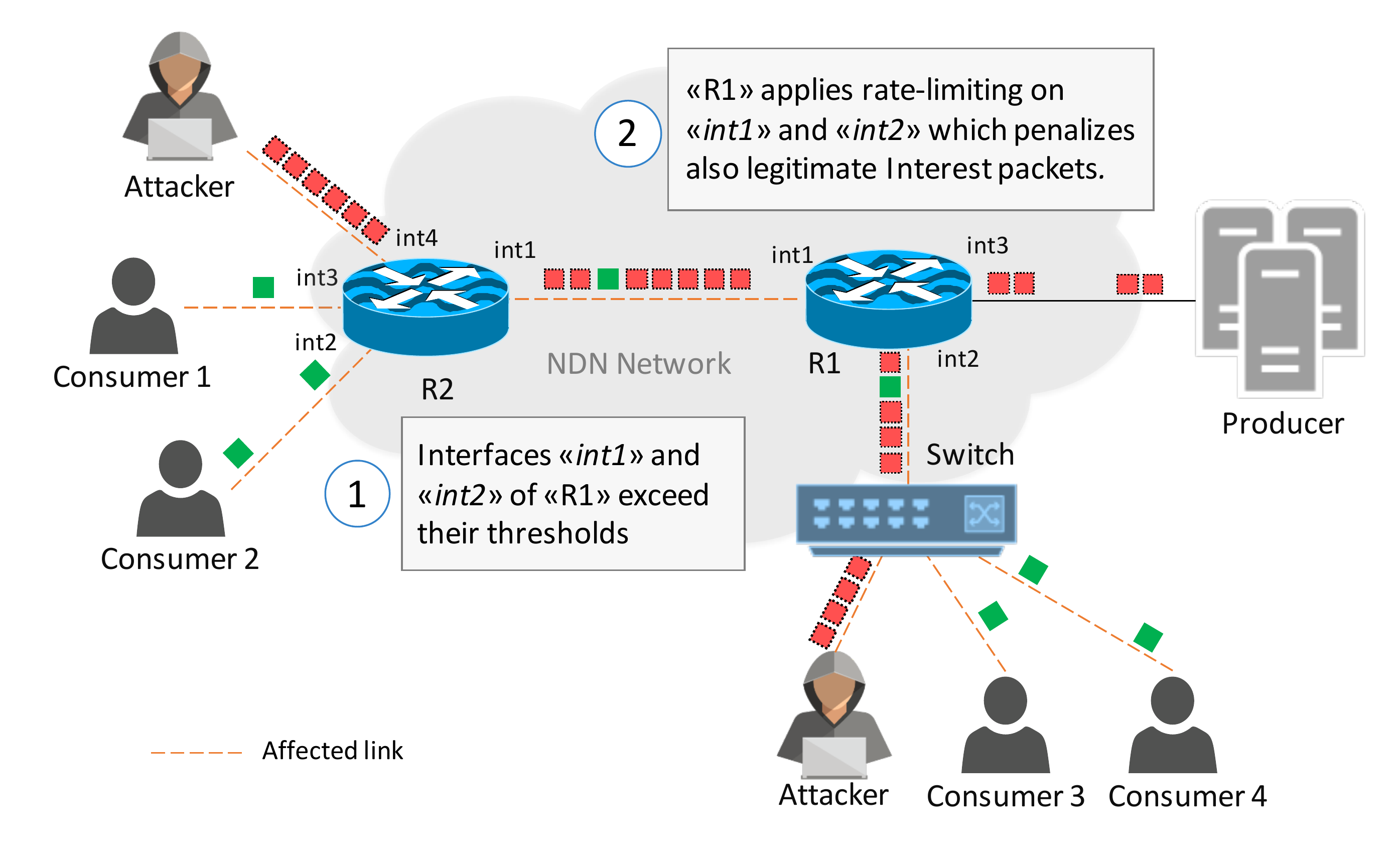}
    \label{fig:non-coop-scenario1}}
    \end{subfigure}
\begin{subfigure}[Attacking scenario against distant consumers in non-cooperative solutions]{
    \includegraphics[width=.48\linewidth] {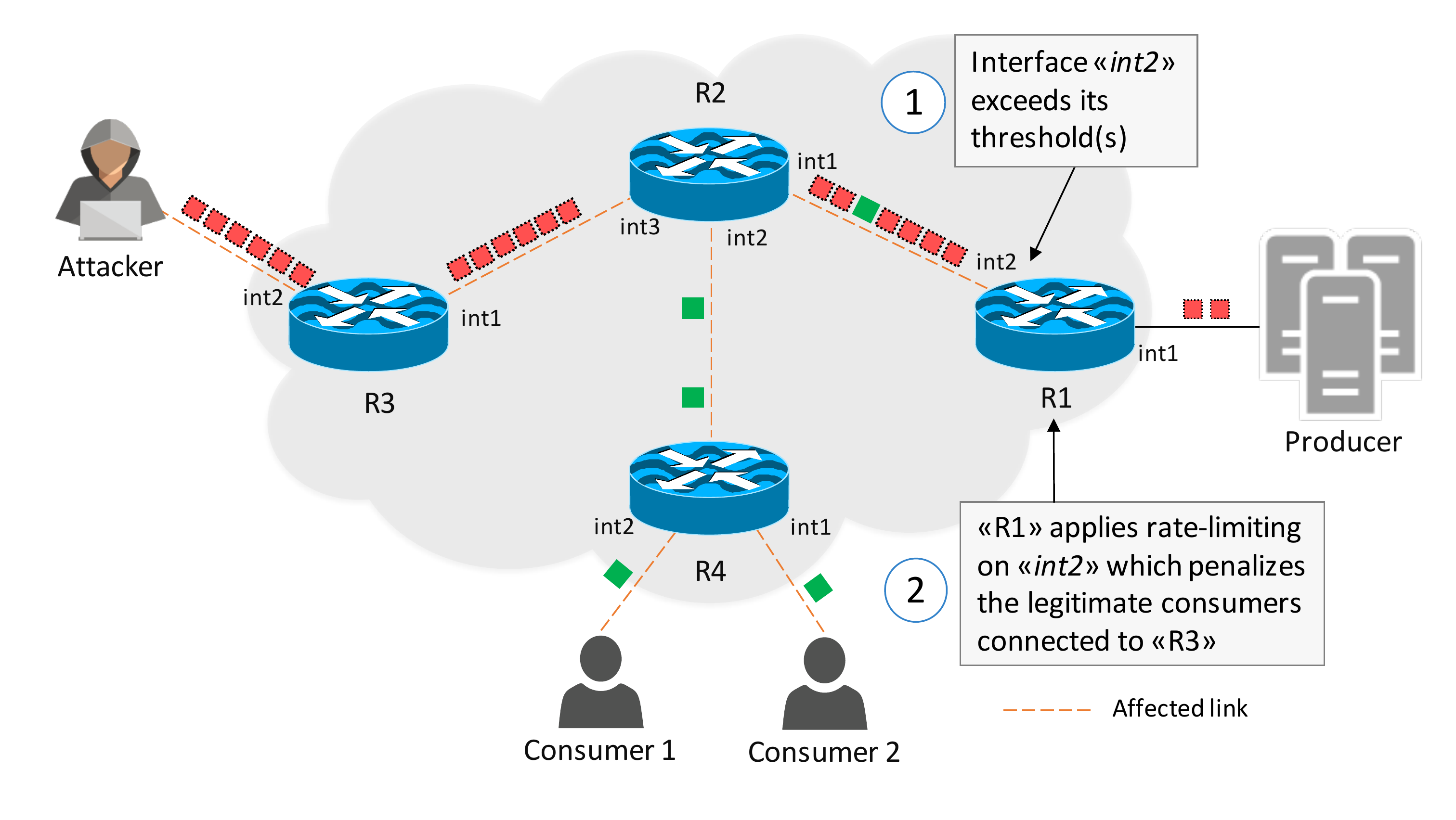}
    \label{fig:non-coop-scenario2}}
    \end{subfigure}
\caption{IFA Scenarios against non-cooperative solutions}
\end{figure}

Similarly, the scenario in Fig.~\ref{fig:non-coop-scenario2} shows that attackers can also affect a distant legitimate consumer in a case of a non-cooperative solution. In this scenario, The router \textit{R2} applied rate limiting on its interface \textit{int2} in response to the malicious generated by the attacker, which lead to affecting the legitimate consumers. The attacker was able to penalize the distant \textit{consumer1} and \textit{consumer2}.

\subsection{Attacking scenarios against cooperative solutions}
This subsection groups several attacking scenarios that target cooperative IFA solutions.

\subsubsection{Countering alert-based solutions with a compromised edge router}
This attacking scenario counters solutions that rely on edge routers to stop attackers (e.i., edge routers are the only mitigation actors). Attackers can get around by taking control of the edge router that they are connected to. Adversary nodes will have the ability to continue flooding the network because the edge router will ignore all received solution-based alerts, as shown in \ref{fig:comp-rt-scneario1}. Routers could reduce the impact of the attack when they take defensive actions. However, in a solution where the edge router is the only mitigation actor, attackers will continue flooding the network.

\subsubsection{Targeting legitimate consumers with alert messages}
Some IFA solutions use alert messages to exchange information and action decisions. Attackers could take advantage of this feature to target legitimate consumers as shown in Fig.~\ref{fig:coop-alert-scenario}. In this scenario, the attacker forges an alert message and sends it to Router \textit{R2} to push it to take defensive action against the legitimate consumers connected to it. The attacker can conduct such attacks only if the solution uses non-signed alert messages. 

\subsubsection{Targeting routers resources with forged alert messages}
\begin{figure}
\centering
\begin{subfigure}[Countering alert-based solutions with a compromised edge router]{
    \includegraphics[width=.48\linewidth] {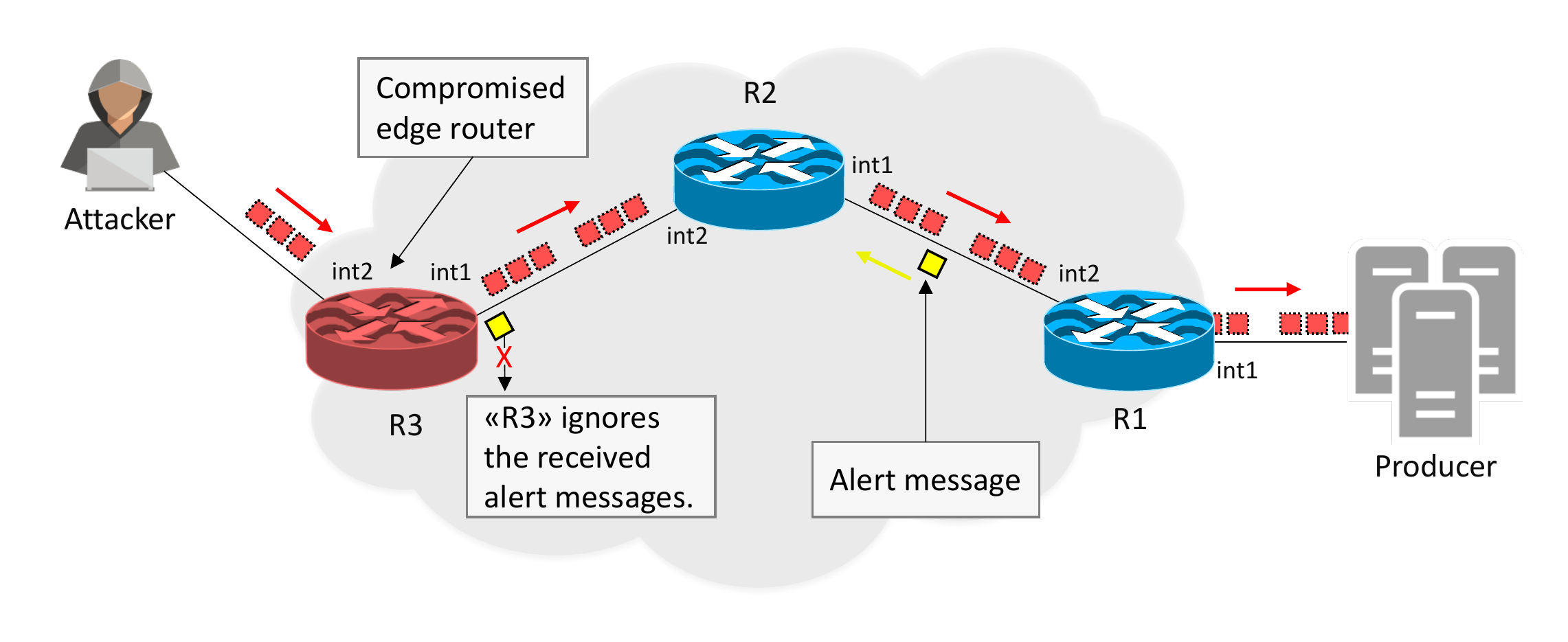}
    \label{fig:comp-rt-scneario1}}
    \end{subfigure}
\begin{subfigure}[Attacking scenario against legitimate consumers with alert message]{
    \includegraphics[width=.48\linewidth] {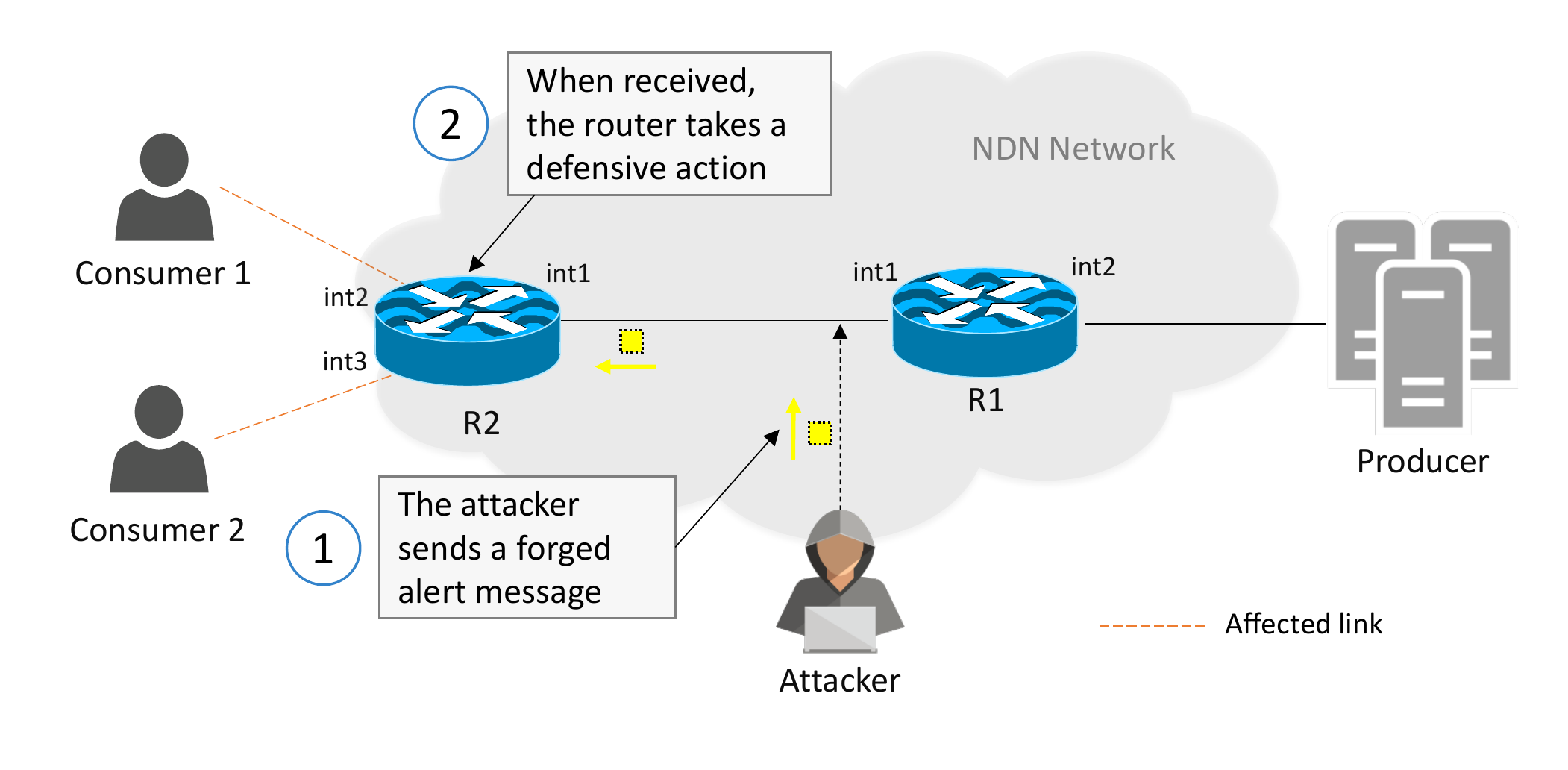}
    \label{fig:coop-alert-scenario}}
    \end{subfigure}
\begin{subfigure}[Targeting routers with forged alert messages]{
    \includegraphics[width=.48\linewidth] {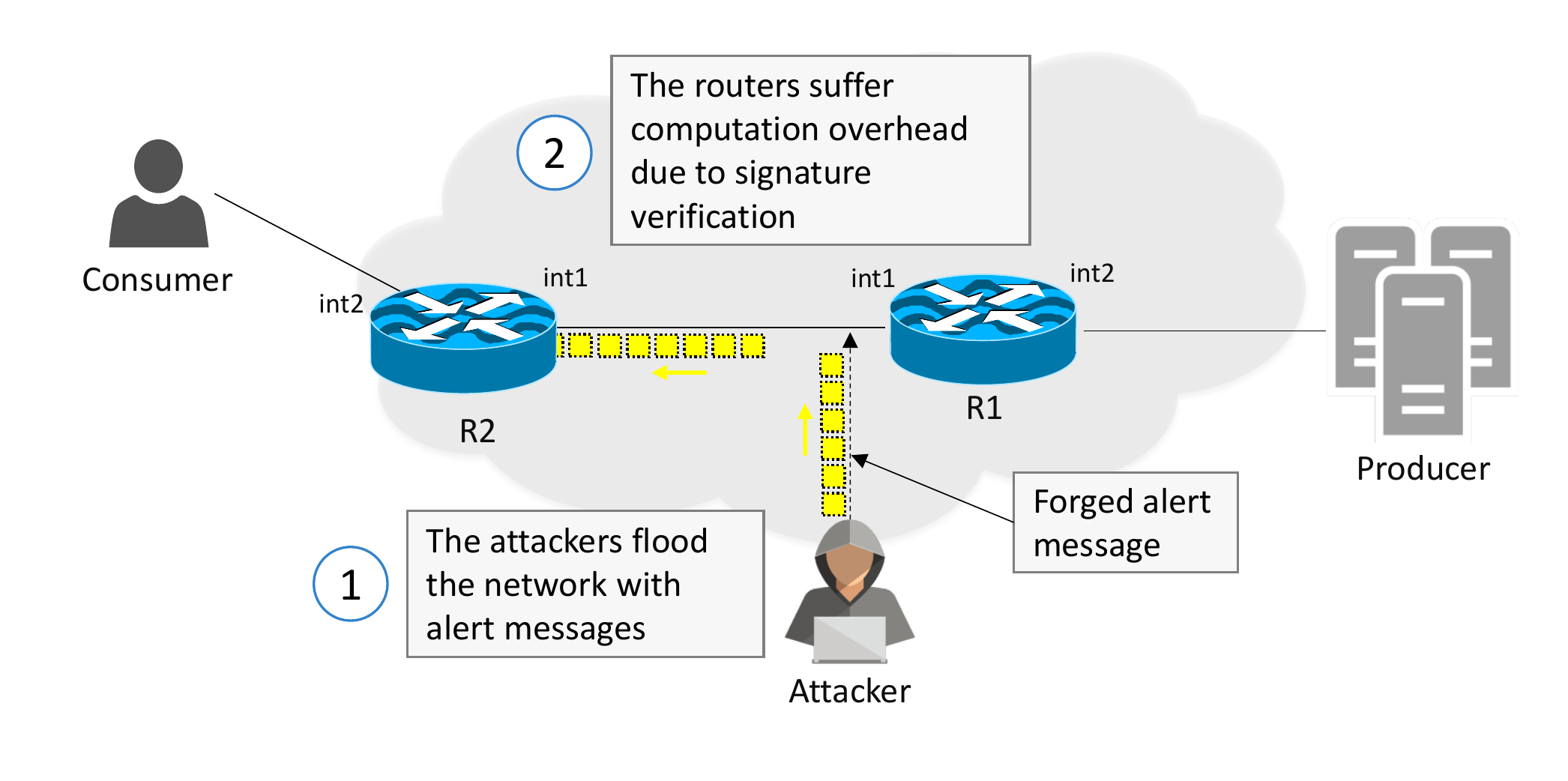}
    \label{fig:rt-res-alert-scenario}}
    \end{subfigure}
\caption{IFA scenarios against alert messages based solutions}
\end{figure}

Another way of using solution-based alert messages is to target routers' resources. As shown in Fig.~\ref{fig:rt-res-alert-scenario}, the attacker continually sends forged alert messages to Router \textit{R2} to stress it with signature verification and leads it to computation overhead.  

\subsubsection{Targeting routers resources with forged NACK packets}
Similar to the previous scenario, attackers can also lunch attacks against routers that rely on NACK packets to send solution-based messages like \cite{liu2018accuracy, dong2020interestfence}. To do so, attackers flood the targeted router with forged NACK packets to penalize him with computation overhead due to signature verification. Figure~\ref{fig:rt-res-nack-scenario} illustrates this attacking scenario.

\subsubsection{Flooding the network with solution-based spoofed \textit{Data} packets}
\begin{figure}
\centering
\begin{subfigure}[Targeting routers with forged NACK packets]{
    \includegraphics[width=.48\linewidth] {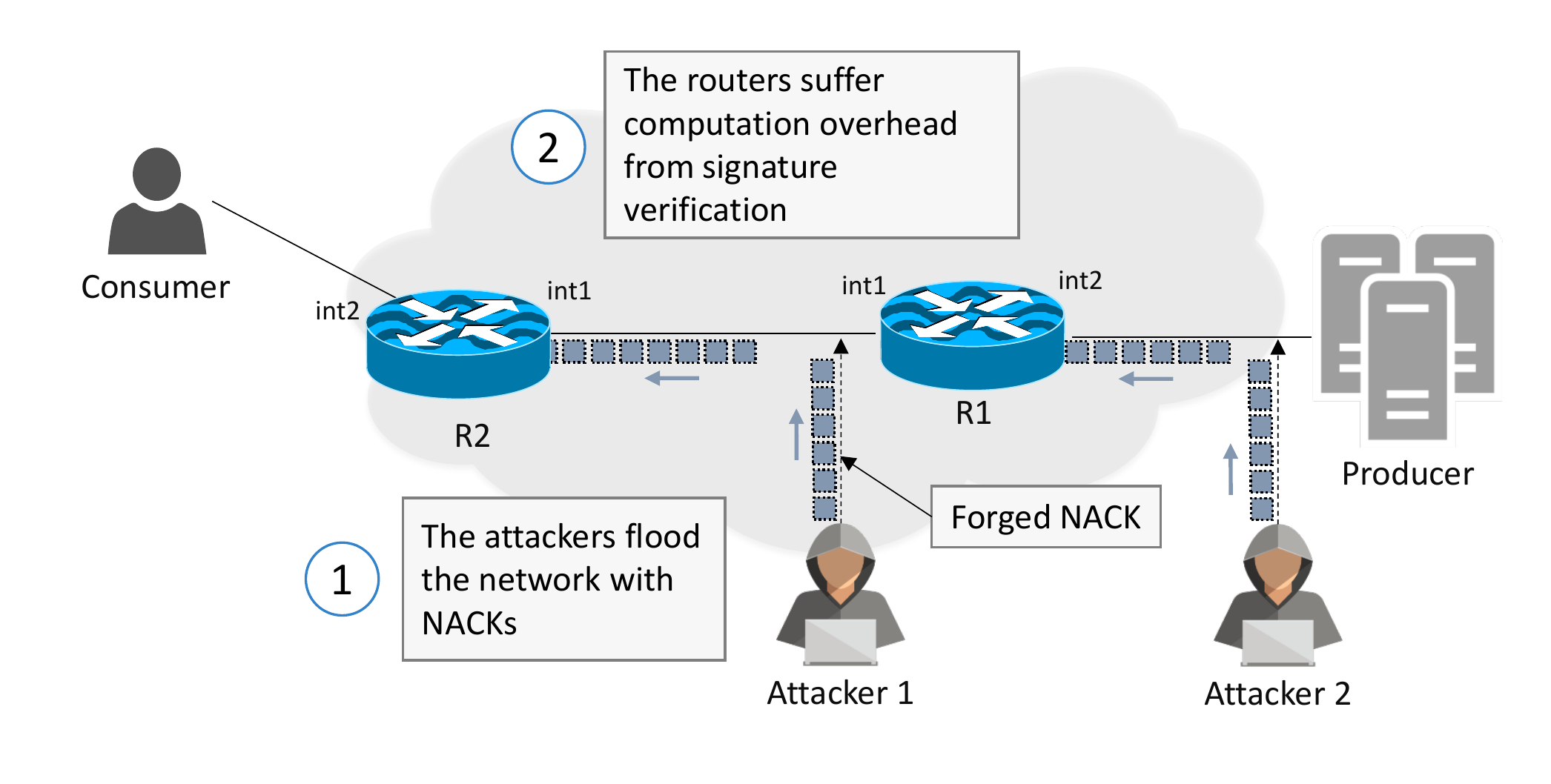}
    \label{fig:rt-res-nack-scenario}}
    \end{subfigure}
    \hfill
\begin{subfigure}[Flooding the network with solution-based spoofed \textit{Data} packets]{
    \includegraphics[width=.48\linewidth] {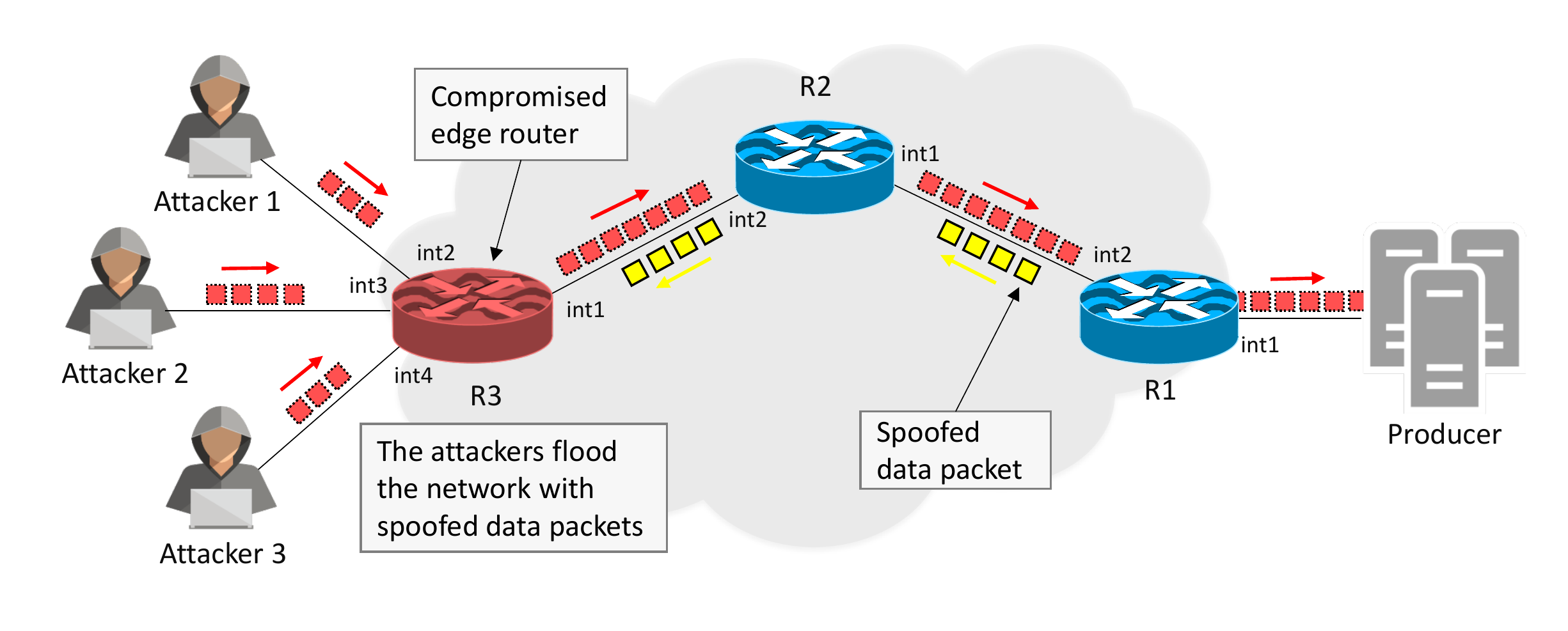}
    \label{fig:ns-spoofed-scenario}}
    \end{subfigure}
    \hfil
\caption{IFA scenarios with forged packets}
\end{figure}

Some solutions like \cite{dai2013mitigate} use spoofed \textit{Data} packets to counter malicious nodes. Attackers can take advantage of this feature to flood the network as shown in Fig.~\ref{fig:ns-spoofed-scenario}. In this scenario, the attackers who are in control of the edge router flood the network with forged \textit{Interest} packets. Router \textit{R1} sends back spoofed \textit{Data} packets to edge Router \textit{R3} to stop adversary nodes. Router \textit{R3} will take no action against the malicious nodes as it is controlled.

\subsubsection{Countering prefix-based solutions}
As mitigation action, some solutions like \cite{tang2013identifying,chen2019isolation} apply rate-limiting or blocking on name prefixes that routers consider as malicious or under attack. However, attackers can easily overcome this restriction by changing the prefix of forged \textit{Interest} packets to keep flooding the network, as shown in Fig.~\ref{fig:ns-scneario}.

\subsubsection{Affecting legitimate traffic in prefix-based solutions}
\begin{figure}
\centering
\begin{subfigure}[Targeting prefix-based solutions]{
    \includegraphics[width=.48\linewidth] {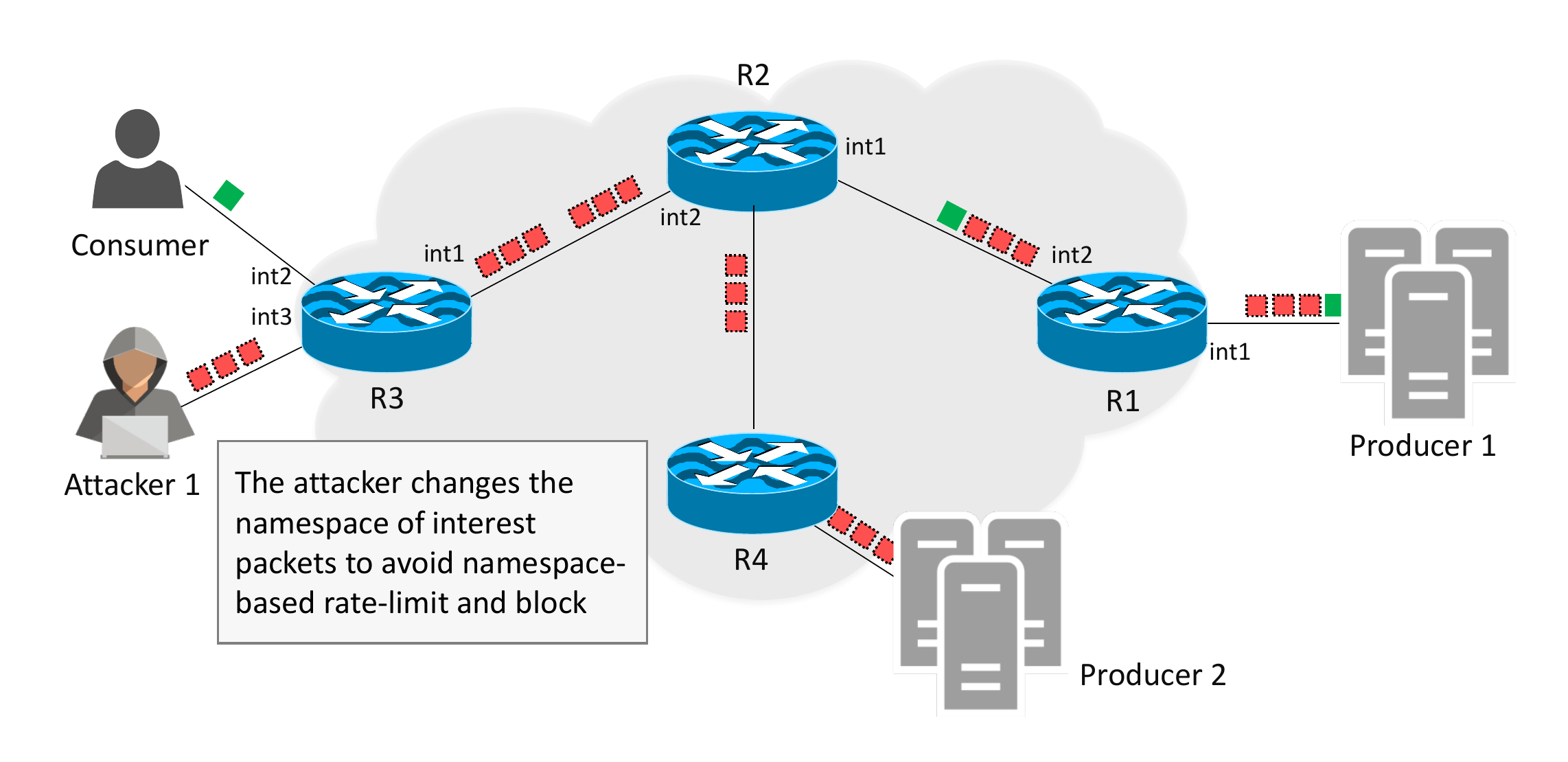}
    \label{fig:ns-scneario}}
    \end{subfigure}
    \hfill
\begin{subfigure}[Targeting legitimate traffic in prefix-based solutions]{
    \includegraphics[width=.48\linewidth] {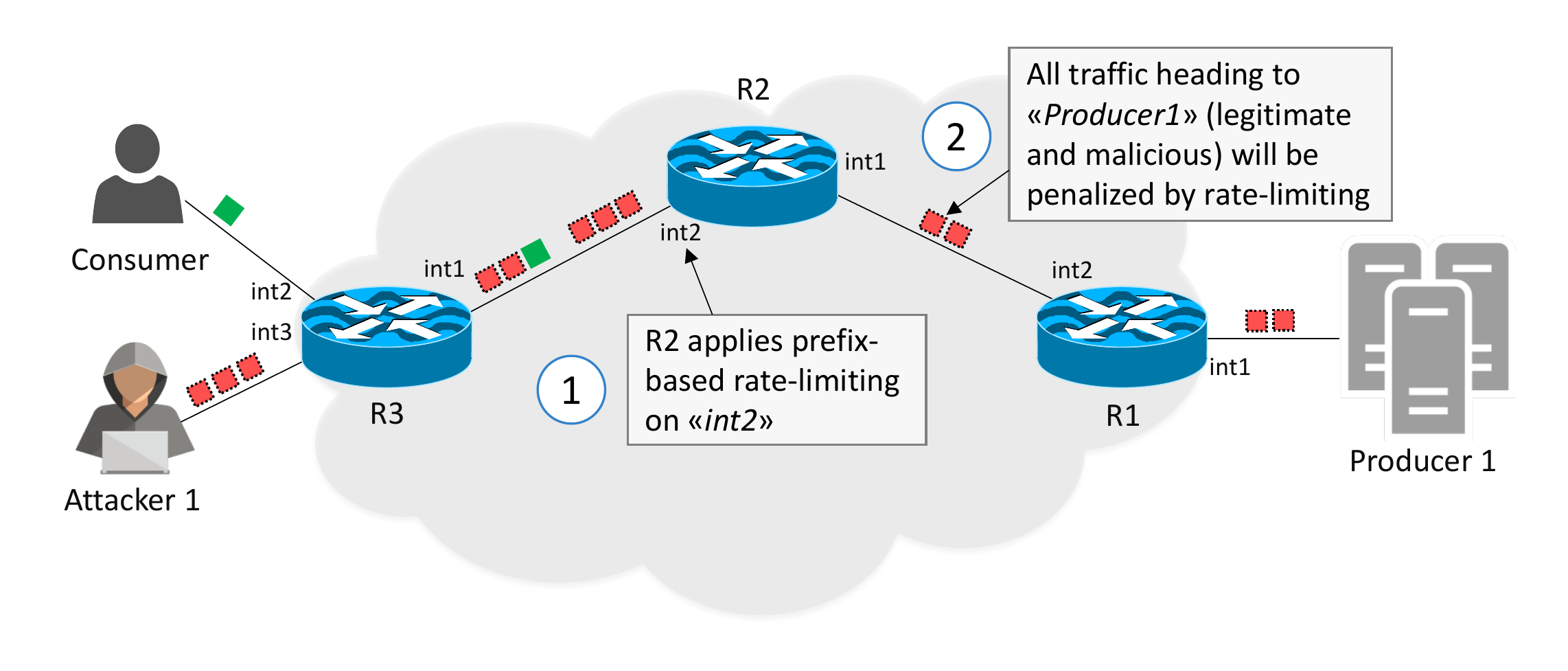}
    \label{fig:ns-scneario-2}}
    \end{subfigure}
    \hfil
\caption{IFA scenarios against prefix-based solutions}
\end{figure}

Another attacking scenario against prefix-based solutions is illustrated in Fig.~\ref{fig:ns-scneario-2}. The goal of the attacker in this scenario is to affect legitimate traffic heading to a specific producer. To do so, the attacker flood the network with forged \textit{Interest} packets with the targeted prefix to push routers to take defensive actions against this prefix, which penalizes also legitimate traffic. 

\subsubsection{Targeting the network with a distributed collusive attack}
Figure~\ref{fig:col-dist-scenario} illustrates a scenario of a distributed collusive attack. Compared to the collusive attack presented in the literature, in this scenario, attackers work with a distributed group of malicious producers to overwhelm the network. Additionally, adversary nodes send with regular rates. Existing solutions rely on metrics like PIT usage and traffic rate, which are linked to the aggressive nature of adversary nodes. That makes it difficult for existing solutions to detect this attacking scenario. 
Because of its distributed nature, this attacking scenario generates high traffic and introduces important delays to the network even with regular sending rates.

\subsubsection{Targeting the network with a low-rate distributed collusive attack} 
\begin{figure}
\centering
\begin{subfigure}[Targeting the network with a distributed collusive attack]{
    \includegraphics[width=.48\linewidth] {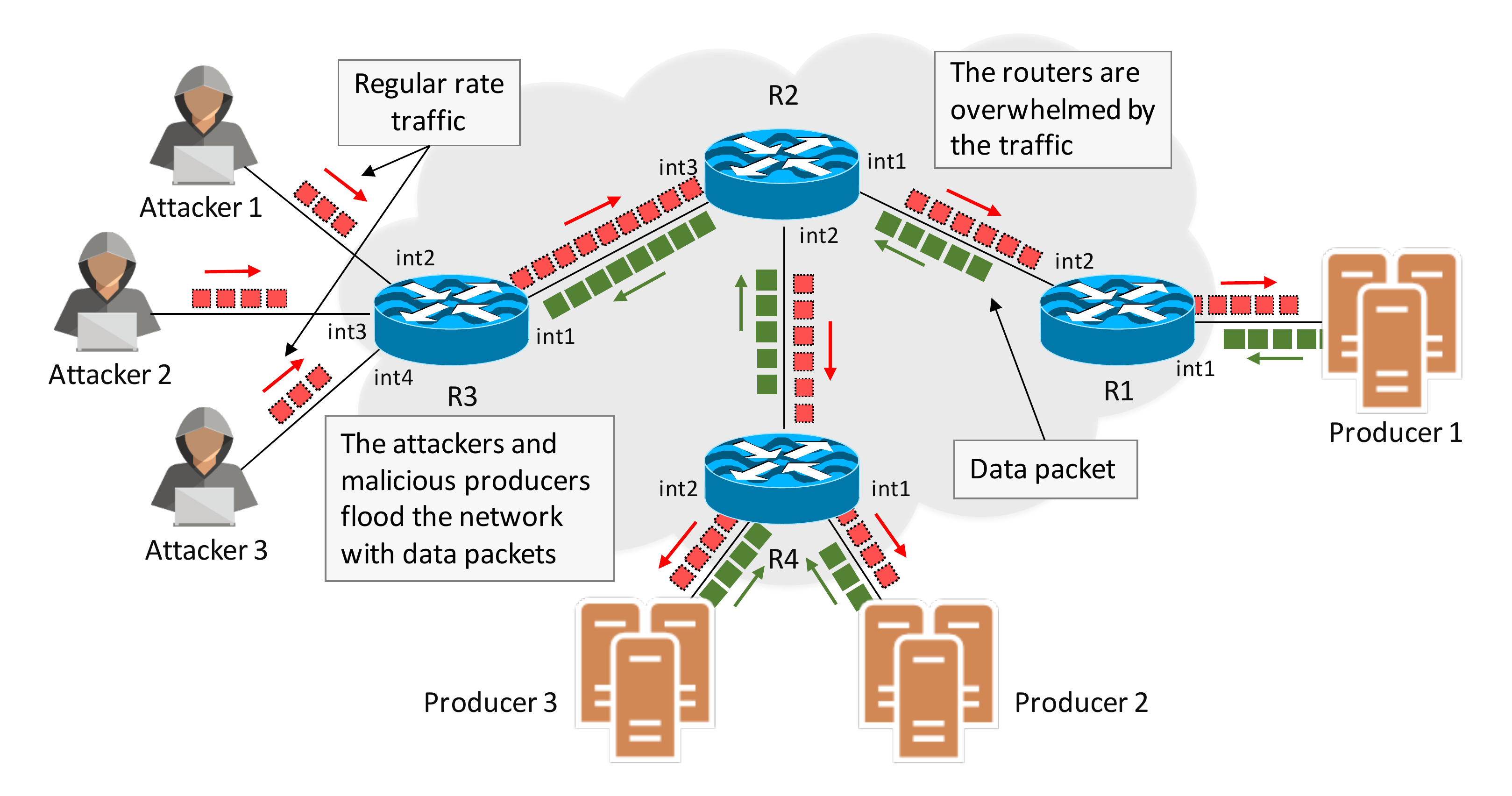}
    \label{fig:col-dist-scenario}}
    \end{subfigure}
    \hfill
\begin{subfigure}[Targeting the network with a low-rate distributed collusive attack]{
    \includegraphics[width=.48\linewidth] {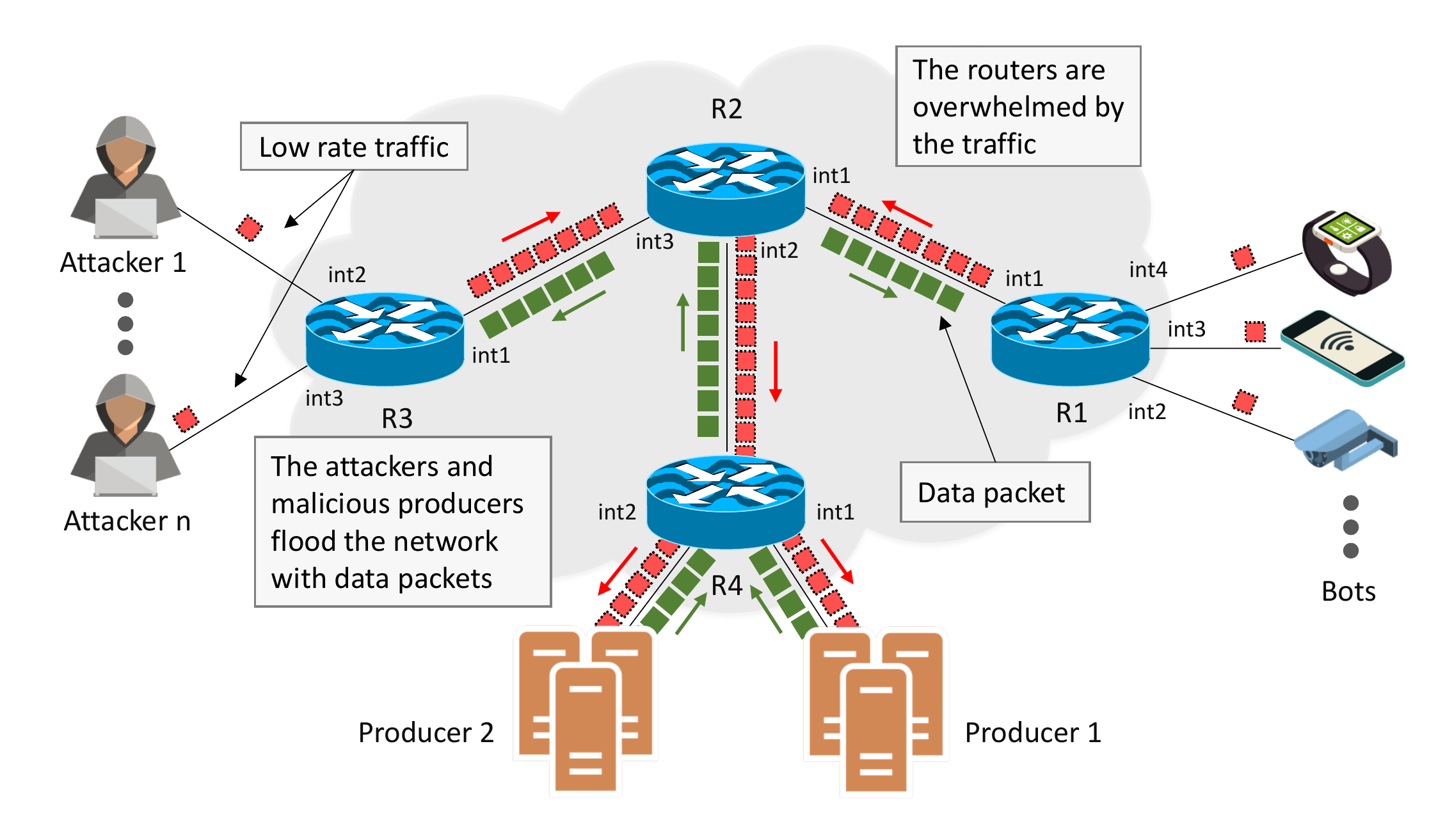}
    \label{fig:col-low-dist-scenario}}
    \end{subfigure}
    \hfil
\caption{Distributed collusive IFA scenarios}
\end{figure}

Another variant of the distributed collusive attack is the low-rate distributed collusive IFA. In this scenario, a large distributed number of attackers or infected bots request \textit{Data} packets from malicious producers with low sending rates, as shown in Fig.~\ref{fig:col-low-dist-scenario}. Similar to the previous scenario, existing collusive solutions rely on high traffic metrics to detect collusive attacks. It makes them inefficient against this attacking scenario.  

\subsubsection{Targeting the network with low-rate distributed IFA}
The attacking scenario depicted in \ref{fig:dist-low-ifa-scenario} represent a distributed IFA with low sending rate. This attacking solution works with all solutions that use the traffic rate and PIT usage as detection metrics. Attacker overcome these solution by adopting a low sending rate to keep flooding the network with malicious \textit{Interest} packets. 

\subsubsection{Targeting the network with low-rate and mixed distributed IFA}
\begin{figure}
\centering
\begin{subfigure}[Targeting the network with low-rate distributed IFA]{
    \includegraphics[width=.48\linewidth] {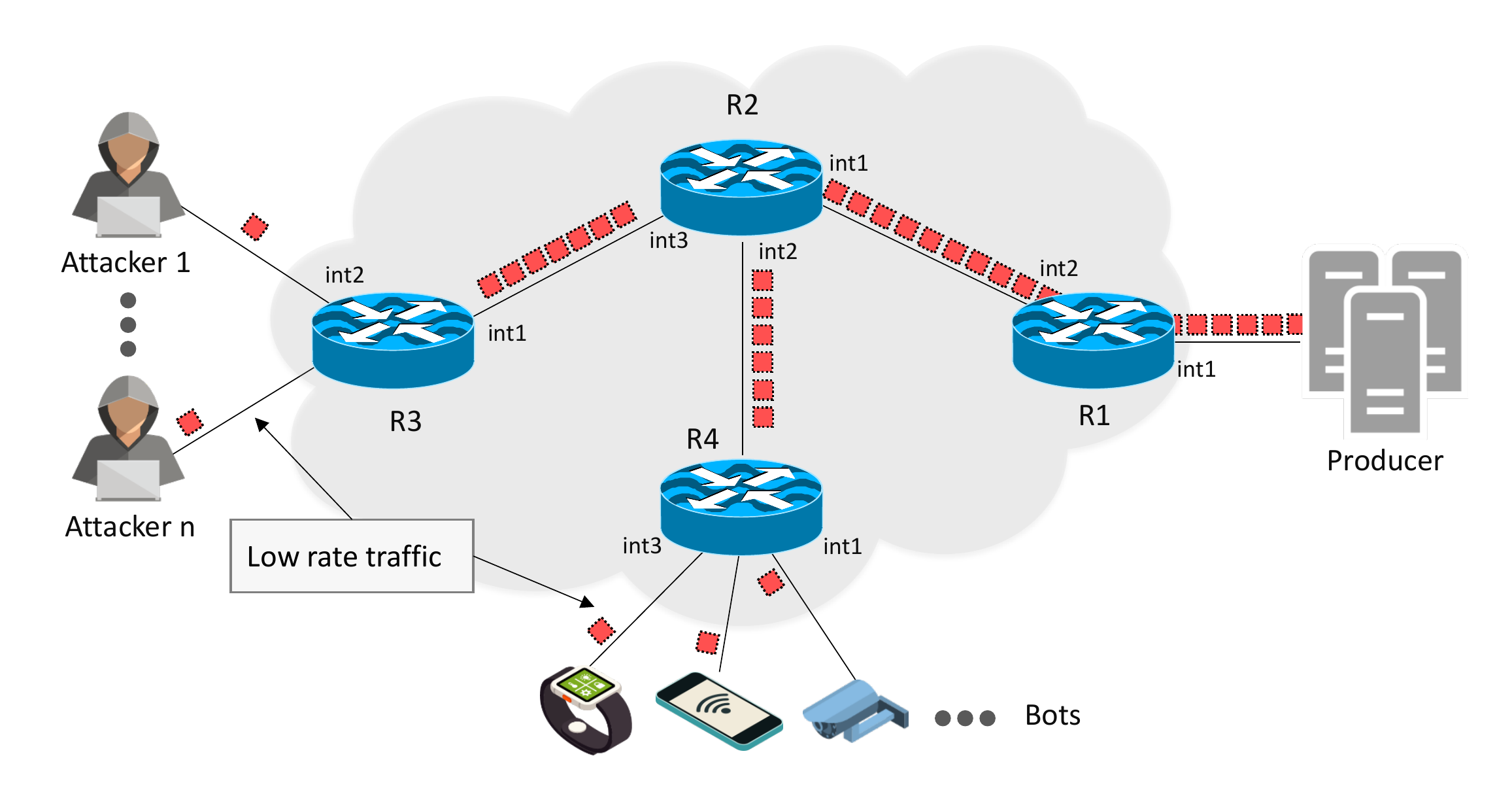}
    \label{fig:dist-low-ifa-scenario}}
    \end{subfigure}
    \hfill
\begin{subfigure}[Targeting the network with low-rate mixed distributed IFA]{
    \includegraphics[width=.48\linewidth] {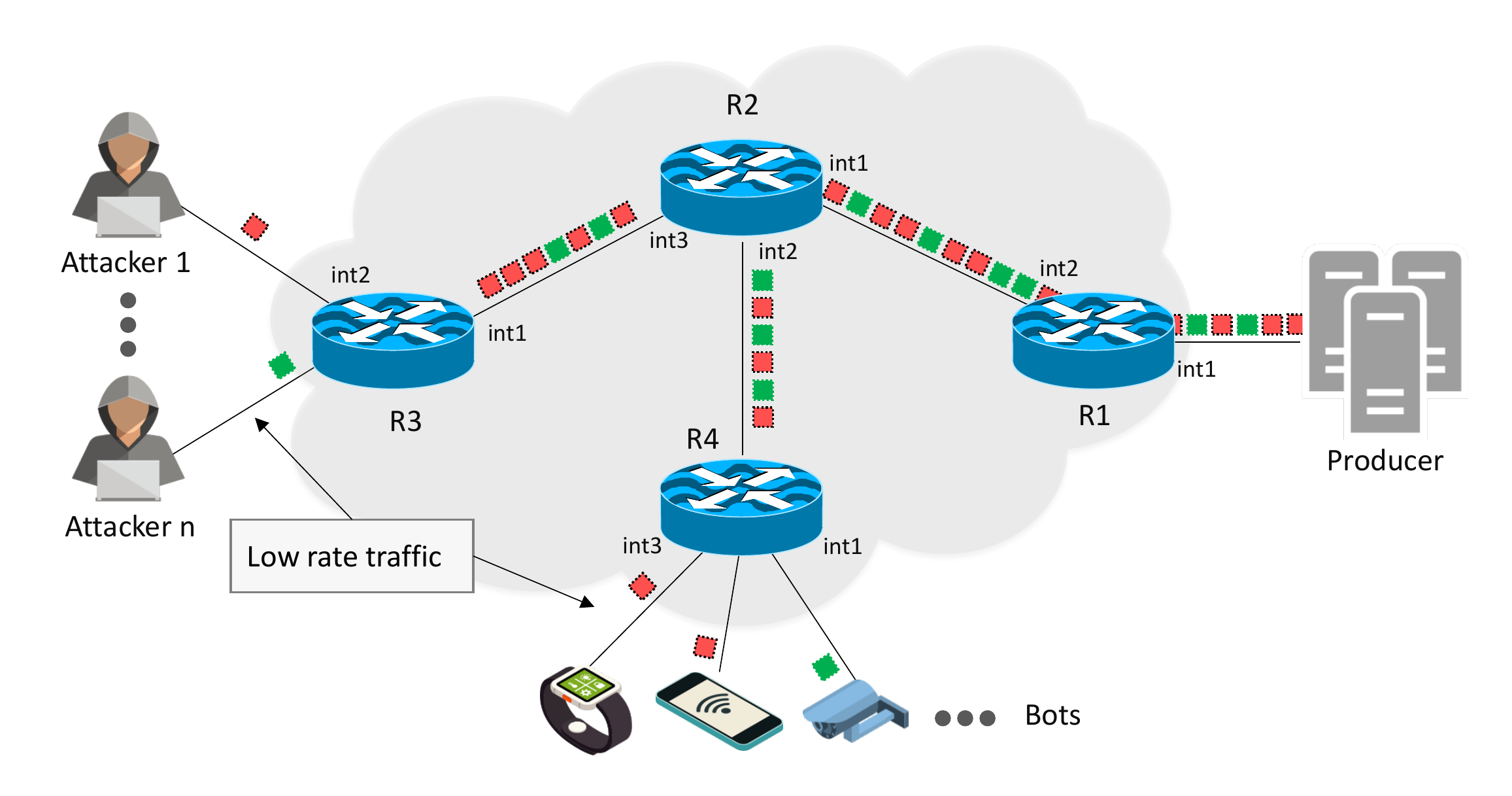}
    \label{fig:dist-low-mixed-ifa-scenario}}
    \end{subfigure}
    \hfil
\caption{Low-rate distributed IFA scenarios}
\end{figure}

An even more hard-to-detect attacking scenario than the previous one is illustrated in \ref{fig:dist-low-mixed-ifa-scenario}. In these particular scenarios attackers and controlled nodes flood the network with valid and invalid \textit{Interest} packets. This permits to attacker to counter other detection metrics like the satisfaction ratio and timed-out \textit{Interest} packets.

\subsubsection{Scenario of a Smart IFA}
\begin{figure}
\centering
\begin{subfigure}[Local IFA: case of smartly behaving attackers]{
    \includegraphics[width=.48\linewidth] {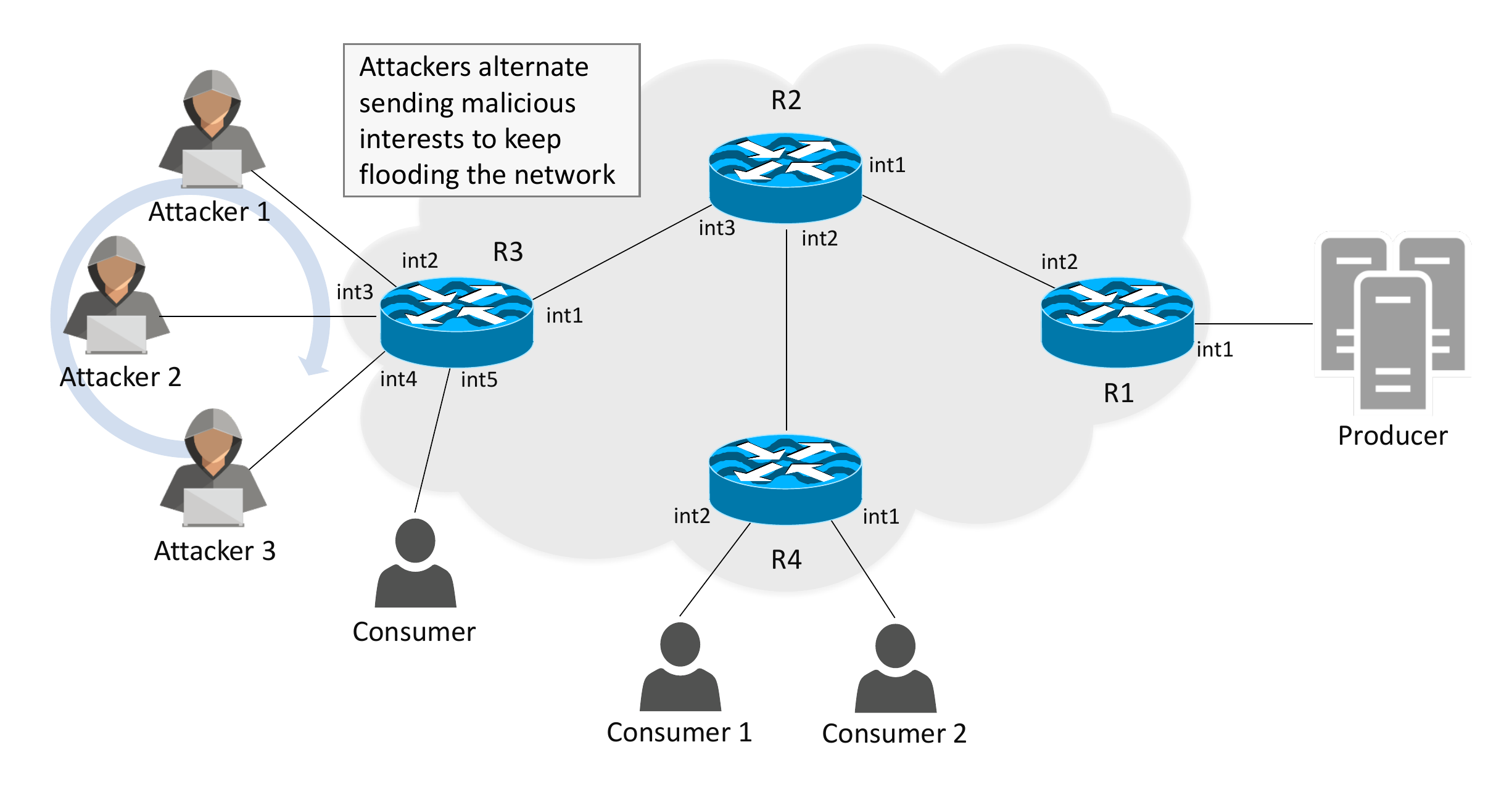}
    \label{fig:ifa-smart-scenario}}
    \end{subfigure}
    \hfill
\begin{subfigure}[Distributed IFA: case of smartly behaving attackers]{
    \includegraphics[width=.48\linewidth] {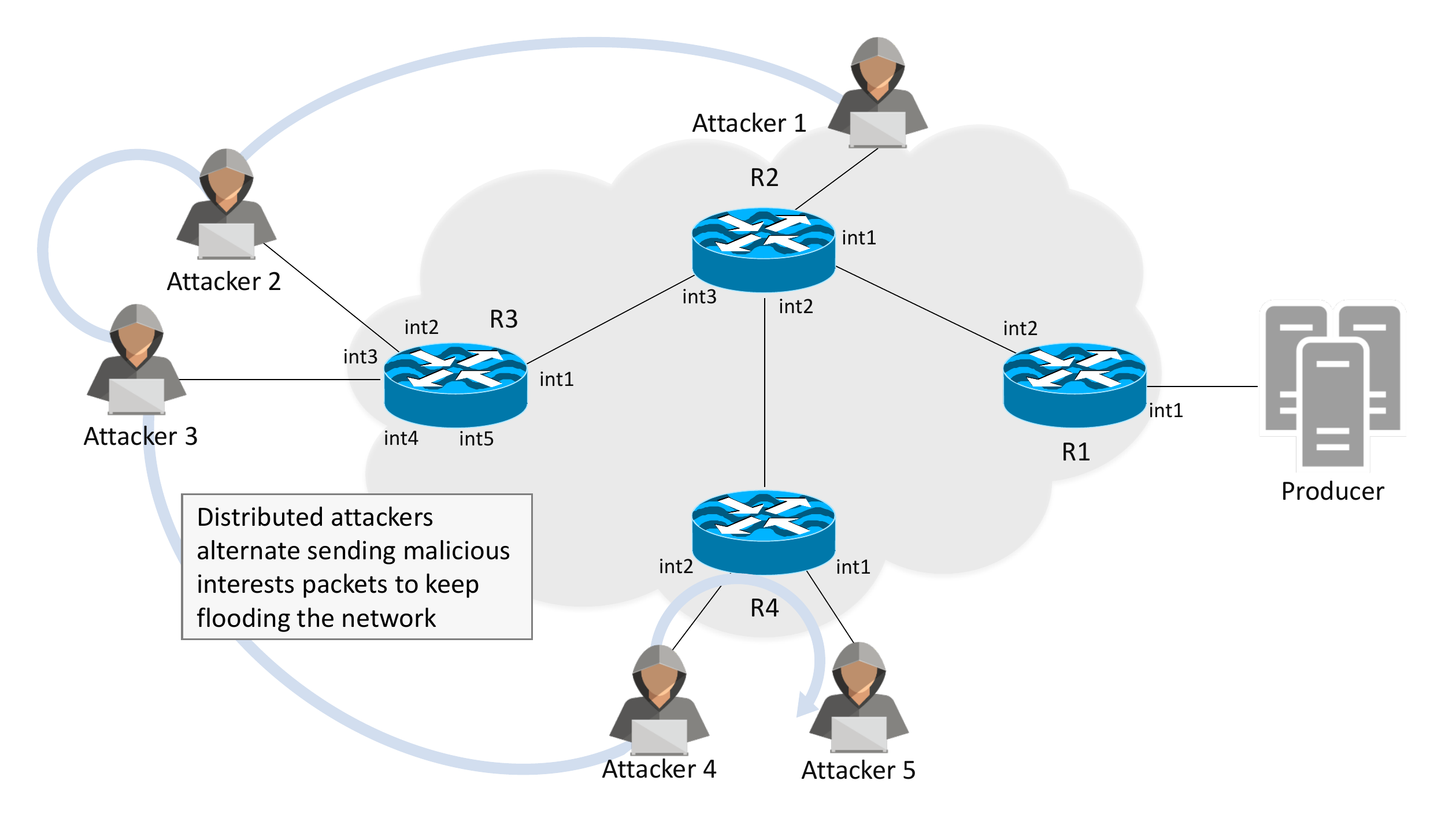}
    \label{fig:dist-ifa-smart-scenario}}
    \end{subfigure}
    \hfil
\caption{Smart IFA Scenarios}
\end{figure}

In this scenario, attackers adopt a smart behaviour when launching IFA to avoid mitigation and keep flooding the network. The adversary nodes cooperate by sending malicious traffic one after another. The goal of attackers is to keep flooding the network in case an attacker gets mitigated. Another more sophisticated scenario is when an attacker keeps sending forged \textit{Interest} packets and before reaching solution's thresholds it stops and another attacker continue the attack. This attacking scenarios could be local as showed in Fig.~\ref{fig:ifa-smart-scenario} or distributed as illustrated in Fig.~\ref{fig:dist-ifa-smart-scenario}.

\section{Lessons Learned and Future Directions}
\label{sec:future}
In this survey, we reviewed the literature on IFA depending on the information they collect, the metrics and parameters they use, and the actions they take. Based on our analysis, we found that IFA still represents a threat to NDN networks. The research community needs to conduct significant work to cover all its aspects. In this section, we list the key lessons learned and highlight research directions.

\subsection{Lessons Learned}
In the following, we list the fundamental lessons learned from this survey that could potentially help security researchers: 

\begin{enumerate}
    \item \textbf{Start by considering other IFA variants instead of conventional scenarios}. Existing works generally focus on a specific type of IFA. However, as outlined in section~\ref{sec:issues}, multiple IFA scenarios are still untreated by state-of-the-art works. Solutions need to widen their field of action and go off the beaten path to cover the full spectrum of the attack.
    
    \item \textbf{Consider additional parameters to enhance the detection accuracy}. As stated in this survey, existing solutions may classify legitimate requests as malicious, which leads to penalizing honest consumers. To this end, solutions need to consider additional parameters and proprieties during the detection process.
    
    \item \textbf{Rethink the solution’s algorithm to adapt to behavioral changes}. As outlined in Sec.~\ref{sec:issues}, existing solutions are incapable of detecting attackers when they adopt different behaviors from those expected by the algorithm. 
\end{enumerate}

\subsection{Future Research Directions}
Accurate detection, embedded security, and autonomous adaptation are some of the research goals that need to be achieved while designing IFA solutions. In the following, we outline potential research directions:

\begin{itemize}

    \item \textbf{Accurate detection and mitigation - }
    Many effective countermeasure solutions are available. However, authored solutions are explicit and generally detect specific attacking scenarios. Additionally, these solutions cannot even stop the scenario that they are intended to detect when some conditions change, which leads to flooding the network with malicious traffic. In addition, false-positive rates are tightly linked to detection conditions, which leads to unfairly blocking consumers and their legitimate traffic, as shown in Sec.\ref{sec:issues}. Newly designed solutions need to be more reliable. They need to revisit detection parameters and adopt intelligent approaches to cope with different IFA scenarios. For example, broad cooperation between network nodes could be considered, and concepts like federated learning and hybrid-based solutions could be of help.  
    
    \item \textbf{Embedded security mechanisms - }
    This research pathway suggests that solutions should embed security mechanisms to ensure reliable and undisrupted interconnectivity. The following prerequisites need to be part of a designed solution: (1) The solution-based information that nodes exchange needs to be securely validated and verified. (2) The solution needs to guarantee a secure and reliable transmission mechanism to ensure the integrity and authenticity of this information. (3) The solution needs to implement trust policies between participating entities. The implementation of these requirements will ensure a healthy environment among system nodes.
    
    \item \textbf{Autonomous adaptation - } The existing solutions can face a stable attacking scenario. However, an intelligent attacker can alter between legitimate and malicious behavior to avoid detection. Hence, this research direction recommends that newly designed solutions should be able to autonomously adapt to different situations, including the intelligent attacker scenarios. Additionally, solutions need to adapt their parameters following network situations, especially in high traffic peaks. 
     
    \item \textbf{Hybrid approaches - }  
    The following research pathway proposes to use hybrid solutions to detect and mitigate IFA. For example, a router may combine two solutions: one for conventional situations, in which it applies predefined rules, and another for non-conventional situations to detect behavioral changes. Another example of a hybrid approach consists of a modular solution. In this hybrid approach, tasks are distributed among the nodes of a system. Each node represents a module of the solution. Nevertheless, implementing hybrid solutions is challenging. It requires coordination between system nodes, which rises interoperability concerns. Another challenge consists of finding a balance between performance and resource consumption. 
     
    \item \textbf{Deployment-friendly - } 
    Another aspect of a reliable IFA solution relies on its deployment. Newly designed solutions need to be able to scale at any level without affecting the overall system. Additionally, solutions need to offer easy deployment mechanisms for recently added nodes without adding significant load on nodes' resources. A solution, as good as it is, needs to have a minimum impact on a device's resources so it can work smoothly even in cases of high traffic peaks. Newly designed solutions need to have a low resource consumption fingerprint so attackers cannot take advantage of it to take down network nodes.
    
\end{itemize}

\section{Conclusion}
\label{sec:conclusion}
Many solutions were authored since the proposal of the first IFA countermeasure mechanism. However, the proposed mechanisms still do not cover the full spectrum of potential IFAs. This survey broadly discussed IFA. Following that, it detailed, classified and compared the state-of-the-art solutions. After that, the survey discussed open issues through the presentation and analysis of several non-conventional attacking scenarios that were not considered before. Finally, the survey discussed the lessons learned and research directions by providing the requirements for a more robust and efficient IFA solution.

\section{Appendix}
\label{sec:simulation}
In this section, we compare existing IFA solutions in terms of the parameters used during the simulation process, along with the evaluation metrics used to validate the proposed solution.

\subsection{Simulation Parameters}
In this subsection, we list and explain all the simulation parameters used to evaluate the proposed solutions. Table \ref{tab:sim_params} summarizes the simulation environment adopted by the existing IFA solutions.  

\begin{itemize}
    \item \textit{Simulator}: The most common simulator used to evaluate IFA solutions is the NDN network simulator (ndnSIM)~\cite{mastorakis2017ndnsim}. ndnSIM is an open-source network simulator based on ns-3 and designed to conduct NDN based simulation studies. Some solutions evaluated their work using Matlab. On the other, one existing IFA solution was evaluated using OMNeT++, which is another network simulator. 
    
    \item \textit{Network topology}: Several network topologies were used by existing works during the evaluation part. The most commonly used topologies are the Rocketfuel topologies\cite{spring2002measuring}. 
    
    \item \textit{Links bandwidth}: This metric represents the bandwidth chosen for the links during the simulation. 
    \item \textit{Network delay}: It represents the network delay values chosen for during the evaluation. 
    
    \item \textit{Forwarding strategy}: It shows the strategy adopted by routers to forward interest packets. 
    
    \item \textit{Dishonesty ratio}: This metric equals the number of malicious nodes deployed during the simulation to the total number of nodes. 
    
    \item \textit{Number of producers}: It represents the number of producers deployed in the network. 
    
    \item \textit{Rate of consumers}: This parameters equals the frequency at which legitimate consumers send interest packets.
    
    \item \textit{Rate of attackers}: It represents the number of malicious interest packets sent by attacker in a period. 
    
    \item \textit{Nature of malicious interest}: it shows the nature of interest packets used by malicious nodes during the attack. 
    
    \item \textit{PIT size}: This parameter denotes the size of routers PIT adopted during the simulations. It is represented in memory size or number of entries. %In recent versions, NFD does not apply size-limiting on PIT. The size of the PIT keeps increasing until the memory runs out \cite{redmine-nfd}.
    
    \item \textit{Interest lifetime}: It represent the value of the  \textit{InterestLifetime} field chosen for the evaluation. 
    
    \item \textit{Intermediate Cache}: This simulation parameter states whether or not routers use the local CS to satisfy incoming interest packets. 
    
    \item \textit{CS size}: Equals the capacity of a router's CS in terms of the number of entries (i.e., data) that it supports.
    
    \item \textit{CS strategy}: This parameter represents the cache replacement strategy adopted by routers. 
    
    \item \textit{Data size}: It represents the size of the \textit{Data} packet chosen during the simulation.
    
\end{itemize}

\begin{table*}
\setlength{\tabcolsep}{4pt}
\centering
\caption{Simulation Parameters}
\label{tab:sim_params}
\begin{adjustbox}{width=\textwidth}
\begin{tabular}{ccccccccccccccccc}
\hline
Ref & \makecell{Simulator} & \makecell{Network\\topology} & \makecell{Links\\bandwidth} & \makecell{Network\\delay} & \makecell{Forwarding\\strategy} & \makecell{Dishonesty\\ratio} & \makecell{Number of\\producers} &  \makecell{Rate of\\consumers} & \makecell{Rate of\\attackers} & \makecell{Nature of\\malicious interest}&\makecell{PIT size} & \makecell{Interest\\lifetime} & \makecell{Intermediate\\Cache} & \makecell{CS size} & \makecell{CS\\strategy} & \makecell{Data size}  \\
\hline

\cite{afanasyev2013interest} & ndnSIM & \makecell{Binary tree\\Rocketfuels\\AS 7018} & 10Mbps & $80-330ms$ & - & \makecell{$6-50\%$\\$40\%$} & 01 & - & - & Non-existent & - & - & No & - & - & 1100Bytes \\ \hline

\cite{compagno2013poseidon} & ndnSIM & \makecell{Rocketfuel\\AS 7018} & - & - & - & $50\%$ & 01 & 30ipps & 01int/min & Non-existent & 120KB & $4s$ & - & - & - & - \\ \hline

\cite{dai2013mitigate} & - & \makecell{Rocketfuel\\AS 1755} & - & - & - & - & 01 & - & 1000ipps & Non-existent & - & $500ms-8s$ & - & - & - & - \\ \hline 

\cite{vassilakis2015mitigating} & ndnSIM & Random & - & - & - & $5\%$ & - & 20ipps & 1000ipps & Non-existent & - & $200ms-1s$ & - & - & - & 1KB \\ \hline

\cite{wang2014cooperative} & ndnSIM & \makecell{Rocketfuel\\AS 7018} & 1-100Mbps & $5-70ms$ & Best route & $40\%$ & 01 & 20ipps & 400ipps & Non-existent & \makecell{100\\entries} & $1s$ & Yes & \makecell{500\\entries} & LRU & 1KB \\ \hline

\cite{salah2015coordination} & ndnSIM & \makecell{Rocketfuel\\AS 3967} & - & - & - & $25\%$ & - & - & \makecell{500-\\10000ipps} & Non-existent & \makecell{5000\\entries} & - & No & - & - & 1100Bytes \\ \hline

\cite{salah2016evaluating} & ndnSIM & \makecell{Rocketfuel\\AS 3257} & - & - & - & $25\%$ & \makecell{01\\malicious} & 100ipps & \makecell{200-\\5000ipps} & Non-existent & \makecell{5000\\entries} & $2s$ & No & - & - & 1100Bytes \\ \hline

\cite{wang2013decoupling} & ndnSIM & \makecell{Rocketfuel\\AS 7018} & 1-100Mbps & $5-70ms$ & Best route  & $40\%$ & 01 & 10-1000ipps & 100ipps & Non-existent & Unlimited & - & Yes & \makecell{500\\entries} & LRU & 1KB \\ \hline

\cite{wang2014detecting} & Matlab & - & - & - & - & - & - & \makecell{Poisson\\distribution} & \makecell{Poisson\\distribution} & Non-existent & - & - & - & - & - & - \\ \hline

\cite{karami2015hybrid} & \makecell{ndnSIM\\Matlab} & \makecell{DFN-like\\Rocketfuel\\AS 7018} & - & - & - & $20-50\%$ & 02-05 & 100-600ipps & \makecell{400-\\3000ipps} & \makecell{Existent and\\non-existent} & - & - & - & - & - & - \\ \hline

\cite{ding2016cooperative} & ndnSIM &\makecell{Small tree\\Rocketfuel\\AS 7018} & 1-100Mbps & $5-70ms$ & - & $16-25\%$ & 01 & 67-1000ipps & \makecell{2000-\\20000ipps} & Non-existent & \makecell{10000\\entries} & $1s$ & No & - & - & -\\ \hline 

\cite{nguyen2015detection} & \makecell{ndnSIM\\Matlab} & Binary tree & - & - & - & - & 01 & \makecell{Poisson\\distribution} & - & - & - & - & - & - & - & - \\ \hline

\cite{xin2017detection} & ndnSIM & \makecell{China\\Telecom} & - & - & - & $15\%$ & 01 & 50ipps & \makecell{Random\\distribution} & - & \makecell{200\\entries} & $4s$ & No & - & - & - \\ \hline

\cite{xin2016novel} & ndnSIM & Small tree & 100Mbps & $10ms$ & Best route & $75\%$ & 01 & 200ipps & 20ipps & Non-existent & \makecell{50\\entries} & $1s$ & No & - & - & 1KB \\ \hline

\cite{zhi2018gini} & ndnSIM & Small tree & - & - & - & $25\%$ & 01 & 50ipps & 100ipps & Non-existent & \makecell{200\\entries} & $1s$ & - & - & - & - \\ \hline

\cite{nakatsuka2018frog} & ndnSIM & \makecell{Rocketfuel\\AS 1221} & 1-100Mbps & $5-70ms$ & - & $95\%$ & 10 & 100ipps & 100ipps & Existent & - & - & - & - & - & 1KB \\ \hline 

\cite{benarfa2019chokifa},\cite{benarfa2020chokifap} & ndnSIM & \makecell{Small\\topology\\Rocketfuel\\AS 7018} & - & - & - & $6-50\%$ & 01 & 30ipps & \makecell{100-\\10000ipps} & Non-existent & 600KB & - & - & - & - & - \\ \hline 

\cite{zhang2019expect} & ndnSIM & \makecell{Meshed\\topology} & - & - & - & $20\%$ & 01 & 40ipps & \makecell{100-\\6000ipps} & \makecell{Existent and\\non-existent} & - & - & Yes & \makecell{200\\entries} & - & - \\ \hline

\cite{hou2019theil} & ndnSIM & Binary tree & - & - & - & $25\%$ & 01 & 200ipps & \makecell{1000-\\2000ipps} & - & \makecell{50\\entries} & $1s$ & - & - & - & - \\ \hline

\cite{pu2019self} & OMNeT++ & Small tree & - & - & - & - & - & - & - & - & - & - & - & - & - & - \\ \hline

\cite{zhang2019ari} & ndnSIM & \makecell{Small Tree\\Large\\network} & - & - & - & $35-40\%$ & 01 & - & - & - & - & - & - & - & - & - \\ \hline

\cite{alston2016neutralizing} & - & \makecell{Internet-\\like} & 300Mbps & - & - & $30\%$ & - & 5ipps & - & - & - & - & - & - & - & 500Bytes \\ \hline 

\cite{cheng2019detecting} & ndnSIM & \makecell{Modified\\Rocketfuel\\AS 7018} & - & $300ms$ & Best route & $40\%$ & 01 & 40ipps & \makecell{13ipps then\\ $3-10\%$\\higher} & Existent & \makecell{2000\\entries} & $1s$ & - & - & - & 1100Bytes \\ \hline  

\cite{liu2018accuracy} & ndnSIM & GEANT & \makecell{155Mbps-\\20Gbps} & $10ms$ & Best route & $50\%$ & 01 & 250ipps & 200ipps & - & - & - & Yes & \makecell{1000\\entries} & - & 4KB \\ \hline

\cite{liu2018blam} & - & GEANT & \makecell{100Mbps-\\20Gbps} & - & - & $40\%$ & 01 & 20ipps & 200ipps & - & - & - & - & - & - & 4KB \\ \hline

\cite{chen2019isolation} & - & Binary tree & - & - & - & $12\%$ & 01 & 100ipps & 2000ipps & Non-existent & \makecell{300\\entries} & $500ms$ & - & - & - & - \\ \hline

\cite{zhi2019resist} & ndnSIM & Binary tree & 10Mbps & $10ms$ & - & $75\%$ & 01 & 500ipps & 100ipps & Non-existent & \makecell{200\\entries} & $1s$ & - & - & - & - \\ \hline

\cite{yin2019controller} & ndnSIM & \makecell{Binary tree\\Small\\topology} & - & $10ms$ & - & $45-50\%$ & 02 & 5-10ipps & 100ipps & Non-existent & \makecell{100\\entries} & - & - & - & - & - \\ \hline

\cite{wang2019analyzing} & Matlab & \makecell{Rocketfuel\\AS 7018} & 1-100Mbps &$5-70ms$ & - & - & - & - & - & - & - & - & - & - & - & - \\ \hline

\cite{benmoussa2019novel} & ndnSIM & \makecell{25 nodes\\topology} & 1-100Mbps & $10-100ms$ & Best route & $10-25\%$ & 09 & \makecell{100-\\500ipps} & 10000ipps & Non-existent & - & - & Yes & \makecell{100\\entries} & LRU & 1KB \\ \hline

\cite{dong2020interestfence} & - & \makecell{Small\\topology} & \makecell{100Mbps-\\1Gbps} & $20-30ms$ & - & $10\%$ & 01 & 100ipps & 1000ipps & Non-existent & - & $2s$ & - & - & - & - \\ \hline

\cite{zhi2020reputation} & ndnSIM & \makecell{25 nodes\\topology} & 1-100Mbps & $5-66ms$ & - & $55\%$ & - & 500ipps & \makecell{500-\\2000ipps} & - & \makecell{100\\entries} & $2s$ & - & - & - & 512Bytes \\ \hline

\cite{shinohara2016cache} & ndnSIM & Binary tree & 10Mbps & $10ms$ & - & $25\%$ & 01 & 1000ipps & 10000ipps & Non-existent & \makecell{275\\entries} & $1s$ & - & - & - & - \\ \hline

\cite{shigeyasu2018distributed} & - & \makecell{12 nodes\\topology} & 1Mbps & $1ms$ & - & $60\%$ & 02 & 10ipps & \makecell{320-\\2030ipps} & Non-existent & \makecell{50\\entries} & - & Yes & \makecell{10\\entries} & FIFO & 1KB \\ \hline

\cite{tang2013identifying} & ndnSIM &\makecell{100-to-1\\topology} & 10Mbps & $1-10ms$ & - & $10-90\%$ & 01 & 200ipps & 200ipps & Non-existent & Unlimited & $2s$ & - & - & - & - \\ \hline

\cite{benmoussa2020msidn} & ndnSIM & \makecell{18 nodes\\topology} & \makecell{10Mbps-\\1Gbps} & $1-10ms$ & Best route & $66\%$ & 03 & 100ipps & \makecell{1500ipps-\\10000ipps} & \makecell{Existent and\\non-existent} & Unlimited & - & Yes & \makecell{100\\entries} & LRU & 1KB \\ \hline
\end{tabular}
\end{adjustbox}
\end{table*}

\subsection{Evaluation Metrics}
In this subsection, we detail all the evaluation metrics used by existing IFA works to validate the proposed solutions. Table \ref{tab:sim_metric} summarizes the evaluation metrics used by each existing solution.

\begin{itemize}
    \item \textit{Satisfaction ratio}: It equals the number of \textit{Data} packet returned to the total number of interest packets issued. 
    
    \item \textit{PIT usage}: This metric is also called PIT occupancy. The PIT usage of an interface represents the number of pending \textit{Interest} packets sourced by this interface to the total capacity of the PIT.
    
    \item \textit{Number of PIT entries}: This metric represents the number of pending PIT entries of an interface. Some solution use this metric to detect abnormal traffics. This metric is compared to the usual activity of an interface to detect a malicious traffic.
    
    \item \textit{Number of \textit{Interest} packets}: Unlike the number of PIT entries metric, which includes only pending \textit{Interests}, this metric represents the total number of \textit{Interest} packets issued by an interface.
    
    \item \textit{Number of \textit{Data} packets}: It represents the total number of \textit{Data} packets received by an interface in a period. 
    
    \item \textit{Number of satisfied \textit{Interest} packets}: It represents the total number of \textit{Interest} packets that resulted in a \textit{Data} packet. 
    
    \item \textit{Number of dropped packets}: This metric represents the total number of dropped packet recorded for a given interface.
    
    \item \textit{Number of timed-out \textit{Interest} packets}: It represents the number of timed-out \textit{Interest} packet of a given interface. An \textit{Interest} packet times-out when its \textit{InterestLifetime} reaches zero before a response comes back. 
    
    \item \textit{Number of NACK packets}: Represents the number of NACK packets recorded by a router for one or all its interfaces. 
    
    \item \textit{Traffic rate}: As already defined, traffic rate on an interface equals the frequency of incoming \textit{Interest} packets.
    
    \item \textit{Interest drop rate}: This metric represents the number of dropped \textit{Interest} packets, due to PIT saturation, to the total number of \textit{Interest} packets issued (by an interface) or received by a router. 
    
    \item \textit{Delay}: It represents the time that takes an \textit{Interest} packet to reach a destination.
    
    \item \textit{False positive ratio}: This metric represents the number of wrongly classified events as malicious to the total number of events recorded. 
\end{itemize}

\begin{table*}
\setlength{\tabcolsep}{4pt}
\centering
\caption{Simulation Evaluation Metrics}
\label{tab:sim_metric}
\begin{adjustbox}{width=\textwidth}
\begin{tabular}{cccccccccccccc}
\hline
Ref & \makecell{Satisfaction\\ratio} & \makecell{PIT usage} & \makecell{\# of PIT\\entries} & \makecell{\# of\\\textit{Interests}} & \makecell{\# of\\\textit{Data}} & \makecell{\# of satisfied\\\textit{Interests}} &  \makecell{Dropped\\packets} & \makecell{\# of timed-out\\\textit{Interests}} & \makecell{\# of NACK\\packets} & \makecell{Traffic\\rate} &  \makecell{\textit{Interests} drop\\ratio} & \makecell{Delay} & \makecell{False positive\\ratio} \\
\hline
\cite{afanasyev2013interest} & \cmark &  &  &  &  &  &  &  &  &  &  &  &  \\ \hline

\cite{compagno2013poseidon} &  & \cmark &  &  & \cmark &  &  &  &  &  &  &  &  \\ \hline

\cite{dai2013mitigate} &  & \cmark & \cmark &  &  &  &  &  &  &  &  &  &  \\ \hline

\cite{vassilakis2015mitigating} &  & \cmark &  &  &  &  & \cmark &  &  &  &  &  &  \\ \hline

\cite{wang2014cooperative} & \cmark & \cmark &  &  &  &  &  &  &  &  &  & \cmark &  \\ \hline

\cite{salah2015coordination},\cite{salah2016evaluating} &  & \cmark &  &  &  & \cmark &  &  &  &  &  &  &  \\ \hline

\cite{wang2013decoupling} &  & \cmark &  &  &  &  &  &  &  &  &  &  &  \\ \hline

\cite{wang2014detecting} &  &  &  & \cmark &  &  &  &  &  &  &  &  &  \\ \hline

\cite{karami2015hybrid} & \cmark & \cmark &  &  & \cmark &  &  &  &  &  &  &  & \cmark \\ \hline

\cite{ding2016cooperative} & \cmark & \cmark &  &  &  &  &  &  &  &  &  &  &  \\ \hline

\cite{xin2017detection} &  & \cmark &  &  &  &  &  &  &  & \cmark &  &  &  \\ \hline

\cite{xin2016novel} &  & \cmark &  &  &  &  &  &  &  &  &  &  &  \\ \hline

\cite{zhi2018gini} &  &  & \cmark &  &  &  &  &  &  &  &  &  & \cmark \\ \hline

\cite{nakatsuka2018frog} & \cmark &  &  &  &  &  &  &  &  &  &  &  &  \\ \hline

\cite{benarfa2019chokifa}, \cite{benarfa2020chokifap} & \cmark & \cmark &  &  &  &  & \cmark &  &  &  &  &  &  \\ \hline

\cite{zhang2019expect} &  &  &  &  &  &  &  &  &  & \cmark &  &  &  \\ \hline  

\cite{hou2019theil} &  & \cmark &  &  & \cmark &  &  &  &  &  &  &  &  \\ \hline

\cite{pu2019self} &  & \cmark &  &  &  & \cmark &  &  &  &  &  &  &  \\ \hline

\cite{zhang2019ari} & \cmark &  & \cmark &  &  &  &  &  &  &  &  &  &  \\ \hline

\cite{alston2016neutralizing} &  &  &  &  &  &  &  &  &  &  & \cmark & \cmark &  \\ \hline

\cite{cheng2019detecting} & \cmark &  & \cmark &  &  &  &  &  &  &  &  & \cmark &  \\ \hline

\cite{liu2018accuracy} &  &  & \cmark &  &  &  &  &  &  &  &  &  & \cmark \\ \hline 

\cite{liu2018blam} &  & \cmark &  &  &  &  &  &  &  &  &  & \cmark & \cmark \\ \hline

\cite{chen2019isolation} &  & \cmark &  &  & \cmark &  &  &  &  &  &  &  &  \\ \hline

\cite{zhi2019resist} & \cmark & \cmark &  & \cmark &  & \cmark & \cmark &  &  &  &  &  &  \\ \hline

\cite{yin2019controller} & \cmark &  &  & \cmark & \cmark &  &  &  & \cmark &  &  &  &  \\ \hline

\cite{benmoussa2019novel} & \cmark &  &  & \cmark &  &  & \cmark & \cmark & \cmark &  &  &  &  \\ \hline

\cite{zhi2020reputation} & \cmark & \cmark &  &  &  &  &  &  &  &  & \cmark & \cmark &  \\ \hline

\cite{shinohara2016cache} &  &  &  & \cmark & \cmark &  & \cmark &  &  &  &  &  &  \\ \hline

\cite{shigeyasu2018distributed} & \cmark &  &  &  &  &  &  &  &  &  &  &  &  \\ \hline

\cite{shigeyasu2018distributed} & \cmark &  &  &  &  &  &  &  &  &  &  &  &  \\ \hline

\cite{benmoussa2020msidn} & \cmark &  &  &  &  & \cmark & \cmark & \cmark & \cmark &  &  &  &  \\ \hline
\end{tabular}
\end{adjustbox}
\end{table*}

\section*{Acknowledgments}
This work is partially supported by National Science Foundation awards CNS-2104700, CNS-2016714, and CBET-2124918, the National Institutes of Health (NIGMS/P20GM109090), the Nebraska University Collaboration Initiative, and the Nebraska Tobacco Settlement Biomedical Research Development Funds.

\bibliographystyle{ACM-Reference-Format}
\bibliography{Survey}


\begin{thebibliography}{130}

%%% the Web address.

\ifx \showCODEN    \undefined \def \showCODEN     #1{\unskip}     \fi
\ifx \showDOI      \undefined \def \showDOI       #1{#1}\fi
\ifx \showISBNx    \undefined \def \showISBNx     #1{\unskip}     \fi
\ifx \showISBNxiii \undefined \def \showISBNxiii  #1{\unskip}     \fi
\ifx \showISSN     \undefined \def \showISSN      #1{\unskip}     \fi
\ifx \showLCCN     \undefined \def \showLCCN      #1{\unskip}     \fi
\ifx \shownote     \undefined \def \shownote      #1{#1}          \fi
\ifx \showarticletitle \undefined \def \showarticletitle #1{#1}   \fi
\ifx \showURL      \undefined \def \showURL       {\relax}        \fi
% invisible to TeX
\providecommand\bibfield[2]{#2}
\providecommand\bibinfo[2]{#2}
\providecommand\natexlab[1]{#1}
\providecommand\showeprint[2][]{arXiv:#2}

\bibitem[\protect\citeauthoryear{??}{cis}{2023}]%
        {ciscorep}
 \bibinfo{year}{Cisco Visual Networking Index: Forecast and Trends
  2018–2023}\natexlab{}.
\newblock \bibinfo{booktitle}{\emph{2020}}.
\newblock
\urldef\tempurl%
\url{https://www.cisco.com/c/en/us/solutions/collateral/executive-perspectives/annual-internet-report/white-paper-c11-741490.html}
\showURL{%
Retrieved Oct 2, 2020 from \tempurl}


\bibitem[\protect\citeauthoryear{??}{dat}{cket}]%
        {datapkt}
 \bibinfo{year}{NDN Data packet}\natexlab{}.
\newblock \bibinfo{booktitle}{\emph{2020}}.
\newblock
\urldef\tempurl%
\url{https://named-data.net/doc/NDN-packet-spec/current/data.html}
\showURL{%
Retrieved Oct 3, 2020 from \tempurl}


\bibitem[\protect\citeauthoryear{??}{ndn}{n 03}]%
        {ndnpktspec}
 \bibinfo{year}{NDN Packet Format Specification version 0.3}\natexlab{}.
\newblock \bibinfo{booktitle}{\emph{2020}}.
\newblock
\urldef\tempurl%
\url{https://named-data.net/doc/NDN-packet-spec/current/interest.html}
\showURL{%
Retrieved Sep 17, 2020 from \tempurl}


\bibitem[\protect\citeauthoryear{??}{sig}{cket}]%
        {signinterest}
 \bibinfo{year}{Signed Interest Packet}\natexlab{}.
\newblock \bibinfo{booktitle}{\emph{2020}}.
\newblock
\urldef\tempurl%
\url{https://named-data.net/doc/NDN-packet-spec/current/signed-interest.html}
\showURL{%
Retrieved Sep 17, 2020 from \tempurl}


\bibitem[\protect\citeauthoryear{Aamir and Zaidi}{Aamir and Zaidi}{2015}]%
        {aamir2015denial}
\bibfield{author}{\bibinfo{person}{Muhammad Aamir} {and} \bibinfo{person}{Syed
  Mustafa~Ali Zaidi}.} \bibinfo{year}{2015}\natexlab{}.
\newblock \showarticletitle{Denial-of-service in content centric (named data)
  networking: a tutorial and state-of-the-art survey}.
\newblock \bibinfo{journal}{\emph{Security and Communication Networks}}
  \bibinfo{volume}{8}, \bibinfo{number}{11} (\bibinfo{year}{2015}),
  \bibinfo{pages}{2037--2059}.
\newblock


\bibitem[\protect\citeauthoryear{AbdAllah, Hassanein, and Zulkernine}{AbdAllah
  et~al\mbox{.}}{2015}]%
        {abdallah2015survey}
\bibfield{author}{\bibinfo{person}{Eslam~G AbdAllah}, \bibinfo{person}{Hossam~S
  Hassanein}, {and} \bibinfo{person}{Mohammad Zulkernine}.}
  \bibinfo{year}{2015}\natexlab{}.
\newblock \showarticletitle{A survey of security attacks in information-centric
  networking}.
\newblock \bibinfo{journal}{\emph{IEEE Communications Surveys \& Tutorials}}
  \bibinfo{volume}{17}, \bibinfo{number}{3} (\bibinfo{year}{2015}),
  \bibinfo{pages}{1441--1454}.
\newblock


\bibitem[\protect\citeauthoryear{Abraham and Crowley}{Abraham and
  Crowley}{2017}]%
        {abraham2017controlling}
\bibfield{author}{\bibinfo{person}{Hila~Ben Abraham} {and}
  \bibinfo{person}{Patrick Crowley}.} \bibinfo{year}{2017}\natexlab{}.
\newblock \showarticletitle{Controlling strategy retransmissions in named data
  networking}. In \bibinfo{booktitle}{\emph{2017 ACM/IEEE Symposium on
  Architectures for Networking and Communications Systems (ANCS)}}. IEEE,
  \bibinfo{pages}{70--81}.
\newblock


\bibitem[\protect\citeauthoryear{Afanasyev, Halderman, Ruoti, Seamons, Yu,
  Zappala, and Zhang}{Afanasyev et~al\mbox{.}}{2016}]%
        {afanasyev2016content}
\bibfield{author}{\bibinfo{person}{Alexander Afanasyev},
  \bibinfo{person}{J~Alex Halderman}, \bibinfo{person}{Scott Ruoti},
  \bibinfo{person}{Kent Seamons}, \bibinfo{person}{Yingdi Yu},
  \bibinfo{person}{Daniel Zappala}, {and} \bibinfo{person}{Lixia Zhang}.}
  \bibinfo{year}{2016}\natexlab{}.
\newblock \showarticletitle{Content-based security for the web}. In
  \bibinfo{booktitle}{\emph{Proceedings of the 2016 New Security Paradigms
  Workshop}}. \bibinfo{pages}{49--60}.
\newblock


\bibitem[\protect\citeauthoryear{Afanasyev, Jiang, Yu, Tan, Xia, Mankin, and
  Zhang}{Afanasyev et~al\mbox{.}}{2017}]%
        {afanasyev2017ndns}
\bibfield{author}{\bibinfo{person}{Alexander Afanasyev},
  \bibinfo{person}{Xiaoke Jiang}, \bibinfo{person}{Yingdi Yu},
  \bibinfo{person}{Jiewen Tan}, \bibinfo{person}{Yumin Xia},
  \bibinfo{person}{Allison Mankin}, {and} \bibinfo{person}{Lixia Zhang}.}
  \bibinfo{year}{2017}\natexlab{}.
\newblock \showarticletitle{NDNS: A DNS-like name service for NDN}. In
  \bibinfo{booktitle}{\emph{2017 26th International Conference on Computer
  Communication and Networks (ICCCN)}}. IEEE, \bibinfo{pages}{1--9}.
\newblock


\bibitem[\protect\citeauthoryear{Afanasyev, Mahadevan, Moiseenko, Uzun, and
  Zhang}{Afanasyev et~al\mbox{.}}{2013}]%
        {afanasyev2013interest}
\bibfield{author}{\bibinfo{person}{Alexander Afanasyev}, \bibinfo{person}{Priya
  Mahadevan}, \bibinfo{person}{Ilya Moiseenko}, \bibinfo{person}{Ersin Uzun},
  {and} \bibinfo{person}{Lixia Zhang}.} \bibinfo{year}{2013}\natexlab{}.
\newblock \showarticletitle{Interest flooding attack and countermeasures in
  Named Data Networking}. In \bibinfo{booktitle}{\emph{2013 IFIP Networking
  Conference}}. IEEE, \bibinfo{pages}{1--9}.
\newblock


\bibitem[\protect\citeauthoryear{Ahlgren, Dannewitz, Imbrenda, Kutscher, and
  Ohlman}{Ahlgren et~al\mbox{.}}{2012}]%
        {ahlgren2012survey}
\bibfield{author}{\bibinfo{person}{Bengt Ahlgren}, \bibinfo{person}{Christian
  Dannewitz}, \bibinfo{person}{Claudio Imbrenda}, \bibinfo{person}{Dirk
  Kutscher}, {and} \bibinfo{person}{Borje Ohlman}.}
  \bibinfo{year}{2012}\natexlab{}.
\newblock \showarticletitle{A survey of information-centric networking}.
\newblock \bibinfo{journal}{\emph{IEEE Communications Magazine}}
  \bibinfo{volume}{50}, \bibinfo{number}{7} (\bibinfo{year}{2012}),
  \bibinfo{pages}{26--36}.
\newblock


\bibitem[\protect\citeauthoryear{Al~Azad and Mastorakis}{Al~Azad and
  Mastorakis}{2022a}]%
        {al2022promise}
\bibfield{author}{\bibinfo{person}{Md~Washik Al~Azad} {and}
  \bibinfo{person}{Spyridon Mastorakis}.} \bibinfo{year}{2022}\natexlab{a}.
\newblock \showarticletitle{The promise and challenges of computation
  deduplication and reuse at the network edge}.
\newblock \bibinfo{journal}{\emph{IEEE Wireless Communications}}
  (\bibinfo{year}{2022}).
\newblock


\bibitem[\protect\citeauthoryear{Al~Azad and Mastorakis}{Al~Azad and
  Mastorakis}{2022b}]%
        {al2022reservoir}
\bibfield{author}{\bibinfo{person}{Md~Washik Al~Azad} {and}
  \bibinfo{person}{Spyridon Mastorakis}.} \bibinfo{year}{2022}\natexlab{b}.
\newblock \showarticletitle{Reservoir: Named Data for Pervasive Computation
  Reuse at the Network Edge}. In \bibinfo{booktitle}{\emph{2022 IEEE
  International Conference on Pervasive Computing and Communications
  (PerCom)}}. IEEE, \bibinfo{pages}{141--151}.
\newblock


\bibitem[\protect\citeauthoryear{Al~Azad, Shannigrahi, Stergiou, Ortega, and
  Mastorakis}{Al~Azad et~al\mbox{.}}{2021}]%
        {al2021cledge}
\bibfield{author}{\bibinfo{person}{Md~Washik Al~Azad}, \bibinfo{person}{Susmit
  Shannigrahi}, \bibinfo{person}{Nicholas Stergiou},
  \bibinfo{person}{Francisco~R Ortega}, {and} \bibinfo{person}{Spyridon
  Mastorakis}.} \bibinfo{year}{2021}\natexlab{}.
\newblock \showarticletitle{CLEDGE: A Hybrid Cloud-Edge Computing Framework
  over Information Centric Networking}. In \bibinfo{booktitle}{\emph{2021 IEEE
  46th Conference on Local Computer Networks (LCN)}}. IEEE,
  \bibinfo{pages}{589--596}.
\newblock


\bibitem[\protect\citeauthoryear{Al-Sheikh, W{\"a}hlisch, and
  Schmidt}{Al-Sheikh et~al\mbox{.}}{2015}]%
        {al2015revisiting}
\bibfield{author}{\bibinfo{person}{Samir Al-Sheikh}, \bibinfo{person}{Matthias
  W{\"a}hlisch}, {and} \bibinfo{person}{Thomas~C Schmidt}.}
  \bibinfo{year}{2015}\natexlab{}.
\newblock \showarticletitle{Revisiting countermeasures against ndn interest
  flooding}. In \bibinfo{booktitle}{\emph{Proceedings of the 2nd ACM Conference
  on Information-Centric Networking}}. \bibinfo{pages}{195--196}.
\newblock


\bibitem[\protect\citeauthoryear{Alhisnawi and Ahmadi}{Alhisnawi and
  Ahmadi}{2020}]%
        {alhisnawidetecting}
\bibfield{author}{\bibinfo{person}{Mohammad Alhisnawi} {and}
  \bibinfo{person}{Mahmood Ahmadi}.} \bibinfo{year}{2020}\natexlab{}.
\newblock \showarticletitle{Detecting and Mitigating DDoS Attack in Named Data
  Networking}.
\newblock \bibinfo{journal}{\emph{Journal of Network and Systems Management}}
  (\bibinfo{year}{2020}), \bibinfo{pages}{1343--1356}.
\newblock


\bibitem[\protect\citeauthoryear{Alston and Refaei}{Alston and Refaei}{2016}]%
        {alston2016neutralizing}
\bibfield{author}{\bibinfo{person}{Aubrey Alston} {and} \bibinfo{person}{Tamer
  Refaei}.} \bibinfo{year}{2016}\natexlab{}.
\newblock \showarticletitle{Neutralizing interest flooding attacks in named
  data networks using cryptographic route tokens}. In
  \bibinfo{booktitle}{\emph{2016 IEEE 15th International Symposium on Network
  Computing and Applications (NCA)}}. IEEE, \bibinfo{pages}{85--88}.
\newblock


\bibitem[\protect\citeauthoryear{Ambrosin, Compagno, Conti, Ghali, and
  Tsudik}{Ambrosin et~al\mbox{.}}{2018}]%
        {ambrosin2018security}
\bibfield{author}{\bibinfo{person}{Moreno Ambrosin}, \bibinfo{person}{Alberto
  Compagno}, \bibinfo{person}{Mauro Conti}, \bibinfo{person}{Cesar Ghali},
  {and} \bibinfo{person}{Gene Tsudik}.} \bibinfo{year}{2018}\natexlab{}.
\newblock \showarticletitle{Security and privacy analysis of national science
  foundation future internet architectures}.
\newblock \bibinfo{journal}{\emph{IEEE Communications Surveys \& Tutorials}}
  \bibinfo{volume}{20}, \bibinfo{number}{2} (\bibinfo{year}{2018}),
  \bibinfo{pages}{1418--1442}.
\newblock


\bibitem[\protect\citeauthoryear{Anand, Dogar, Han, Li, Lim, Machado, Wu,
  Akella, Andersen, Byers, et~al\mbox{.}}{Anand et~al\mbox{.}}{2011}]%
        {anand2011xia}
\bibfield{author}{\bibinfo{person}{Ashok Anand}, \bibinfo{person}{Fahad Dogar},
  \bibinfo{person}{Dongsu Han}, \bibinfo{person}{Boyan Li},
  \bibinfo{person}{Hyeontaek Lim}, \bibinfo{person}{Michel Machado},
  \bibinfo{person}{Wenfei Wu}, \bibinfo{person}{Aditya Akella},
  \bibinfo{person}{David~G Andersen}, \bibinfo{person}{John~W Byers},
  {et~al\mbox{.}}} \bibinfo{year}{2011}\natexlab{}.
\newblock \showarticletitle{XIA: An architecture for an evolvable and
  trustworthy Internet}. In \bibinfo{booktitle}{\emph{Proceedings of the 10th
  ACM Workshop on Hot Topics in Networks}}. \bibinfo{pages}{1--6}.
\newblock


\bibitem[\protect\citeauthoryear{Anderson, Birman, Broberg, Caesar, Comer,
  Cotton, Freedman, Haeberlen, Ives, Krishnamurthy, et~al\mbox{.}}{Anderson
  et~al\mbox{.}}{2013}]%
        {anderson2013nebula}
\bibfield{author}{\bibinfo{person}{Tom Anderson}, \bibinfo{person}{Ken Birman},
  \bibinfo{person}{Robert Broberg}, \bibinfo{person}{Matthew Caesar},
  \bibinfo{person}{Douglas Comer}, \bibinfo{person}{Chase Cotton},
  \bibinfo{person}{Michael~J Freedman}, \bibinfo{person}{Andreas Haeberlen},
  \bibinfo{person}{Zachary~G Ives}, \bibinfo{person}{Arvind Krishnamurthy},
  {et~al\mbox{.}}} \bibinfo{year}{2013}\natexlab{}.
\newblock \showarticletitle{The nebula future internet architecture}. In
  \bibinfo{booktitle}{\emph{The Future Internet Assembly}}. Springer,
  \bibinfo{pages}{16--26}.
\newblock


\bibitem[\protect\citeauthoryear{Benarfa, Hassan, , Losiouk, Compagno, Yagoubi,
  and Conti}{Benarfa et~al\mbox{.}}{2020}]%
        {benarfa2020chokifap}
\bibfield{author}{\bibinfo{person}{Abdelmadjid Benarfa},
  \bibinfo{person}{Muhammad Hassan}, \bibinfo{person}{},
  \bibinfo{person}{Eleonora Losiouk}, \bibinfo{person}{Alberto Compagno},
  \bibinfo{person}{Mohamed~Bachir Yagoubi}, {and} \bibinfo{person}{Mauro
  Conti}.} \bibinfo{year}{2020}\natexlab{}.
\newblock \showarticletitle{ChoKIFA+: an early detection and mitigation
  approach against interest flooding attacks in NDN}.
\newblock \bibinfo{journal}{\emph{International Journal of Information
  Security}} (\bibinfo{year}{2020}).
\newblock


\bibitem[\protect\citeauthoryear{Benarfa, Hassan, Compagno, Losiouk, Yagoubi,
  and Conti}{Benarfa et~al\mbox{.}}{2019}]%
        {benarfa2019chokifa}
\bibfield{author}{\bibinfo{person}{Abdelmadjid Benarfa},
  \bibinfo{person}{Muhammad Hassan}, \bibinfo{person}{Alberto Compagno},
  \bibinfo{person}{Eleonora Losiouk}, \bibinfo{person}{Mohamed~Bachir Yagoubi},
  {and} \bibinfo{person}{Mauro Conti}.} \bibinfo{year}{2019}\natexlab{}.
\newblock \showarticletitle{ChoKIFA: A New Detection and Mitigation Approach
  Against Interest Flooding Attacks in NDN}. In
  \bibinfo{booktitle}{\emph{International Conference on Wired/Wireless Internet
  Communication}}. Springer, \bibinfo{pages}{53--65}.
\newblock


\bibitem[\protect\citeauthoryear{Benmoussa, el~Karim~Tahari, Kerrache, Lagraa,
  Lakas, Hussain, and Ahmad}{Benmoussa et~al\mbox{.}}{2020}]%
        {benmoussa2020msidn}
\bibfield{author}{\bibinfo{person}{Ahmed Benmoussa}, \bibinfo{person}{Abdou el
  Karim~Tahari}, \bibinfo{person}{Chaker~Abdelaziz Kerrache},
  \bibinfo{person}{Nasreddine Lagraa}, \bibinfo{person}{Abderrahmane Lakas},
  \bibinfo{person}{Rasheed Hussain}, {and} \bibinfo{person}{Farhan Ahmad}.}
  \bibinfo{year}{2020}\natexlab{}.
\newblock \showarticletitle{MSIDN: Mitigation of sophisticated interest
  flooding-based DDoS attacks in named data networking}.
\newblock \bibinfo{journal}{\emph{Future Generation Computer Systems}}
  \bibinfo{volume}{107} (\bibinfo{year}{2020}), \bibinfo{pages}{293--306}.
\newblock


\bibitem[\protect\citeauthoryear{Benmoussa, el~Karim~Tahari, Lagaa, Lakas,
  Ahmad, Hussain, Kerrache, and Kurugollu}{Benmoussa et~al\mbox{.}}{2019}]%
        {benmoussa2019novel}
\bibfield{author}{\bibinfo{person}{Ahmed Benmoussa}, \bibinfo{person}{Abdou el
  Karim~Tahari}, \bibinfo{person}{Nasreddine Lagaa},
  \bibinfo{person}{Abderrahmane Lakas}, \bibinfo{person}{Farhan Ahmad},
  \bibinfo{person}{Rasheed Hussain}, \bibinfo{person}{Chaker~Abdelaziz
  Kerrache}, {and} \bibinfo{person}{Fatih Kurugollu}.}
  \bibinfo{year}{2019}\natexlab{}.
\newblock \showarticletitle{A Novel Congestion-Aware Interest Flooding Attacks
  Detection Mechanism in Named Data Networking}. In
  \bibinfo{booktitle}{\emph{2019 28th International Conference on Computer
  Communication and Networks (ICCCN)}}. IEEE, \bibinfo{pages}{1--6}.
\newblock


\bibitem[\protect\citeauthoryear{Buragohain and Nandi}{Buragohain and
  Nandi}{2020}]%
        {buragohaindemystifying}
\bibfield{author}{\bibinfo{person}{Madhurima Buragohain} {and}
  \bibinfo{person}{Sukumar Nandi}.} \bibinfo{year}{2020}\natexlab{}.
\newblock \showarticletitle{Demystifying Security on NDN: A Survey of Existing
  Attacks and Open Research Challenges}.
\newblock In \bibinfo{booktitle}{\emph{The" Essence" of Network Security: An
  End-to-End Panorama}}. \bibinfo{publisher}{Springer},
  \bibinfo{pages}{241--261}.
\newblock


\bibitem[\protect\citeauthoryear{Cao, Pei, Zhang, Zhang, and Zhao}{Cao
  et~al\mbox{.}}{2016}]%
        {cao2016fetching}
\bibfield{author}{\bibinfo{person}{Jianxun Cao}, \bibinfo{person}{Dan Pei},
  \bibinfo{person}{Xiaoping Zhang}, \bibinfo{person}{Beichuan Zhang}, {and}
  \bibinfo{person}{Youjian Zhao}.} \bibinfo{year}{2016}\natexlab{}.
\newblock \showarticletitle{Fetching popular data from the nearest replica in
  NDN}. In \bibinfo{booktitle}{\emph{2016 25th International Conference on
  Computer Communication and Networks (ICCCN)}}. IEEE, \bibinfo{pages}{1--9}.
\newblock


\bibitem[\protect\citeauthoryear{Chan, Ko, Mastorakis, Afanasyev, and
  Zhang}{Chan et~al\mbox{.}}{2017}]%
        {chan2017fuzzy}
\bibfield{author}{\bibinfo{person}{Kevin Chan}, \bibinfo{person}{Bongjun Ko},
  \bibinfo{person}{Spyridon Mastorakis}, \bibinfo{person}{Alexander Afanasyev},
  {and} \bibinfo{person}{Lixia Zhang}.} \bibinfo{year}{2017}\natexlab{}.
\newblock \showarticletitle{Fuzzy interest forwarding}. In
  \bibinfo{booktitle}{\emph{Proceedings of the Asian Internet Engineering
  Conference}}. \bibinfo{pages}{31--37}.
\newblock


\bibitem[\protect\citeauthoryear{Chapin, Clark, Braden, Hobby, and Cerf}{Chapin
  et~al\mbox{.}}{1991}]%
        {rfc1287}
\bibfield{author}{\bibinfo{person}{Lyman Chapin}, \bibinfo{person}{Dr. David~D.
  Clark}, \bibinfo{person}{Robert~T. Braden}, \bibinfo{person}{Russ Hobby},
  {and} \bibinfo{person}{Dr. Vinton~G. Cerf}.} \bibinfo{year}{1991}\natexlab{}.
\newblock \bibinfo{title}{{Towards the Future Internet Architecture}}.
\newblock \bibinfo{howpublished}{RFC 1287}.
\newblock
\urldef\tempurl%
\url{https://rfc-editor.org/rfc/rfc1287.txt}
\showURL{%
\tempurl}


\bibitem[\protect\citeauthoryear{Chen, Xing, Cui, Huo, and Hou}{Chen
  et~al\mbox{.}}{2019}]%
        {chen2019isolation}
\bibfield{author}{\bibinfo{person}{Jing Chen}, \bibinfo{person}{Guanglin Xing},
  \bibinfo{person}{Mengtian Cui}, \bibinfo{person}{Hong Huo}, {and}
  \bibinfo{person}{Rui Hou}.} \bibinfo{year}{2019}\natexlab{}.
\newblock \showarticletitle{Isolation Forest Based Interest Flooding Attack
  Detection Mechanism in NDN}. In \bibinfo{booktitle}{\emph{2019 2nd
  International Conference on Hot Information-Centric Networking (HotICN)}}.
  IEEE, \bibinfo{pages}{58--62}.
\newblock


\bibitem[\protect\citeauthoryear{Chen and Mizero}{Chen and Mizero}{2015}]%
        {chen2015survey}
\bibfield{author}{\bibinfo{person}{Shuoshuo Chen} {and}
  \bibinfo{person}{Fabrice Mizero}.} \bibinfo{year}{2015}\natexlab{}.
\newblock \showarticletitle{A survey on security in named data networking}.
\newblock \bibinfo{journal}{\emph{arXiv preprint arXiv:1512.04127}}
  (\bibinfo{year}{2015}).
\newblock


\bibitem[\protect\citeauthoryear{Cheng, Zhao, Hu, Zheng, Wu, Li, and Fan}{Cheng
  et~al\mbox{.}}{2019}]%
        {cheng2019detecting}
\bibfield{author}{\bibinfo{person}{Guang Cheng}, \bibinfo{person}{Lixia Zhao},
  \bibinfo{person}{Xiaoyan Hu}, \bibinfo{person}{Shaoqi Zheng},
  \bibinfo{person}{Hua Wu}, \bibinfo{person}{Ruidong Li}, {and}
  \bibinfo{person}{Chengyu Fan}.} \bibinfo{year}{2019}\natexlab{}.
\newblock \showarticletitle{Detecting and Mitigating A Sophisticated Interest
  Flooding Attack in NDN from the Network-Wide View}. In
  \bibinfo{booktitle}{\emph{2019 IEEE First International Workshop on Network
  Meets Intelligent Computations (NMIC)}}. IEEE, \bibinfo{pages}{7--12}.
\newblock


\bibitem[\protect\citeauthoryear{Choi, Kim, Kim, and Roh}{Choi
  et~al\mbox{.}}{2013}]%
        {choi2013threat}
\bibfield{author}{\bibinfo{person}{Seungoh Choi}, \bibinfo{person}{Kwangsoo
  Kim}, \bibinfo{person}{Seongmin Kim}, {and} \bibinfo{person}{Byeong-hee
  Roh}.} \bibinfo{year}{2013}\natexlab{}.
\newblock \showarticletitle{Threat of DoS by interest flooding attack in
  content-centric networking}. In \bibinfo{booktitle}{\emph{The International
  Conference on Information Networking 2013 (ICOIN)}}. IEEE,
  \bibinfo{pages}{315--319}.
\newblock


\bibitem[\protect\citeauthoryear{Compagno, Conti, Gasti, and Tsudik}{Compagno
  et~al\mbox{.}}{2013}]%
        {compagno2013poseidon}
\bibfield{author}{\bibinfo{person}{Alberto Compagno}, \bibinfo{person}{Mauro
  Conti}, \bibinfo{person}{Paolo Gasti}, {and} \bibinfo{person}{Gene Tsudik}.}
  \bibinfo{year}{2013}\natexlab{}.
\newblock \showarticletitle{Poseidon: Mitigating interest flooding DDoS attacks
  in named data networking}. In \bibinfo{booktitle}{\emph{38th annual IEEE
  conference on local computer networks}}. IEEE, \bibinfo{pages}{630--638}.
\newblock


\bibitem[\protect\citeauthoryear{Compagno, Conti, Ghali, and Tsudik}{Compagno
  et~al\mbox{.}}{2015}]%
        {compagno2015nack}
\bibfield{author}{\bibinfo{person}{Alberto Compagno}, \bibinfo{person}{Mauro
  Conti}, \bibinfo{person}{Cesar Ghali}, {and} \bibinfo{person}{Gene Tsudik}.}
  \bibinfo{year}{2015}\natexlab{}.
\newblock \showarticletitle{To NACK or not to NACK? negative acknowledgments in
  information-centric networking}. In \bibinfo{booktitle}{\emph{2015 24th
  International Conference on Computer Communication and Networks (ICCCN)}}.
  IEEE, \bibinfo{pages}{1--10}.
\newblock


\bibitem[\protect\citeauthoryear{Dai, Wang, Fan, and Liu}{Dai
  et~al\mbox{.}}{2013}]%
        {dai2013mitigate}
\bibfield{author}{\bibinfo{person}{Huichen Dai}, \bibinfo{person}{Yi Wang},
  \bibinfo{person}{Jindou Fan}, {and} \bibinfo{person}{Bin Liu}.}
  \bibinfo{year}{2013}\natexlab{}.
\newblock \showarticletitle{Mitigate ddos attacks in ndn by interest
  traceback}. In \bibinfo{booktitle}{\emph{2013 IEEE Conference on Computer
  Communications Workshops (INFOCOM WKSHPS)}}. IEEE, \bibinfo{pages}{381--386}.
\newblock


\bibitem[\protect\citeauthoryear{Dierks and Rescorla}{Dierks and
  Rescorla}{2008}]%
        {dierks2008rfc}
\bibfield{author}{\bibinfo{person}{Tim Dierks} {and} \bibinfo{person}{Eric
  Rescorla}.} \bibinfo{year}{2008}\natexlab{}.
\newblock \showarticletitle{RFC 5246-the transport layer security (TLS)
  protocol version 1.2}.
\newblock \bibinfo{journal}{\emph{The Internet Engineering Task Force (IETF)}}
  (\bibinfo{year}{2008}).
\newblock


\bibitem[\protect\citeauthoryear{Din, Hassan, Khan, Guizani, Ghazali, and
  Habbal}{Din et~al\mbox{.}}{2017}]%
        {din2017caching}
\bibfield{author}{\bibinfo{person}{Ikram~Ud Din}, \bibinfo{person}{Suhaidi
  Hassan}, \bibinfo{person}{Muhammad~Khurram Khan}, \bibinfo{person}{Mohsen
  Guizani}, \bibinfo{person}{Osman Ghazali}, {and} \bibinfo{person}{Adib
  Habbal}.} \bibinfo{year}{2017}\natexlab{}.
\newblock \showarticletitle{Caching in information-centric networking:
  Strategies, challenges, and future research directions}.
\newblock \bibinfo{journal}{\emph{IEEE Communications Surveys \& Tutorials}}
  \bibinfo{volume}{20}, \bibinfo{number}{2} (\bibinfo{year}{2017}),
  \bibinfo{pages}{1443--1474}.
\newblock


\bibitem[\protect\citeauthoryear{Ding, Liu, Cho, Chao, and Shih}{Ding
  et~al\mbox{.}}{2016}]%
        {ding2016cooperative}
\bibfield{author}{\bibinfo{person}{Kun Ding}, \bibinfo{person}{Yun Liu},
  \bibinfo{person}{Hsin-Hung Cho}, \bibinfo{person}{Han-Chieh Chao}, {and}
  \bibinfo{person}{Timothy~K Shih}.} \bibinfo{year}{2016}\natexlab{}.
\newblock \showarticletitle{Cooperative detection and protection for interest
  flooding attacks in named data networking}.
\newblock \bibinfo{journal}{\emph{International Journal of Communication
  Systems}} \bibinfo{volume}{29}, \bibinfo{number}{13} (\bibinfo{year}{2016}),
  \bibinfo{pages}{1968--1980}.
\newblock


\bibitem[\protect\citeauthoryear{Dong, Wang, Quan, and Yin}{Dong
  et~al\mbox{.}}{2020}]%
        {dong2020interestfence}
\bibfield{author}{\bibinfo{person}{Jiaqing Dong}, \bibinfo{person}{Kai Wang},
  \bibinfo{person}{Wei Quan}, {and} \bibinfo{person}{Hao Yin}.}
  \bibinfo{year}{2020}\natexlab{}.
\newblock \showarticletitle{InterestFence: Simple but efficient way to counter
  interest flooding attack}.
\newblock \bibinfo{journal}{\emph{Computers \& Security}}  \bibinfo{volume}{88}
  (\bibinfo{year}{2020}), \bibinfo{pages}{101628}.
\newblock


\bibitem[\protect\citeauthoryear{Fotiou, Nikander, Trossen, and Polyzos}{Fotiou
  et~al\mbox{.}}{2010}]%
        {fotiou2010developing}
\bibfield{author}{\bibinfo{person}{Nikos Fotiou}, \bibinfo{person}{Pekka
  Nikander}, \bibinfo{person}{Dirk Trossen}, {and} \bibinfo{person}{George~C
  Polyzos}.} \bibinfo{year}{2010}\natexlab{}.
\newblock \showarticletitle{Developing information networking further: From
  PSIRP to PURSUIT}. In \bibinfo{booktitle}{\emph{International Conference on
  Broadband Communications, Networks and Systems}}. Springer,
  \bibinfo{pages}{1--13}.
\newblock


\bibitem[\protect\citeauthoryear{Garc{\'\i}a, Beben, Ram{\'o}n, Maeso, Psaras,
  Pavlou, Wang, {\'S}liwi{\'n}ski, Spirou, Soursos, et~al\mbox{.}}{Garc{\'\i}a
  et~al\mbox{.}}{2011}]%
        {garcia2011comet}
\bibfield{author}{\bibinfo{person}{Gerardo Garc{\'\i}a},
  \bibinfo{person}{Andrzej Beben}, \bibinfo{person}{Francisco~J Ram{\'o}n},
  \bibinfo{person}{Adri{\'a}n Maeso}, \bibinfo{person}{Ioannis Psaras},
  \bibinfo{person}{George Pavlou}, \bibinfo{person}{Ning Wang},
  \bibinfo{person}{Jaros{\l}aw {\'S}liwi{\'n}ski}, \bibinfo{person}{Spiros
  Spirou}, \bibinfo{person}{Sergios Soursos}, {et~al\mbox{.}}}
  \bibinfo{year}{2011}\natexlab{}.
\newblock \showarticletitle{COMET: Content mediator architecture for
  content-aware networks}. In \bibinfo{booktitle}{\emph{2011 Future Network \&
  Mobile Summit}}. IEEE, \bibinfo{pages}{1--8}.
\newblock


\bibitem[\protect\citeauthoryear{Gasti and Tsudik}{Gasti and Tsudik}{2018}]%
        {gasti2018content}
\bibfield{author}{\bibinfo{person}{Paolo Gasti} {and} \bibinfo{person}{Gene
  Tsudik}.} \bibinfo{year}{2018}\natexlab{}.
\newblock \showarticletitle{Content-Centric and Named-Data Networking Security:
  The Good, The Bad and The Rest}. In \bibinfo{booktitle}{\emph{2018 IEEE
  International Symposium on Local and Metropolitan Area Networks (LANMAN)}}.
  IEEE, \bibinfo{pages}{1--6}.
\newblock


\bibitem[\protect\citeauthoryear{Gasti, Tsudik, Uzun, and Zhang}{Gasti
  et~al\mbox{.}}{2013}]%
        {gasti2013and}
\bibfield{author}{\bibinfo{person}{Paolo Gasti}, \bibinfo{person}{Gene Tsudik},
  \bibinfo{person}{Ersin Uzun}, {and} \bibinfo{person}{Lixia Zhang}.}
  \bibinfo{year}{2013}\natexlab{}.
\newblock \showarticletitle{DoS and DDoS in named data networking}. In
  \bibinfo{booktitle}{\emph{2013 22nd International Conference on Computer
  Communication and Networks (ICCCN)}}. IEEE, \bibinfo{pages}{1--7}.
\newblock


\bibitem[\protect\citeauthoryear{Ghali, Tsudik, Uzun, and Wood}{Ghali
  et~al\mbox{.}}{2017}]%
        {ghali2017closing}
\bibfield{author}{\bibinfo{person}{Cesar Ghali}, \bibinfo{person}{Gene Tsudik},
  \bibinfo{person}{Ersin Uzun}, {and} \bibinfo{person}{Christopher~A Wood}.}
  \bibinfo{year}{2017}\natexlab{}.
\newblock \showarticletitle{Closing the floodgate with stateless
  content-centric networking}. In \bibinfo{booktitle}{\emph{2017 26th
  International Conference on Computer Communication and Networks (ICCCN)}}.
  IEEE, \bibinfo{pages}{1--10}.
\newblock


\bibitem[\protect\citeauthoryear{Ghasemi, Yousefi, Shin, and Zhang}{Ghasemi
  et~al\mbox{.}}{2018}]%
        {ghasemi2018muca}
\bibfield{author}{\bibinfo{person}{Chavoosh Ghasemi}, \bibinfo{person}{Hamed
  Yousefi}, \bibinfo{person}{Kang~G Shin}, {and} \bibinfo{person}{Beichuan
  Zhang}.} \bibinfo{year}{2018}\natexlab{}.
\newblock \showarticletitle{MUCA: New routing for named data networking}. In
  \bibinfo{booktitle}{\emph{2018 IFIP Networking Conference (IFIP Networking)
  and Workshops}}. IEEE, \bibinfo{pages}{289--297}.
\newblock


\bibitem[\protect\citeauthoryear{Hou, Han, Chen, Hu, Tan, Luo, and Ma}{Hou
  et~al\mbox{.}}{2019}]%
        {hou2019theil}
\bibfield{author}{\bibinfo{person}{Rui Hou}, \bibinfo{person}{Min Han},
  \bibinfo{person}{Jing Chen}, \bibinfo{person}{Wenbin Hu},
  \bibinfo{person}{Xiaobin Tan}, \bibinfo{person}{Jiangtao Luo}, {and}
  \bibinfo{person}{Maode Ma}.} \bibinfo{year}{2019}\natexlab{}.
\newblock \showarticletitle{Theil-Based Countermeasure against Interest
  Flooding Attacks for Named Data Networks}.
\newblock \bibinfo{journal}{\emph{IEEE Network}} \bibinfo{volume}{33},
  \bibinfo{number}{3} (\bibinfo{year}{2019}), \bibinfo{pages}{116--121}.
\newblock


\bibitem[\protect\citeauthoryear{Jacobson, Mosko, Smetters, and
  Garcia-Luna-Aceves}{Jacobson et~al\mbox{.}}{2007}]%
        {jacobson2007content}
\bibfield{author}{\bibinfo{person}{Van Jacobson}, \bibinfo{person}{Marc Mosko},
  \bibinfo{person}{D Smetters}, {and} \bibinfo{person}{Jose
  Garcia-Luna-Aceves}.} \bibinfo{year}{2007}\natexlab{}.
\newblock \showarticletitle{Content-centric networking}.
\newblock \bibinfo{journal}{\emph{Whitepaper, Palo Alto Research Center}}
  (\bibinfo{year}{2007}), \bibinfo{pages}{2--4}.
\newblock


\bibitem[\protect\citeauthoryear{Jacobson, Smetters, Thornton, Plass, Briggs,
  and Braynard}{Jacobson et~al\mbox{.}}{2009}]%
        {jacobson2009networking}
\bibfield{author}{\bibinfo{person}{Van Jacobson}, \bibinfo{person}{Diana~K
  Smetters}, \bibinfo{person}{James~D Thornton}, \bibinfo{person}{Michael~F
  Plass}, \bibinfo{person}{Nicholas~H Briggs}, {and} \bibinfo{person}{Rebecca~L
  Braynard}.} \bibinfo{year}{2009}\natexlab{}.
\newblock \showarticletitle{Networking named content}. In
  \bibinfo{booktitle}{\emph{Proceedings of the 5th international conference on
  Emerging networking experiments and technologies}}. ACM,
  \bibinfo{pages}{1--12}.
\newblock


\bibitem[\protect\citeauthoryear{Karami and Guerrero-Zapata}{Karami and
  Guerrero-Zapata}{2015}]%
        {karami2015hybrid}
\bibfield{author}{\bibinfo{person}{Amin Karami} {and} \bibinfo{person}{Manel
  Guerrero-Zapata}.} \bibinfo{year}{2015}\natexlab{}.
\newblock \showarticletitle{A hybrid multiobjective rbf-pso method for
  mitigating dos attacks in named data networking}.
\newblock \bibinfo{journal}{\emph{Neurocomputing}}  \bibinfo{volume}{151}
  (\bibinfo{year}{2015}), \bibinfo{pages}{1262--1282}.
\newblock


\bibitem[\protect\citeauthoryear{Khelifi, Luo, Nour, and Shah}{Khelifi
  et~al\mbox{.}}{2018}]%
        {khelifi2018security}
\bibfield{author}{\bibinfo{person}{Hakima Khelifi}, \bibinfo{person}{Senlin
  Luo}, \bibinfo{person}{Boubakr Nour}, {and} \bibinfo{person}{Sayed~Chhattan
  Shah}.} \bibinfo{year}{2018}\natexlab{}.
\newblock \showarticletitle{Security and privacy issues in vehicular named data
  networks: An overview}.
\newblock \bibinfo{journal}{\emph{Mobile Information Systems}}
  \bibinfo{volume}{2018} (\bibinfo{year}{2018}).
\newblock


\bibitem[\protect\citeauthoryear{Koponen, Chawla, Chun, Ermolinskiy, Kim,
  Shenker, and Stoica}{Koponen et~al\mbox{.}}{2007}]%
        {koponen2007data}
\bibfield{author}{\bibinfo{person}{Teemu Koponen}, \bibinfo{person}{Mohit
  Chawla}, \bibinfo{person}{Byung-Gon Chun}, \bibinfo{person}{Andrey
  Ermolinskiy}, \bibinfo{person}{Kye~Hyun Kim}, \bibinfo{person}{Scott
  Shenker}, {and} \bibinfo{person}{Ion Stoica}.}
  \bibinfo{year}{2007}\natexlab{}.
\newblock \showarticletitle{A data-oriented (and beyond) network architecture}.
  In \bibinfo{booktitle}{\emph{Proceedings of the 2007 conference on
  Applications, technologies, architectures, and protocols for computer
  communications}}. \bibinfo{pages}{181--192}.
\newblock


\bibitem[\protect\citeauthoryear{Krishnan and Frankel}{Krishnan and
  Frankel}{2011}]%
        {krishnan2011ip}
\bibfield{author}{\bibinfo{person}{Suresh Krishnan} {and}
  \bibinfo{person}{Sheila Frankel}.} \bibinfo{year}{2011}\natexlab{}.
\newblock \showarticletitle{IP Security (IPsec) and Internet Key Exchange (IKE)
  Document Roadmap}.
\newblock \bibinfo{journal}{\emph{IETF RFC 6071, Ericsson}}
  (\bibinfo{year}{2011}).
\newblock


\bibitem[\protect\citeauthoryear{Kumar, Singh, Aleem, and Srivastava}{Kumar
  et~al\mbox{.}}{2019a}]%
        {kumar2019security}
\bibfield{author}{\bibinfo{person}{Naveen Kumar},
  \bibinfo{person}{Ashutosh~Kumar Singh}, \bibinfo{person}{Abdul Aleem}, {and}
  \bibinfo{person}{Shashank Srivastava}.} \bibinfo{year}{2019}\natexlab{a}.
\newblock \showarticletitle{Security Attacks in Named Data Networking: A Review
  and Research Directions}.
\newblock \bibinfo{journal}{\emph{Journal of Computer Science and Technology}}
  \bibinfo{volume}{34}, \bibinfo{number}{6} (\bibinfo{year}{2019}),
  \bibinfo{pages}{1319--1350}.
\newblock


\bibitem[\protect\citeauthoryear{Kumar, Singh, and Srivastava}{Kumar
  et~al\mbox{.}}{2019b}]%
        {kumar2019feature}
\bibfield{author}{\bibinfo{person}{Naveen Kumar},
  \bibinfo{person}{Ashutosh~Kumar Singh}, {and} \bibinfo{person}{Shashank
  Srivastava}.} \bibinfo{year}{2019}\natexlab{b}.
\newblock \showarticletitle{Feature selection for interest flooding attack in
  named data networking}.
\newblock \bibinfo{journal}{\emph{International Journal of Computers and
  Applications}} (\bibinfo{year}{2019}), \bibinfo{pages}{1--10}.
\newblock


\bibitem[\protect\citeauthoryear{Lee, Zhang, Tu, Afanasyev, and Zhang}{Lee
  et~al\mbox{.}}{2018}]%
        {lee2018supporting}
\bibfield{author}{\bibinfo{person}{Craig~A Lee}, \bibinfo{person}{Zhiyi Zhang},
  \bibinfo{person}{Yukai Tu}, \bibinfo{person}{Alex Afanasyev}, {and}
  \bibinfo{person}{Lixia Zhang}.} \bibinfo{year}{2018}\natexlab{}.
\newblock \showarticletitle{Supporting virtual organizations using
  attribute-based encryption in named data networking}. In
  \bibinfo{booktitle}{\emph{2018 IEEE 4th International Conference on
  Collaboration and Internet Computing (CIC)}}. IEEE,
  \bibinfo{pages}{188--196}.
\newblock


\bibitem[\protect\citeauthoryear{Lehman, Hoque, Yu, Wang, Zhang, and
  Zhang}{Lehman et~al\mbox{.}}{2016}]%
        {lehman2016secure}
\bibfield{author}{\bibinfo{person}{Vince Lehman}, \bibinfo{person}{AKM~Mahmudul
  Hoque}, \bibinfo{person}{Yingdi Yu}, \bibinfo{person}{Lan Wang},
  \bibinfo{person}{Beichuan Zhang}, {and} \bibinfo{person}{Lixia Zhang}.}
  \bibinfo{year}{2016}\natexlab{}.
\newblock \showarticletitle{A secure link state routing protocol for NDN}.
\newblock \bibinfo{journal}{\emph{Tech. Rep. NDN-0037}} (\bibinfo{year}{2016}).
\newblock


\bibitem[\protect\citeauthoryear{Li, Kong, Mastorakis, and Zhang}{Li
  et~al\mbox{.}}{2019a}]%
        {li2019distributed}
\bibfield{author}{\bibinfo{person}{Tianxiang Li}, \bibinfo{person}{Zhaoning
  Kong}, \bibinfo{person}{Spyridon Mastorakis}, {and} \bibinfo{person}{Lixia
  Zhang}.} \bibinfo{year}{2019}\natexlab{a}.
\newblock \showarticletitle{Distributed Dataset Synchronization in Disruptive
  Networks}. In \bibinfo{booktitle}{\emph{2019 IEEE 16th International
  Conference on Mobile Ad Hoc and Sensor Systems (MASS)}}. IEEE,
  \bibinfo{pages}{428--437}.
\newblock


\bibitem[\protect\citeauthoryear{Li, Zhang, Wang, Lu, Zhang, and Zhang}{Li
  et~al\mbox{.}}{2019b}]%
        {li2019secure}
\bibfield{author}{\bibinfo{person}{Yanbiao Li}, \bibinfo{person}{Zhiyi Zhang},
  \bibinfo{person}{Xin Wang}, \bibinfo{person}{Edward Lu},
  \bibinfo{person}{Dafang Zhang}, {and} \bibinfo{person}{Lixia Zhang}.}
  \bibinfo{year}{2019}\natexlab{b}.
\newblock \showarticletitle{A secure sign-on protocol for smart homes over
  named data networking}.
\newblock \bibinfo{journal}{\emph{IEEE Communications Magazine}}
  \bibinfo{volume}{57}, \bibinfo{number}{7} (\bibinfo{year}{2019}),
  \bibinfo{pages}{62--68}.
\newblock


\bibitem[\protect\citeauthoryear{Li and Bi}{Li and Bi}{2014}]%
        {li2014interest}
\bibfield{author}{\bibinfo{person}{Zhaogeng Li} {and} \bibinfo{person}{Jun
  Bi}.} \bibinfo{year}{2014}\natexlab{}.
\newblock \showarticletitle{Interest cash: an application-based countermeasure
  against interest flooding for dynamic content in named data networking}. In
  \bibinfo{booktitle}{\emph{Proceedings of The Ninth International Conference
  on Future Internet Technologies}}. \bibinfo{pages}{1--6}.
\newblock


\bibitem[\protect\citeauthoryear{Li, Xu, Zhang, Yan, and Liu}{Li
  et~al\mbox{.}}{2018}]%
        {li2018packet}
\bibfield{author}{\bibinfo{person}{Zhuo Li}, \bibinfo{person}{Yaping Xu},
  \bibinfo{person}{Beichuan Zhang}, \bibinfo{person}{Liu Yan}, {and}
  \bibinfo{person}{Kaihua Liu}.} \bibinfo{year}{2018}\natexlab{}.
\newblock \showarticletitle{Packet forwarding in named data networking
  requirements and survey of solutions}.
\newblock \bibinfo{journal}{\emph{IEEE Communications Surveys \& Tutorials}}
  \bibinfo{volume}{21}, \bibinfo{number}{2} (\bibinfo{year}{2018}),
  \bibinfo{pages}{1950--1987}.
\newblock


\bibitem[\protect\citeauthoryear{Liu, Quan, Cheng, Feng, Zhang, and Shen}{Liu
  et~al\mbox{.}}{2018a}]%
        {liu2018blam}
\bibfield{author}{\bibinfo{person}{Gang Liu}, \bibinfo{person}{Wei Quan},
  \bibinfo{person}{Nan Cheng}, \bibinfo{person}{Bohao Feng},
  \bibinfo{person}{Hongke Zhang}, {and} \bibinfo{person}{Xuemin~Sherman Shen}.}
  \bibinfo{year}{2018}\natexlab{a}.
\newblock \showarticletitle{BLAM: Lightweight Bloom-filter based DDoS
  mitigation for information-centric IoT}. In \bibinfo{booktitle}{\emph{2018
  IEEE Global Communications Conference (GLOBECOM)}}. IEEE,
  \bibinfo{pages}{1--7}.
\newblock


\bibitem[\protect\citeauthoryear{Liu, Quan, Cheng, Wang, and Zhang}{Liu
  et~al\mbox{.}}{2018b}]%
        {liu2018accuracy}
\bibfield{author}{\bibinfo{person}{Gang Liu}, \bibinfo{person}{Wei Quan},
  \bibinfo{person}{Nan Cheng}, \bibinfo{person}{Kai Wang}, {and}
  \bibinfo{person}{Hongke Zhang}.} \bibinfo{year}{2018}\natexlab{b}.
\newblock \showarticletitle{Accuracy or delay? A game in detecting interest
  flooding attacks}.
\newblock \bibinfo{journal}{\emph{Internet Technology Letters}}
  \bibinfo{volume}{1}, \bibinfo{number}{2} (\bibinfo{year}{2018}),
  \bibinfo{pages}{e31}.
\newblock


\bibitem[\protect\citeauthoryear{Lutz}{Lutz}{2016}]%
        {lutz2016security}
\bibfield{author}{\bibinfo{person}{Roman Lutz}.}
  \bibinfo{year}{2016}\natexlab{}.
\newblock \showarticletitle{Security and privacy in future internet
  architectures-benefits and challenges of content centric networks}.
\newblock \bibinfo{journal}{\emph{arXiv preprint arXiv:1601.01278}}
  (\bibinfo{year}{2016}).
\newblock


\bibitem[\protect\citeauthoryear{Mannes and Maziero}{Mannes and
  Maziero}{2019}]%
        {mannes2019naming}
\bibfield{author}{\bibinfo{person}{Elisa Mannes} {and} \bibinfo{person}{Carlos
  Maziero}.} \bibinfo{year}{2019}\natexlab{}.
\newblock \showarticletitle{Naming content on the network layer: a security
  analysis of the information-centric network model}.
\newblock \bibinfo{journal}{\emph{ACM Computing Surveys (CSUR)}}
  \bibinfo{volume}{52}, \bibinfo{number}{3} (\bibinfo{year}{2019}),
  \bibinfo{pages}{1--28}.
\newblock


\bibitem[\protect\citeauthoryear{Mastorakis, Afanasyev, Yu, and
  Zhang}{Mastorakis et~al\mbox{.}}{2017b}]%
        {mastorakis2017ntorrent}
\bibfield{author}{\bibinfo{person}{Spyridon Mastorakis},
  \bibinfo{person}{Alexander Afanasyev}, \bibinfo{person}{Yingdi Yu}, {and}
  \bibinfo{person}{Lixia Zhang}.} \bibinfo{year}{2017}\natexlab{b}.
\newblock \showarticletitle{ntorrent: Peer-to-peer file sharing in named data
  networking}. In \bibinfo{booktitle}{\emph{2017 26th International Conference
  on Computer Communication and Networks (ICCCN)}}. IEEE,
  \bibinfo{pages}{1--10}.
\newblock


\bibitem[\protect\citeauthoryear{Mastorakis, Afanasyev, and Zhang}{Mastorakis
  et~al\mbox{.}}{2017a}]%
        {mastorakis2017ndnsim}
\bibfield{author}{\bibinfo{person}{Spyridon Mastorakis},
  \bibinfo{person}{Alexander Afanasyev}, {and} \bibinfo{person}{Lixia Zhang}.}
  \bibinfo{year}{2017}\natexlab{a}.
\newblock \showarticletitle{On the Evolution of {ndnSIM}: an Open-Source
  Simulator for {NDN} Experimentation}.
\newblock \bibinfo{journal}{\emph{ACM Computer Communication Review}}
  (\bibinfo{date}{July} \bibinfo{year}{2017}).
\newblock


\bibitem[\protect\citeauthoryear{Mastorakis, Li, and Zhang}{Mastorakis
  et~al\mbox{.}}{2020a}]%
        {mastorakis2020dapes}
\bibfield{author}{\bibinfo{person}{Spyridon Mastorakis},
  \bibinfo{person}{Tianxiang Li}, {and} \bibinfo{person}{Lixia Zhang}.}
  \bibinfo{year}{2020}\natexlab{a}.
\newblock \showarticletitle{DAPES: Named Data for Off-the-Grid File Sharing
  with Peer-to-Peer Interactions}.
\newblock \bibinfo{journal}{\emph{arXiv preprint arXiv:2006.01651}}
  (\bibinfo{year}{2020}).
\newblock


\bibitem[\protect\citeauthoryear{Mastorakis, Mtibaa, Lee, and Misra}{Mastorakis
  et~al\mbox{.}}{2020b}]%
        {mastorakis2020icedge}
\bibfield{author}{\bibinfo{person}{Spyridon Mastorakis},
  \bibinfo{person}{Abderrahmen Mtibaa}, \bibinfo{person}{Jonathan Lee}, {and}
  \bibinfo{person}{Satyajayant Misra}.} \bibinfo{year}{2020}\natexlab{b}.
\newblock \showarticletitle{ICedge: When Edge Computing Meets
  Information-Centric Networking}.
\newblock \bibinfo{journal}{\emph{IEEE Internet of Things Journal}}
  \bibinfo{volume}{7}, \bibinfo{number}{5} (\bibinfo{year}{2020}),
  \bibinfo{pages}{4203--4217}.
\newblock


\bibitem[\protect\citeauthoryear{Mtibaa and Mastorakis}{Mtibaa and
  Mastorakis}{2020}]%
        {mtibaa2020ndntp}
\bibfield{author}{\bibinfo{person}{Abderrahmen Mtibaa} {and}
  \bibinfo{person}{Spyridon Mastorakis}.} \bibinfo{year}{2020}\natexlab{}.
\newblock \showarticletitle{NDNTP: A Named Data Networking Time Protocol}.
\newblock \bibinfo{journal}{\emph{arXiv preprint arXiv:2007.07807}}
  (\bibinfo{year}{2020}).
\newblock


\bibitem[\protect\citeauthoryear{Nakatsuka, Wijekoon, and Nishi}{Nakatsuka
  et~al\mbox{.}}{2018}]%
        {nakatsuka2018frog}
\bibfield{author}{\bibinfo{person}{Yoshimichi Nakatsuka},
  \bibinfo{person}{Janaka~L Wijekoon}, {and} \bibinfo{person}{Hiroaki Nishi}.}
  \bibinfo{year}{2018}\natexlab{}.
\newblock \showarticletitle{FROG: A Packet Hop Count based DDoS Countermeasure
  in NDN}. In \bibinfo{booktitle}{\emph{2018 IEEE Symposium on Computers and
  Communications (ISCC)}}. IEEE, \bibinfo{pages}{00492--00497}.
\newblock


\bibitem[\protect\citeauthoryear{Newberry and Zhang}{Newberry and
  Zhang}{2019}]%
        {newberry2019power}
\bibfield{author}{\bibinfo{person}{Eric Newberry} {and}
  \bibinfo{person}{Beichuan Zhang}.} \bibinfo{year}{2019}\natexlab{}.
\newblock \showarticletitle{On the Power of In-Network Caching in the Hadoop
  Distributed File System}. In \bibinfo{booktitle}{\emph{Proceedings of the 6th
  ACM Conference on Information-Centric Networking}}. \bibinfo{pages}{89--99}.
\newblock


\bibitem[\protect\citeauthoryear{Nguyen, Mai, Cogranne, Doyen, Mallouli,
  Nguyen, El~Aoun, De~Oca, and Festor}{Nguyen et~al\mbox{.}}{2019}]%
        {nguyen2019reliable}
\bibfield{author}{\bibinfo{person}{Tan Nguyen}, \bibinfo{person}{Hoang-Long
  Mai}, \bibinfo{person}{R{\'e}mi Cogranne}, \bibinfo{person}{Guillaume Doyen},
  \bibinfo{person}{Wissam Mallouli}, \bibinfo{person}{Luong Nguyen},
  \bibinfo{person}{Moustapha El~Aoun}, \bibinfo{person}{Edgardo~Montes De~Oca},
  {and} \bibinfo{person}{Olivier Festor}.} \bibinfo{year}{2019}\natexlab{}.
\newblock \showarticletitle{Reliable detection of interest flooding attack in
  real deployment of named data networking}.
\newblock \bibinfo{journal}{\emph{IEEE Transactions on Information Forensics
  and Security}} \bibinfo{volume}{14}, \bibinfo{number}{9}
  (\bibinfo{year}{2019}), \bibinfo{pages}{2470--2485}.
\newblock


\bibitem[\protect\citeauthoryear{Nguyen, Marchal, Doyen, Cholez, and
  Cogranne}{Nguyen et~al\mbox{.}}{2017}]%
        {nguyen2017content}
\bibfield{author}{\bibinfo{person}{Tan Nguyen}, \bibinfo{person}{Xavier
  Marchal}, \bibinfo{person}{Guillaume Doyen}, \bibinfo{person}{Thibault
  Cholez}, {and} \bibinfo{person}{R{\'e}mi Cogranne}.}
  \bibinfo{year}{2017}\natexlab{}.
\newblock \showarticletitle{Content poisoning in named data networking:
  Comprehensive characterization of real deployment}. In
  \bibinfo{booktitle}{\emph{2017 IFIP/IEEE Symposium on Integrated Network and
  Service Management (IM)}}. IEEE, \bibinfo{pages}{72--80}.
\newblock


\bibitem[\protect\citeauthoryear{Nguyen, Cogranne, Doyen, and Retraint}{Nguyen
  et~al\mbox{.}}{2015}]%
        {nguyen2015detection}
\bibfield{author}{\bibinfo{person}{Tan~N Nguyen}, \bibinfo{person}{R{\'e}mi
  Cogranne}, \bibinfo{person}{Guillaume Doyen}, {and} \bibinfo{person}{Florent
  Retraint}.} \bibinfo{year}{2015}\natexlab{}.
\newblock \showarticletitle{Detection of interest flooding attacks in named
  data networking using hypothesis testing}. In \bibinfo{booktitle}{\emph{2015
  IEEE International Workshop on Information Forensics and Security (WIFS)}}.
  IEEE, \bibinfo{pages}{1--6}.
\newblock


\bibitem[\protect\citeauthoryear{Nour, Khelifi, Hussain, Mastorakis, and
  Moungla}{Nour et~al\mbox{.}}{2021}]%
        {nour2021access}
\bibfield{author}{\bibinfo{person}{Boubakr Nour}, \bibinfo{person}{Hakima
  Khelifi}, \bibinfo{person}{Rasheed Hussain}, \bibinfo{person}{Spyridon
  Mastorakis}, {and} \bibinfo{person}{Hassine Moungla}.}
  \bibinfo{year}{2021}\natexlab{}.
\newblock \showarticletitle{Access control mechanisms in named data networks: A
  comprehensive survey}.
\newblock \bibinfo{journal}{\emph{ACM Computing Surveys (CSUR)}}
  \bibinfo{volume}{54}, \bibinfo{number}{3} (\bibinfo{year}{2021}),
  \bibinfo{pages}{1--35}.
\newblock


\bibitem[\protect\citeauthoryear{Pang, Li, Zhang, Shi, and Huang}{Pang
  et~al\mbox{.}}{2017}]%
        {pang2017research}
\bibfield{author}{\bibinfo{person}{Bin Pang}, \bibinfo{person}{Ru Li},
  \bibinfo{person}{Xin Zhang}, \bibinfo{person}{Jinshan Shi}, {and}
  \bibinfo{person}{Manxin Huang}.} \bibinfo{year}{2017}\natexlab{}.
\newblock \showarticletitle{Research on interest flooding attack analysis in
  conspiracy with content providers}. In \bibinfo{booktitle}{\emph{2017 7th
  IEEE International Conference on Electronics Information and Emergency
  Communication (ICEIEC)}}. IEEE, \bibinfo{pages}{543--547}.
\newblock


\bibitem[\protect\citeauthoryear{Psaras, Ascigil, Rene, Pavlou, Afanasyev, and
  Zhang}{Psaras et~al\mbox{.}}{2018}]%
        {psaras2018mobile}
\bibfield{author}{\bibinfo{person}{Ioannis Psaras}, \bibinfo{person}{Onur
  Ascigil}, \bibinfo{person}{Sergi Rene}, \bibinfo{person}{George Pavlou},
  \bibinfo{person}{Alex Afanasyev}, {and} \bibinfo{person}{Lixia Zhang}.}
  \bibinfo{year}{2018}\natexlab{}.
\newblock \showarticletitle{Mobile data repositories at the edge}. In
  \bibinfo{booktitle}{\emph{$\{$USENIX$\}$ Workshop on Hot Topics in Edge
  Computing (HotEdge 18)}}.
\newblock


\bibitem[\protect\citeauthoryear{Pu, Payne, and Brown}{Pu
  et~al\mbox{.}}{2019}]%
        {pu2019self}
\bibfield{author}{\bibinfo{person}{Cong Pu}, \bibinfo{person}{Nathaniel Payne},
  {and} \bibinfo{person}{Jacqueline Brown}.} \bibinfo{year}{2019}\natexlab{}.
\newblock \showarticletitle{Self-Adjusting Share-Based Countermeasure to
  Interest Flooding Attack in Named Data Networking}. In
  \bibinfo{booktitle}{\emph{2019 International Conference on Internet of Things
  (iThings) and IEEE Green Computing and Communications (GreenCom) and IEEE
  Cyber, Physical and Social Computing (CPSCom) and IEEE Smart Data
  (SmartData)}}. IEEE, \bibinfo{pages}{142--147}.
\newblock


\bibitem[\protect\citeauthoryear{Rai and Dhakal}{Rai and Dhakal}{2018}]%
        {rai2018survey}
\bibfield{author}{\bibinfo{person}{Sandesh Rai} {and} \bibinfo{person}{Dependra
  Dhakal}.} \bibinfo{year}{2018}\natexlab{}.
\newblock \showarticletitle{A survey on detection and mitigation of interest
  flooding attack in named data networking}.
\newblock In \bibinfo{booktitle}{\emph{Advanced Computational and Communication
  Paradigms}}. \bibinfo{publisher}{Springer}, \bibinfo{pages}{523--531}.
\newblock


\bibitem[\protect\citeauthoryear{Rai, Sharma, and Dhakal}{Rai
  et~al\mbox{.}}{2019}]%
        {rai2019survey}
\bibfield{author}{\bibinfo{person}{Sandesh Rai}, \bibinfo{person}{Kalpana
  Sharma}, {and} \bibinfo{person}{Dependra Dhakal}.}
  \bibinfo{year}{2019}\natexlab{}.
\newblock \showarticletitle{A survey on detection and mitigation of distributed
  denial-of-service attack in named data networking}.
\newblock In \bibinfo{booktitle}{\emph{Advances in communication, cloud, and
  big data}}. \bibinfo{publisher}{Springer}, \bibinfo{pages}{163--171}.
\newblock


\bibitem[\protect\citeauthoryear{Ramani, Tourani, Torres, Misra, and
  Afanasyev}{Ramani et~al\mbox{.}}{2019}]%
        {ramani2019ndn}
\bibfield{author}{\bibinfo{person}{Sanjeev~Kaushik Ramani},
  \bibinfo{person}{Reza Tourani}, \bibinfo{person}{George Torres},
  \bibinfo{person}{Satyajayant Misra}, {and} \bibinfo{person}{Alexander
  Afanasyev}.} \bibinfo{year}{2019}\natexlab{}.
\newblock \showarticletitle{NDN-ABS: Attribute-Based Signature Scheme for Named
  Data Networking}. In \bibinfo{booktitle}{\emph{Proceedings of the 6th ACM
  Conference on Information-Centric Networking}}. \bibinfo{pages}{123--133}.
\newblock


\bibitem[\protect\citeauthoryear{Salah and Strufe}{Salah and Strufe}{2016}]%
        {salah2016evaluating}
\bibfield{author}{\bibinfo{person}{Hani Salah} {and} \bibinfo{person}{Thorsten
  Strufe}.} \bibinfo{year}{2016}\natexlab{}.
\newblock \showarticletitle{Evaluating and mitigating a collusive version of
  the interest flooding attack in NDN}. In \bibinfo{booktitle}{\emph{2016 IEEE
  Symposium on Computers and Communication (ISCC)}}. IEEE,
  \bibinfo{pages}{938--945}.
\newblock


\bibitem[\protect\citeauthoryear{Salah, Wulfheide, and Strufe}{Salah
  et~al\mbox{.}}{2015}]%
        {salah2015coordination}
\bibfield{author}{\bibinfo{person}{Hani Salah}, \bibinfo{person}{Julian
  Wulfheide}, {and} \bibinfo{person}{Thorsten Strufe}.}
  \bibinfo{year}{2015}\natexlab{}.
\newblock \showarticletitle{Coordination supports security: A new defence
  mechanism against interest flooding in NDN}. In
  \bibinfo{booktitle}{\emph{2015 IEEE 40th Conference on Local Computer
  Networks (LCN)}}. IEEE, \bibinfo{pages}{73--81}.
\newblock


\bibitem[\protect\citeauthoryear{Seskar, Nagaraja, Nelson, and
  Raychaudhuri}{Seskar et~al\mbox{.}}{2011}]%
        {seskar2011mobilityfirst}
\bibfield{author}{\bibinfo{person}{Ivan Seskar}, \bibinfo{person}{Kiran
  Nagaraja}, \bibinfo{person}{Sam Nelson}, {and} \bibinfo{person}{Dipankar
  Raychaudhuri}.} \bibinfo{year}{2011}\natexlab{}.
\newblock \showarticletitle{Mobilityfirst future internet architecture
  project}. In \bibinfo{booktitle}{\emph{Proceedings of the 7th Asian Internet
  Engineering Conference}}. \bibinfo{pages}{1--3}.
\newblock


\bibitem[\protect\citeauthoryear{Shang, Wang, Afanasyev, Burke, and
  Zhang}{Shang et~al\mbox{.}}{2017}]%
        {shang2017breaking}
\bibfield{author}{\bibinfo{person}{Wentao Shang}, \bibinfo{person}{Zhehao
  Wang}, \bibinfo{person}{Alexander Afanasyev}, \bibinfo{person}{Jeff Burke},
  {and} \bibinfo{person}{Lixia Zhang}.} \bibinfo{year}{2017}\natexlab{}.
\newblock \showarticletitle{Breaking out of the cloud: Local trust management
  and rendezvous in Named Data Networking of Things}. In
  \bibinfo{booktitle}{\emph{Proceedings of the Second International Conference
  on Internet-of-Things Design and Implementation}}. \bibinfo{pages}{3--13}.
\newblock


\bibitem[\protect\citeauthoryear{Shang, Yu, Liang, Zhang, and Zhang}{Shang
  et~al\mbox{.}}{2015}]%
        {shang2015ndn}
\bibfield{author}{\bibinfo{person}{Wentao Shang}, \bibinfo{person}{Yingdi Yu},
  \bibinfo{person}{Teng Liang}, \bibinfo{person}{Beichuan Zhang}, {and}
  \bibinfo{person}{Lixia Zhang}.} \bibinfo{year}{2015}\natexlab{}.
\newblock \showarticletitle{Ndn-ace: Access control for constrained
  environments over named data networking}.
\newblock \bibinfo{journal}{\emph{NDN Project, Tech. Rep. NDN-0036, Revision
  1}} (\bibinfo{year}{2015}).
\newblock


\bibitem[\protect\citeauthoryear{Shi}{Shi}{2017}]%
        {shi2017named}
\bibfield{author}{\bibinfo{person}{Junxiao Shi}.}
  \bibinfo{year}{2017}\natexlab{}.
\newblock \emph{\bibinfo{title}{Named data networking in local area networks}}.
\newblock \bibinfo{thesistype}{Ph.D. Dissertation}. \bibinfo{school}{The
  University of Arizona.}
\newblock


\bibitem[\protect\citeauthoryear{Shi, Newberry, and Zhang}{Shi
  et~al\mbox{.}}{2017}]%
        {shi2017broadcast}
\bibfield{author}{\bibinfo{person}{Junxiao Shi}, \bibinfo{person}{Eric
  Newberry}, {and} \bibinfo{person}{Beichuan Zhang}.}
  \bibinfo{year}{2017}\natexlab{}.
\newblock \showarticletitle{On broadcast-based self-learning in named data
  networking}. In \bibinfo{booktitle}{\emph{2017 IFIP Networking Conference
  (IFIP Networking) and Workshops}}. IEEE, \bibinfo{pages}{1--9}.
\newblock


\bibitem[\protect\citeauthoryear{Shigeyasu and Sonoda}{Shigeyasu and
  Sonoda}{2018}]%
        {shigeyasu2018distributed}
\bibfield{author}{\bibinfo{person}{Tetsuya Shigeyasu} {and}
  \bibinfo{person}{Ayaka Sonoda}.} \bibinfo{year}{2018}\natexlab{}.
\newblock \showarticletitle{Distributed Approach for Detecting Collusive
  Interest Flooding Attack on Named Data Networking}. In
  \bibinfo{booktitle}{\emph{International Conference on Network-Based
  Information Systems}}. Springer, \bibinfo{pages}{76--86}.
\newblock


\bibitem[\protect\citeauthoryear{Shinohara, Kamimoto, Sato, and
  Shigeno}{Shinohara et~al\mbox{.}}{2016}]%
        {shinohara2016cache}
\bibfield{author}{\bibinfo{person}{Ryoki Shinohara}, \bibinfo{person}{Takashi
  Kamimoto}, \bibinfo{person}{Kazuya Sato}, {and} \bibinfo{person}{Hiroshi
  Shigeno}.} \bibinfo{year}{2016}\natexlab{}.
\newblock \showarticletitle{Cache control method mitigating packet
  concentration of router caused by interest flooding attack}. In
  \bibinfo{booktitle}{\emph{2016 IEEE Trustcom/BigDataSE/ISPA}}. IEEE,
  \bibinfo{pages}{324--331}.
\newblock


\bibitem[\protect\citeauthoryear{Signorello, Marchal, Fran{\c{c}}ois, Festor,
  and State}{Signorello et~al\mbox{.}}{2017}]%
        {signorello2017advanced}
\bibfield{author}{\bibinfo{person}{Salvatore Signorello},
  \bibinfo{person}{Samuel Marchal}, \bibinfo{person}{J{\'e}r{\^o}me
  Fran{\c{c}}ois}, \bibinfo{person}{Olivier Festor}, {and}
  \bibinfo{person}{Radu State}.} \bibinfo{year}{2017}\natexlab{}.
\newblock \showarticletitle{Advanced interest flooding attacks in named-data
  networking}. In \bibinfo{booktitle}{\emph{2017 IEEE 16th International
  Symposium on Network Computing and Applications (NCA)}}. IEEE,
  \bibinfo{pages}{1--10}.
\newblock


\bibitem[\protect\citeauthoryear{So, Narayanan, and Oran}{So
  et~al\mbox{.}}{2013}]%
        {so2013named}
\bibfield{author}{\bibinfo{person}{Won So}, \bibinfo{person}{Ashok Narayanan},
  {and} \bibinfo{person}{David Oran}.} \bibinfo{year}{2013}\natexlab{}.
\newblock \showarticletitle{Named data networking on a router: Fast and
  DoS-resistant forwarding with hash tables}. In
  \bibinfo{booktitle}{\emph{Architectures for Networking and Communications
  Systems}}. IEEE, \bibinfo{pages}{215--225}.
\newblock


\bibitem[\protect\citeauthoryear{Song, Yuan, Crowley, and Zhang}{Song
  et~al\mbox{.}}{2015}]%
        {song2015scalable}
\bibfield{author}{\bibinfo{person}{Tian Song}, \bibinfo{person}{Haowei Yuan},
  \bibinfo{person}{Patrick Crowley}, {and} \bibinfo{person}{Beichuan Zhang}.}
  \bibinfo{year}{2015}\natexlab{}.
\newblock \showarticletitle{Scalable name-based packet forwarding: From
  millions to billions}. In \bibinfo{booktitle}{\emph{Proceedings of the 2nd
  ACM conference on information-centric networking}}. \bibinfo{pages}{19--28}.
\newblock


\bibitem[\protect\citeauthoryear{Spring, Mahajan, and Wetherall}{Spring
  et~al\mbox{.}}{2002}]%
        {spring2002measuring}
\bibfield{author}{\bibinfo{person}{Neil Spring}, \bibinfo{person}{Ratul
  Mahajan}, {and} \bibinfo{person}{David Wetherall}.}
  \bibinfo{year}{2002}\natexlab{}.
\newblock \showarticletitle{Measuring ISP topologies with Rocketfuel}.
\newblock \bibinfo{journal}{\emph{ACM SIGCOMM Computer Communication Review}}
  \bibinfo{volume}{32}, \bibinfo{number}{4} (\bibinfo{year}{2002}),
  \bibinfo{pages}{133--145}.
\newblock


\bibitem[\protect\citeauthoryear{Tang, Zhang, Liu, and Zhang}{Tang
  et~al\mbox{.}}{2013}]%
        {tang2013identifying}
\bibfield{author}{\bibinfo{person}{Jianqiang Tang}, \bibinfo{person}{Zhongyue
  Zhang}, \bibinfo{person}{Ying Liu}, {and} \bibinfo{person}{Hongke Zhang}.}
  \bibinfo{year}{2013}\natexlab{}.
\newblock \showarticletitle{Identifying interest flooding in named data
  networking}. In \bibinfo{booktitle}{\emph{2013 IEEE International Conference
  on Green Computing and Communications and IEEE Internet of Things and IEEE
  Cyber, Physical and Social Computing}}. IEEE, \bibinfo{pages}{306--310}.
\newblock


\bibitem[\protect\citeauthoryear{Tourani, Misra, Mick, and Panwar}{Tourani
  et~al\mbox{.}}{2017}]%
        {tourani2017security}
\bibfield{author}{\bibinfo{person}{Reza Tourani}, \bibinfo{person}{Satyajayant
  Misra}, \bibinfo{person}{Travis Mick}, {and} \bibinfo{person}{Gaurav
  Panwar}.} \bibinfo{year}{2017}\natexlab{}.
\newblock \showarticletitle{Security, privacy, and access control in
  information-centric networking: A survey}.
\newblock \bibinfo{journal}{\emph{IEEE communications surveys \& tutorials}}
  \bibinfo{volume}{20}, \bibinfo{number}{1} (\bibinfo{year}{2017}),
  \bibinfo{pages}{566--600}.
\newblock


\bibitem[\protect\citeauthoryear{Tourani, Torres, and Misra}{Tourani
  et~al\mbox{.}}{2020}]%
        {tourani2020persia}
\bibfield{author}{\bibinfo{person}{Reza Tourani}, \bibinfo{person}{George
  Torres}, {and} \bibinfo{person}{Satyajayant Misra}.}
  \bibinfo{year}{2020}\natexlab{}.
\newblock \showarticletitle{PERSIA: a PuzzlE-based InteReSt FloodIng Attack
  Countermeasure}. In \bibinfo{booktitle}{\emph{Proceedings of the 7th ACM
  Conference on Information-Centric Networking}}. \bibinfo{pages}{117--128}.
\newblock


\bibitem[\protect\citeauthoryear{Vassilakis, Alohali, Moscholios, and
  Logothetis}{Vassilakis et~al\mbox{.}}{2015}]%
        {vassilakis2015mitigating}
\bibfield{author}{\bibinfo{person}{Vassilios~G Vassilakis},
  \bibinfo{person}{Bashar~A Alohali}, \bibinfo{person}{I Moscholios}, {and}
  \bibinfo{person}{Michael~D Logothetis}.} \bibinfo{year}{2015}\natexlab{}.
\newblock \showarticletitle{Mitigating distributed denial-of-service attacks in
  named data networking}. In \bibinfo{booktitle}{\emph{Proceedings of the 11th
  Advanced International Conference on Telecommunications (AICT), Brussels,
  Belgium}}. \bibinfo{pages}{18--23}.
\newblock


\bibitem[\protect\citeauthoryear{Virgilio, Marchetto, and Sisto}{Virgilio
  et~al\mbox{.}}{2013}]%
        {virgilio2013pit}
\bibfield{author}{\bibinfo{person}{Matteo Virgilio}, \bibinfo{person}{Guido
  Marchetto}, {and} \bibinfo{person}{Riccardo Sisto}.}
  \bibinfo{year}{2013}\natexlab{}.
\newblock \showarticletitle{PIT overload analysis in content centric networks}.
  In \bibinfo{booktitle}{\emph{Proceedings of the 3rd ACM SIGCOMM workshop on
  Information-centric networking}}. \bibinfo{pages}{67--72}.
\newblock


\bibitem[\protect\citeauthoryear{Voitalov, Aldecoa, Wang, and
  Krioukov}{Voitalov et~al\mbox{.}}{2017}]%
        {voitalov2017geohyperbolic}
\bibfield{author}{\bibinfo{person}{Ivan Voitalov}, \bibinfo{person}{Rodrigo
  Aldecoa}, \bibinfo{person}{Lan Wang}, {and} \bibinfo{person}{Dmitri
  Krioukov}.} \bibinfo{year}{2017}\natexlab{}.
\newblock \showarticletitle{Geohyperbolic routing and addressing schemes}.
\newblock \bibinfo{journal}{\emph{ACM SIGCOMM Computer Communication Review}}
  \bibinfo{volume}{47}, \bibinfo{number}{3} (\bibinfo{year}{2017}),
  \bibinfo{pages}{11--18}.
\newblock


\bibitem[\protect\citeauthoryear{Vusirikala, Mastorakis, Afanasyev, and
  Zhang}{Vusirikala et~al\mbox{.}}{2016}]%
        {vusirikala2016hop}
\bibfield{author}{\bibinfo{person}{Satyanarayana Vusirikala},
  \bibinfo{person}{Spyridon Mastorakis}, \bibinfo{person}{Alexander Afanasyev},
  {and} \bibinfo{person}{Lixia Zhang}.} \bibinfo{year}{2016}\natexlab{}.
\newblock \showarticletitle{Hop-by-hop best effort link layer reliability in
  named data networking}.
\newblock \bibinfo{journal}{\emph{NDN, Technical Report NDN-0041}}
  (\bibinfo{year}{2016}).
\newblock


\bibitem[\protect\citeauthoryear{Wang, Guo, and Quan}{Wang
  et~al\mbox{.}}{2019}]%
        {wang2019analyzing}
\bibfield{author}{\bibinfo{person}{Kai Wang}, \bibinfo{person}{Dongchao Guo},
  {and} \bibinfo{person}{Wei Quan}.} \bibinfo{year}{2019}\natexlab{}.
\newblock \showarticletitle{Analyzing NDN NACK on Interest Flooding Attack via
  SIS Epidemic Model}.
\newblock \bibinfo{journal}{\emph{IEEE Systems Journal}}
  (\bibinfo{year}{2019}).
\newblock


\bibitem[\protect\citeauthoryear{Wang, Zhao, Tong, et~al\mbox{.}}{Wang
  et~al\mbox{.}}{2017b}]%
        {wang2017urgency}
\bibfield{author}{\bibinfo{person}{Kai Wang}, \bibinfo{person}{Yude Zhao},
  \bibinfo{person}{Xiangrong Tong}, {et~al\mbox{.}}}
  \bibinfo{year}{2017}\natexlab{b}.
\newblock \showarticletitle{On the urgency of implementing Interest NACK into
  CCN: from the perspective of countering advanced interest flooding attacks}.
\newblock \bibinfo{journal}{\emph{IET Networks}} \bibinfo{volume}{7},
  \bibinfo{number}{3} (\bibinfo{year}{2017}), \bibinfo{pages}{136--140}.
\newblock


\bibitem[\protect\citeauthoryear{Wang, Zhou, Luo, Guan, Qin, and Zhang}{Wang
  et~al\mbox{.}}{2014a}]%
        {wang2014detecting}
\bibfield{author}{\bibinfo{person}{Kai Wang}, \bibinfo{person}{Huachun Zhou},
  \bibinfo{person}{Hongbin Luo}, \bibinfo{person}{Jianfeng Guan},
  \bibinfo{person}{Yajuan Qin}, {and} \bibinfo{person}{Hongke Zhang}.}
  \bibinfo{year}{2014}\natexlab{a}.
\newblock \showarticletitle{Detecting and mitigating interest flooding attacks
  in content-centric network}.
\newblock \bibinfo{journal}{\emph{Security and Communication Networks}}
  \bibinfo{volume}{7}, \bibinfo{number}{4} (\bibinfo{year}{2014}),
  \bibinfo{pages}{685--699}.
\newblock


\bibitem[\protect\citeauthoryear{Wang, Zhou, Qin, Chen, and Zhang}{Wang
  et~al\mbox{.}}{2013}]%
        {wang2013decoupling}
\bibfield{author}{\bibinfo{person}{Kai Wang}, \bibinfo{person}{Huachun Zhou},
  \bibinfo{person}{Yajuan Qin}, \bibinfo{person}{Jia Chen}, {and}
  \bibinfo{person}{Hongke Zhang}.} \bibinfo{year}{2013}\natexlab{}.
\newblock \showarticletitle{Decoupling malicious interests from pending
  interest table to mitigate interest flooding attacks}. In
  \bibinfo{booktitle}{\emph{2013 IEEE Globecom Workshops (GC Wkshps)}}. IEEE,
  \bibinfo{pages}{963--968}.
\newblock


\bibitem[\protect\citeauthoryear{Wang, Zhou, Qin, and Zhang}{Wang
  et~al\mbox{.}}{2014b}]%
        {wang2014cooperative}
\bibfield{author}{\bibinfo{person}{Kai Wang}, \bibinfo{person}{Huachun Zhou},
  \bibinfo{person}{Yajuan Qin}, {and} \bibinfo{person}{Hongke Zhang}.}
  \bibinfo{year}{2014}\natexlab{b}.
\newblock \showarticletitle{Cooperative-Filter: countering Interest flooding
  attacks in named data networking}.
\newblock \bibinfo{journal}{\emph{Soft Computing}} \bibinfo{volume}{18},
  \bibinfo{number}{9} (\bibinfo{year}{2014}), \bibinfo{pages}{1803--1813}.
\newblock


\bibitem[\protect\citeauthoryear{Wang, Pan, Dong, Yu, and Wang}{Wang
  et~al\mbox{.}}{2017a}]%
        {wang2017economic}
\bibfield{author}{\bibinfo{person}{Licheng Wang}, \bibinfo{person}{Yun Pan},
  \bibinfo{person}{Mianxiong Dong}, \bibinfo{person}{Yafang Yu}, {and}
  \bibinfo{person}{Kun Wang}.} \bibinfo{year}{2017}\natexlab{a}.
\newblock \showarticletitle{Economic levers for mitigating interest flooding
  attack in Named Data Networking}.
\newblock \bibinfo{journal}{\emph{Mathematical Problems in Engineering}}
  \bibinfo{volume}{2017} (\bibinfo{year}{2017}).
\newblock


\bibitem[\protect\citeauthoryear{Wu, Feng, Yue, Xu, and Liu}{Wu
  et~al\mbox{.}}{2020}]%
        {wu2020mitigation}
\bibfield{author}{\bibinfo{person}{Zhijun Wu}, \bibinfo{person}{Wenzhi Feng},
  \bibinfo{person}{Meng Yue}, \bibinfo{person}{Xinran Xu}, {and}
  \bibinfo{person}{Liang Liu}.} \bibinfo{year}{2020}\natexlab{}.
\newblock \showarticletitle{Mitigation measures of collusive interest flooding
  attacks in named data networking}.
\newblock \bibinfo{journal}{\emph{Computers \& Security}}  \bibinfo{volume}{97}
  (\bibinfo{year}{2020}), \bibinfo{pages}{101971}.
\newblock


\bibitem[\protect\citeauthoryear{Xin, Li, Wang, Li, and Chen}{Xin
  et~al\mbox{.}}{2016}]%
        {xin2016novel}
\bibfield{author}{\bibinfo{person}{Yonghui Xin}, \bibinfo{person}{Yang Li},
  \bibinfo{person}{Wei Wang}, \bibinfo{person}{Weiyuan Li}, {and}
  \bibinfo{person}{Xin Chen}.} \bibinfo{year}{2016}\natexlab{}.
\newblock \showarticletitle{A novel interest flooding attacks detection and
  countermeasure scheme in NDN}. In \bibinfo{booktitle}{\emph{2016 IEEE Global
  Communications Conference (GLOBECOM)}}. IEEE, \bibinfo{pages}{1--7}.
\newblock


\bibitem[\protect\citeauthoryear{Xin, Li, Wang, Li, and Chen}{Xin
  et~al\mbox{.}}{2017}]%
        {xin2017detection}
\bibfield{author}{\bibinfo{person}{Yonghui Xin}, \bibinfo{person}{Yang Li},
  \bibinfo{person}{Wei Wang}, \bibinfo{person}{Weiyuan Li}, {and}
  \bibinfo{person}{Xin Chen}.} \bibinfo{year}{2017}\natexlab{}.
\newblock \showarticletitle{Detection of collusive interest flooding attacks in
  named data networking using wavelet analysis}. In
  \bibinfo{booktitle}{\emph{MILCOM 2017-2017 IEEE Military Communications
  Conference (MILCOM)}}. IEEE, \bibinfo{pages}{557--562}.
\newblock


\bibitem[\protect\citeauthoryear{Xylomenos, Ververidis, Siris, Fotiou,
  Tsilopoulos, Vasilakos, Katsaros, and Polyzos}{Xylomenos
  et~al\mbox{.}}{2013}]%
        {xylomenos2013survey}
\bibfield{author}{\bibinfo{person}{George Xylomenos},
  \bibinfo{person}{Christopher~N Ververidis}, \bibinfo{person}{Vasilios~A
  Siris}, \bibinfo{person}{Nikos Fotiou}, \bibinfo{person}{Christos
  Tsilopoulos}, \bibinfo{person}{Xenofon Vasilakos},
  \bibinfo{person}{Konstantinos~V Katsaros}, {and} \bibinfo{person}{George~C
  Polyzos}.} \bibinfo{year}{2013}\natexlab{}.
\newblock \showarticletitle{A survey of information-centric networking
  research}.
\newblock \bibinfo{journal}{\emph{IEEE communications surveys \& tutorials}}
  \bibinfo{volume}{16}, \bibinfo{number}{2} (\bibinfo{year}{2013}),
  \bibinfo{pages}{1024--1049}.
\newblock


\bibitem[\protect\citeauthoryear{Yeh, Ho, Cui, Burd, Liu, and Leong}{Yeh
  et~al\mbox{.}}{2014}]%
        {yeh2014vip}
\bibfield{author}{\bibinfo{person}{Edmund Yeh}, \bibinfo{person}{Tracey Ho},
  \bibinfo{person}{Ying Cui}, \bibinfo{person}{Michael Burd},
  \bibinfo{person}{Ran Liu}, {and} \bibinfo{person}{Derek Leong}.}
  \bibinfo{year}{2014}\natexlab{}.
\newblock \showarticletitle{Vip: A framework for joint dynamic forwarding and
  caching in named data networks}. In \bibinfo{booktitle}{\emph{Proceedings of
  the 1st ACM Conference on Information-Centric Networking}}.
  \bibinfo{pages}{117--126}.
\newblock


\bibitem[\protect\citeauthoryear{Yi, Afanasyev, Moiseenko, Wang, Zhang, and
  Zhang}{Yi et~al\mbox{.}}{2013}]%
        {yi2013case}
\bibfield{author}{\bibinfo{person}{Cheng Yi}, \bibinfo{person}{Alexander
  Afanasyev}, \bibinfo{person}{Ilya Moiseenko}, \bibinfo{person}{Lan Wang},
  \bibinfo{person}{Beichuan Zhang}, {and} \bibinfo{person}{Lixia Zhang}.}
  \bibinfo{year}{2013}\natexlab{}.
\newblock \showarticletitle{A case for stateful forwarding plane}.
\newblock \bibinfo{journal}{\emph{Computer Communications}}
  \bibinfo{volume}{36}, \bibinfo{number}{7} (\bibinfo{year}{2013}),
  \bibinfo{pages}{779--791}.
\newblock


\bibitem[\protect\citeauthoryear{Yin, Tang, Zou, Wu, and Li}{Yin
  et~al\mbox{.}}{2019}]%
        {yin2019controller}
\bibfield{author}{\bibinfo{person}{Gubei Yin}, \bibinfo{person}{Junhua Tang},
  \bibinfo{person}{Futai Zou}, \bibinfo{person}{Yue Wu}, {and}
  \bibinfo{person}{Jianhua Li}.} \bibinfo{year}{2019}\natexlab{}.
\newblock \showarticletitle{Controller Based Detection Scheme of Interest
  Flooding Attack in Named Data Networking}. In \bibinfo{booktitle}{\emph{2019
  IEEE 5th International Conference on Computer and Communications (ICCC)}}.
  IEEE, \bibinfo{pages}{1628--1633}.
\newblock


\bibitem[\protect\citeauthoryear{Yu, Afanasyev, Clark, Claffy, Jacobson, and
  Zhang}{Yu et~al\mbox{.}}{2015}]%
        {yu2015schematizing}
\bibfield{author}{\bibinfo{person}{Yingdi Yu}, \bibinfo{person}{Alexander
  Afanasyev}, \bibinfo{person}{David Clark}, \bibinfo{person}{KC Claffy},
  \bibinfo{person}{Van Jacobson}, {and} \bibinfo{person}{Lixia Zhang}.}
  \bibinfo{year}{2015}\natexlab{}.
\newblock \showarticletitle{Schematizing trust in named data networking}. In
  \bibinfo{booktitle}{\emph{Proceedings of the 2nd ACM Conference on
  Information-Centric Networking}}. \bibinfo{pages}{177--186}.
\newblock


\bibitem[\protect\citeauthoryear{Zhang, Li, Zhang, Afanasyev, and Zhang}{Zhang
  et~al\mbox{.}}{2018a}]%
        {zhang2018ndn}
\bibfield{author}{\bibinfo{person}{Haitao Zhang}, \bibinfo{person}{Yanbiao Li},
  \bibinfo{person}{Zhiyi Zhang}, \bibinfo{person}{Alexander Afanasyev}, {and}
  \bibinfo{person}{Lixia Zhang}.} \bibinfo{year}{2018}\natexlab{a}.
\newblock \showarticletitle{NDN host model}.
\newblock \bibinfo{journal}{\emph{ACM SIGCOMM Computer Communication Review}}
  \bibinfo{volume}{48}, \bibinfo{number}{3} (\bibinfo{year}{2018}),
  \bibinfo{pages}{35--41}.
\newblock


\bibitem[\protect\citeauthoryear{Zhang, Wang, Scherb, Marxer, Burke, Zhang, and
  Tschudin}{Zhang et~al\mbox{.}}{2016b}]%
        {zhang2016sharing}
\bibfield{author}{\bibinfo{person}{Haitao Zhang}, \bibinfo{person}{Zhehao
  Wang}, \bibinfo{person}{Christopher Scherb}, \bibinfo{person}{Claudio
  Marxer}, \bibinfo{person}{Jeff Burke}, \bibinfo{person}{Lixia Zhang}, {and}
  \bibinfo{person}{Christian Tschudin}.} \bibinfo{year}{2016}\natexlab{b}.
\newblock \showarticletitle{Sharing mhealth data via named data networking}. In
  \bibinfo{booktitle}{\emph{Proceedings of the 3rd ACM Conference on
  Information-Centric Networking}}. \bibinfo{pages}{142--147}.
\newblock


\bibitem[\protect\citeauthoryear{Zhang, Afanasyev, Burke, Jacobson, Crowley,
  Papadopoulos, Wang, Zhang, et~al\mbox{.}}{Zhang et~al\mbox{.}}{2014}]%
        {zhang2014named}
\bibfield{author}{\bibinfo{person}{Lixia Zhang}, \bibinfo{person}{Alexander
  Afanasyev}, \bibinfo{person}{Jeffrey Burke}, \bibinfo{person}{Van Jacobson},
  \bibinfo{person}{Patrick Crowley}, \bibinfo{person}{Christos Papadopoulos},
  \bibinfo{person}{Lan Wang}, \bibinfo{person}{Beichuan Zhang},
  {et~al\mbox{.}}} \bibinfo{year}{2014}\natexlab{}.
\newblock \showarticletitle{Named data networking}.
\newblock \bibinfo{journal}{\emph{ACM SIGCOMM Computer Communication Review}}
  \bibinfo{volume}{44}, \bibinfo{number}{3} (\bibinfo{year}{2014}),
  \bibinfo{pages}{66--73}.
\newblock


\bibitem[\protect\citeauthoryear{Zhang, Estrin, Burke, Jacobson, Thornton,
  Smetters, Zhang, Tsudik, Massey, Papadopoulos, et~al\mbox{.}}{Zhang
  et~al\mbox{.}}{2010}]%
        {zhang2010named}
\bibfield{author}{\bibinfo{person}{Lixia Zhang}, \bibinfo{person}{Deborah
  Estrin}, \bibinfo{person}{Jeffrey Burke}, \bibinfo{person}{Van Jacobson},
  \bibinfo{person}{James~D Thornton}, \bibinfo{person}{Diana~K Smetters},
  \bibinfo{person}{Beichuan Zhang}, \bibinfo{person}{Gene Tsudik},
  \bibinfo{person}{Dan Massey}, \bibinfo{person}{Christos Papadopoulos},
  {et~al\mbox{.}}} \bibinfo{year}{2010}\natexlab{}.
\newblock \showarticletitle{Named data networking (ndn) project}.
\newblock \bibinfo{journal}{\emph{Relat{\'o}rio T{\'e}cnico NDN-0001, Xerox
  Palo Alto Research Center-PARC}}  \bibinfo{volume}{157}
  (\bibinfo{year}{2010}), \bibinfo{pages}{158}.
\newblock


\bibitem[\protect\citeauthoryear{Zhang, Luo, and Zhang}{Zhang
  et~al\mbox{.}}{2015}]%
        {zhang2015survey}
\bibfield{author}{\bibinfo{person}{Meng Zhang}, \bibinfo{person}{Hongbin Luo},
  {and} \bibinfo{person}{Hongke Zhang}.} \bibinfo{year}{2015}\natexlab{}.
\newblock \showarticletitle{A survey of caching mechanisms in
  information-centric networking}.
\newblock \bibinfo{journal}{\emph{IEEE Communications Surveys \& Tutorials}}
  \bibinfo{volume}{17}, \bibinfo{number}{3} (\bibinfo{year}{2015}),
  \bibinfo{pages}{1473--1499}.
\newblock


\bibitem[\protect\citeauthoryear{Zhang and Li}{Zhang and Li}{2019}]%
        {zhang2019ari}
\bibfield{author}{\bibinfo{person}{Xin Zhang} {and} \bibinfo{person}{Ru Li}.}
  \bibinfo{year}{2019}\natexlab{}.
\newblock \showarticletitle{An ARI-HMM based Interest Flooding Attack
  countermeasure in NDN}. In \bibinfo{booktitle}{\emph{2019 IEEE 23rd
  International Conference on Computer Supported Cooperative Work in Design
  (CSCWD)}}. IEEE, \bibinfo{pages}{10--15}.
\newblock


\bibitem[\protect\citeauthoryear{Zhang, Afanasyev, Burke, and Zhang}{Zhang
  et~al\mbox{.}}{2016a}]%
        {zhang2016survey}
\bibfield{author}{\bibinfo{person}{Yu Zhang}, \bibinfo{person}{Alexander
  Afanasyev}, \bibinfo{person}{Jeff Burke}, {and} \bibinfo{person}{Lixia
  Zhang}.} \bibinfo{year}{2016}\natexlab{a}.
\newblock \showarticletitle{A survey of mobility support in named data
  networking}. In \bibinfo{booktitle}{\emph{2016 IEEE Conference on Computer
  Communications Workshops (INFOCOM WKSHPS)}}. IEEE, \bibinfo{pages}{83--88}.
\newblock


\bibitem[\protect\citeauthoryear{Zhang, Xia, Mastorakis, and Zhang}{Zhang
  et~al\mbox{.}}{2018b}]%
        {zhang2018kite}
\bibfield{author}{\bibinfo{person}{Yu Zhang}, \bibinfo{person}{Zhongda Xia},
  \bibinfo{person}{Spyridon Mastorakis}, {and} \bibinfo{person}{Lixia Zhang}.}
  \bibinfo{year}{2018}\natexlab{b}.
\newblock \showarticletitle{Kite: Producer mobility support in named data
  networking}. In \bibinfo{booktitle}{\emph{Proceedings of the 5th ACM
  Conference on Information-Centric Networking}}. \bibinfo{pages}{125--136}.
\newblock


\bibitem[\protect\citeauthoryear{Zhang, Vasavada, Osterweil, Zhang,
  et~al\mbox{.}}{Zhang et~al\mbox{.}}{2019}]%
        {zhang2019expect}
\bibfield{author}{\bibinfo{person}{Zhiyi Zhang}, \bibinfo{person}{Vishrant
  Vasavada}, \bibinfo{person}{Eric Osterweil}, \bibinfo{person}{Lixia Zhang},
  {et~al\mbox{.}}} \bibinfo{year}{2019}\natexlab{}.
\newblock \showarticletitle{Expect More from the Networking: DDoS Mitigation by
  FITT in Named Data Networking}.
\newblock \bibinfo{journal}{\emph{arXiv preprint arXiv:1902.09033}}
  (\bibinfo{year}{2019}).
\newblock


\bibitem[\protect\citeauthoryear{Zhang, Yu, Afanasyev, and Zhang}{Zhang
  et~al\mbox{.}}{2017}]%
        {zhang2017ndncert}
\bibfield{author}{\bibinfo{person}{Zhiyi Zhang}, \bibinfo{person}{Yingdi Yu},
  \bibinfo{person}{Alex Afanasyev}, {and} \bibinfo{person}{Lixia Zhang}.}
  \bibinfo{year}{2017}\natexlab{}.
\newblock \showarticletitle{NDN certificate management protocol (NDNCERT)}.
\newblock \bibinfo{journal}{\emph{NDN, Technical Report NDN-0050}}
  (\bibinfo{year}{2017}).
\newblock


\bibitem[\protect\citeauthoryear{Zhang, Yu, Ramani, Afanasyev, and Zhang}{Zhang
  et~al\mbox{.}}{2018c}]%
        {zhang2018nac}
\bibfield{author}{\bibinfo{person}{Zhiyi Zhang}, \bibinfo{person}{Yingdi Yu},
  \bibinfo{person}{Sanjeev~Kaushik Ramani}, \bibinfo{person}{Alex Afanasyev},
  {and} \bibinfo{person}{Lixia Zhang}.} \bibinfo{year}{2018}\natexlab{c}.
\newblock \showarticletitle{NAC: Automating access control via Named Data}. In
  \bibinfo{booktitle}{\emph{MILCOM 2018-2018 IEEE Military Communications
  Conference (MILCOM)}}. IEEE, \bibinfo{pages}{626--633}.
\newblock


\bibitem[\protect\citeauthoryear{Zhang, Yu, Zhang, Newberry, Mastorakis, Li,
  Afanasyev, and Zhang}{Zhang et~al\mbox{.}}{2018d}]%
        {zhang2018overview}
\bibfield{author}{\bibinfo{person}{Zhiyi Zhang}, \bibinfo{person}{Yingdi Yu},
  \bibinfo{person}{Haitao Zhang}, \bibinfo{person}{Eric Newberry},
  \bibinfo{person}{Spyridon Mastorakis}, \bibinfo{person}{Yanbiao Li},
  \bibinfo{person}{Alexander Afanasyev}, {and} \bibinfo{person}{Lixia Zhang}.}
  \bibinfo{year}{2018}\natexlab{d}.
\newblock \showarticletitle{An overview of security support in named data
  networking}.
\newblock \bibinfo{journal}{\emph{IEEE Communications Magazine}}
  \bibinfo{volume}{56}, \bibinfo{number}{11} (\bibinfo{year}{2018}),
  \bibinfo{pages}{62--68}.
\newblock


\bibitem[\protect\citeauthoryear{Zhi, Liu, Wang, and Zhang}{Zhi
  et~al\mbox{.}}{2019}]%
        {zhi2019resist}
\bibfield{author}{\bibinfo{person}{Ting Zhi}, \bibinfo{person}{Ying Liu},
  \bibinfo{person}{Jiushuang Wang}, {and} \bibinfo{person}{Hongke Zhang}.}
  \bibinfo{year}{2019}\natexlab{}.
\newblock \showarticletitle{Resist Interest Flooding Attacks via Entropy--SVM
  and Jensen--Shannon Divergence in Information-Centric Networking}.
\newblock \bibinfo{journal}{\emph{IEEE Systems Journal}}
  (\bibinfo{year}{2019}).
\newblock


\bibitem[\protect\citeauthoryear{Zhi, Liu, and Wu}{Zhi et~al\mbox{.}}{2020}]%
        {zhi2020reputation}
\bibfield{author}{\bibinfo{person}{Ting Zhi}, \bibinfo{person}{Ying Liu}, {and}
  \bibinfo{person}{Jun Wu}.} \bibinfo{year}{2020}\natexlab{}.
\newblock \showarticletitle{A Reputation Value-Based Early Detection Mechanism
  Against the Consumer-Provider Collusive Attack in Information-Centric IoT}.
\newblock \bibinfo{journal}{\emph{IEEE Access}}  \bibinfo{volume}{8}
  (\bibinfo{year}{2020}), \bibinfo{pages}{38262--38275}.
\newblock


\bibitem[\protect\citeauthoryear{Zhi, Luo, and Liu}{Zhi et~al\mbox{.}}{2018}]%
        {zhi2018gini}
\bibfield{author}{\bibinfo{person}{Ting Zhi}, \bibinfo{person}{Hongbin Luo},
  {and} \bibinfo{person}{Ying Liu}.} \bibinfo{year}{2018}\natexlab{}.
\newblock \showarticletitle{A Gini impurity-based interest flooding attack
  defence mechanism in NDN}.
\newblock \bibinfo{journal}{\emph{IEEE Communications Letters}}
  \bibinfo{volume}{22}, \bibinfo{number}{3} (\bibinfo{year}{2018}),
  \bibinfo{pages}{538--541}.
\newblock


\end{thebibliography}

\end{document}